\documentclass[twocolumn,aps,prb,showpacs,raggedbottom,nobalancelastpage,amssymb,superscriptaddress]{revtex4-1}

\usepackage{amsmath}
\usepackage{amssymb}
\usepackage{color}
\usepackage{graphicx}
\usepackage{MnSymbol}
\usepackage{dsfont}

\def\im{\mathrm{Im}}

\newcommand{\ket}[1]{\left|{#1}\right>}
\newcommand{\kett}[1]{|{#1}\rangle}

\def\pdagger{{\phantom{\dagger}}}

\newcommand{\rs}{\rm \scriptscriptstyle}
\newcommand{\s}{\scriptscriptstyle}
\newcommand{\D}{\mathrm{d}}
\newcommand{\F}{{\scriptscriptstyle\mathrm{F}}}
\newcommand{\C}{\mathrm{c}}

\newcommand{\rsL}{\rm \scriptscriptstyle L}
\newcommand{\rsR}{\rm \scriptscriptstyle R}
\newcommand{\Nup}{N^{\uparrow}}
\newcommand{\Ndo}{N^{\downarrow}}

\begin{document}

\title{Long-range spin-coherence in a strongly-coupled all electronic 
dot--cavity system}
\author{Michael Sven Ferguson}
\affiliation{Institute for Theoretical Physics, ETH Zurich, 
8093 Z{\"u}rich, Switzerland}
\author{David Oehri}
\affiliation{Institute for Theoretical Physics, ETH Zurich, 
8093 Z{\"u}rich, Switzerland}
\author{Clemens R{\"o}ssler}
\affiliation{Infineon Technologies Austria, Siemensstra{\ss}e 2, 9500 Villach, Austria}
\author{Thomas Ihn}
\affiliation{Solid State Physics Laboratory, ETH Zurich, 
8093 Z{\"u}rich, Switzerland}
\author{Klaus Ensslin}
\affiliation{Solid State Physics Laboratory, ETH Zurich, 
8093 Z{\"u}rich, Switzerland}
\author{Gianni Blatter}
\affiliation{Institute for Theoretical Physics, ETH Zurich, 
8093 Z{\"u}rich, Switzerland}
\author{Oded Zilberberg}
\affiliation{Institute for Theoretical Physics, ETH Zurich, 
8093 Z{\"u}rich, Switzerland}
\date{\today}
\begin{abstract}
We present a theoretical analysis of spin-coherent electronic transport across
a mesoscopic dot--cavity system. Such spin-coherent transport has been
recently demonstrated in an experiment with a dot--cavity hybrid implemented
in a high-mobility two-dimensional electron gas [C. R\"ossler et al., Phys.\ Rev.\ Lett.\ {\bf
115}, 166603 (2015)] and its spectroscopic signatures have been interpreted in
terms of a competition between Kondo-type dot-lead and molecular-type
dot--cavity singlet-formation. Our analysis brings forward all the transport
features observed in the experiments and supports the claim that a
spin-coherent molecular singlet forms across the full extent of the
dot--cavity device.  Our model analysis includes: (i) a single-particle
numerical investigation of the two-dimensional geometry, its quantum-coral--type eigenstates
and associated spectroscopic transport features, (ii) the derivation of an
effective interacting model based on the observations of the numerical and
experimental studies, and (iii) the prediction of transport characteristics
through the device using a combination of a master-equation approach on top of
exact eigenstates of the dot--cavity system, and an equation-of-motion
analysis that includes Kondo physics.  The latter provides additional
temperature scaling predictions for the many-body phase transition between
molecular- and Kondo-singlet formation and its associated transport
signatures.
\end{abstract}
%
%

%
\maketitle
%

\section{Introduction}
%
Mesoscopic physics provides a framework for the study of coherent transport
across engineered controllable systems.  A standard method for obtaining such
quantum coherent devices is by geometrically confining electrons to
effectively low-dimensional structures that are embedded within ultraclean
materials. Typical devices include quantum dots acting as effectively zero-dimensional (0D)
artificial atoms, one-dimensional (1D) quantum wires or quantum point contacts, and electronic
interferometers composed of edges of two-dimensional  (2D) quantum Hall
bars~\cite{ihn2010semiconductor}.  In such devices, electron-electron
interactions play an important role, giving way to numerous interesting
transport phenomena, such as the Coulomb
blockade\cite{kouwenhoven_few-electron_2001} and the Kondo
resonance\cite{goldhaber-gordon_kondo_1998, kouwenhoven_revival_2001} in
quantum dots, the 0.7 conductance plateau in quantum point contacts
~\cite{thomas:96,cronenwett_low-temperature_2002,hirose:03,meir-07}, and noise
signatures of fractionally charged particles in electronic
interferometers\cite{picciotto:97,saminadayar:97,dolev_observation_2008}.

Of particular interest in the present context are 2D coherent standing waves
(quantum corrals) that have led to the observation of fascinating signatures
of coherence and interaction. For example a QPC coupled to a mesoscopic,
\(\rm \mu\)m-size quantum corral displays a modulated tunneling
\cite{katine_point_1997, hersch_diffractive_1999}, while the observed Kondo
mirage \cite{crommie_imaging_1993, manoharan_quantum_2000,agam:01,fiete:03} is
the result of nm-scale coherence and interaction.  Such geometry-induced
complex many-body phenomena can be described theoretically.  When doing so, it
is important to account for both, the spatial structure of the electronic wave
functions imposed by the device geometry as well as interactions.
In search for new phenomena and applications, the coupling of various
mesoscopic devices has led to new implications on both, fundamental questions
in many-body physics, as well as novel quantum engineering prospects.
Examples of the former include the study of many-body quantum phase
transitions in the context of the Kondo effect and competing mechanisms, such
as Ruderman-Kittel-Kasuya-Yoshida (RKKY) interactions~\cite{craig_tunable_2004}, two-channel
Kondo~\cite{potok_observation_2007}, and singlet--triplet switching on a
molecule~\cite{roch_quantum_2008}. On the engineering front, the combination
of several dots into controlled quantum bits (qubits) has been
demonstrated~\cite{hayashi_coherent_2003, petta_coherent_2005}, thus promoting
the next challenge of introducing coherent coupling between distant qubits
without relying on nearest-neighbor exchange.
Recently, we have reported on the transport signatures of a coherent
electronic dot--cavity system in a high-mobility two-dimensional electron
gas~\cite{roessler:2015}. The role of the (\(\rm\mu\)m-size) cavity was played
by a carefully designed electron reservoir of suitable geometrical shape,
similar to that used in mesoscopic quantum
corrals\cite{katine_point_1997,hersch_diffractive_1999}.  The high quality of
the underlying material has allowed for the demonstration of a coherent
spin-singlet formation that spans across the dot and extended cavity states.
This strong hybridization between dot and cavity has quenched the competing
Kondo transport in a controlled way, thus allowing for a systematic tuning of
the device. Such strong hybridization typically occurs in quantum optics
between photons and atoms when the optical cavities have a high quality-factor~\cite{Haroche2013}; quite interestingly, we have seen it transpire
for an electronic cavity that has a mere quality factor of $\sim 5$. This
highlights the important difference between the electronic and optical
platforms: electrons are strongly interacting and the dot--cavity physics
takes place within a many-body interacting Fermi sea of electrons, whereas
photons are weakly interacting and optical cavities have isolated spectral
lines.
In this paper, we provide a detailed account of the theoretical modeling
invoked in the analysis of the dot--cavity experiments~\cite{roessler:2015}.
The work involves analytical and numerical studies of the 2D geometry that has
facilitated the design of an optimized quality-factor for the electronic
cavity. Feeding the spectral properties of the 2D structure into an effective
0D model(coupled to Fermi leads) allows us to introduce electron-electron interaction into the problem
and describe the dot--cavity hybrid as an original realization of a Kondo box
setup~\cite{thimm_kondo_1999,Cornaglia2002,dias_da_silva_zero-field_2006,
dias_da_silva_spin-polarized_2013}. We then employ several methods for the
prediction of transport signatures associated with the effective model: we
make use of a combination of exact-diagonalization-, master-equation-, and
equation-of-motion approaches in order to analyze the complex many-body
signatures observed in the experiment. Moreover, we present a detailed
comparison of our model's predictions with the reported experimental results,
as well as with previously unpublished experimental findings.  The agreement
is remarkable, both when comparing equilibrium and out-of-equilibrium
transport.
The paper is structured as follows: In Sec.~\ref{2dnum}, we analyze the
effects of the 2D geometry of our system on the single-particle transport
across the device. In Sec.~\ref{sec:eff}, we discuss how this geometrical
shaping can be accounted for within an effective model. The many-body
transport properties of this model are studied using exact diagonalization and
a master-equation approach in Sec.~\ref{sec:ed}. Finally, we discuss the
interplay between the dot--lead Kondo-physics and the dot--cavity spin-singlet
formation using an equation-of-motion approach in Sec.~\ref{sec:eom}, followed
by the conclusions and an outlook in Sec.~\ref{sec:con}.
%

\section{Cavity engineering}
\label{2dnum}

%
The interesting physics arising in the dot--cavity
experiment~\cite{roessler:2015} is the result of a deliberate structuring of
the two-dimensional electron gas (2DEG).  The geometrical confinement
generates modes which are related to the standing electron waves discussed in the
context of quantum corrals\cite{fiete:03}.  Here, we analyze the
single-particle effects of shaping the potential landscape. We do this
numerically by considering the transport through a quantum point contact
in the presence of an electronic cavity. We shall see that the (numerically calculated) local density of states (LDOS) of the device is related to the
eigenfunctions of the corresponding closed system, a
half-disk quantum box with hard walls. This relation provides us with a simple understanding of the observed features and allows us
to set up specific design rules for future devices. Additionally this analysis serves as the foundation for the
construction of the effective many-body Hamiltonian in Sec.~\ref{sec:eff} where the QPC will be replaced by a quantum
dot.
\begin{figure}
	\centering 
	\includegraphics[width=7.0cm]{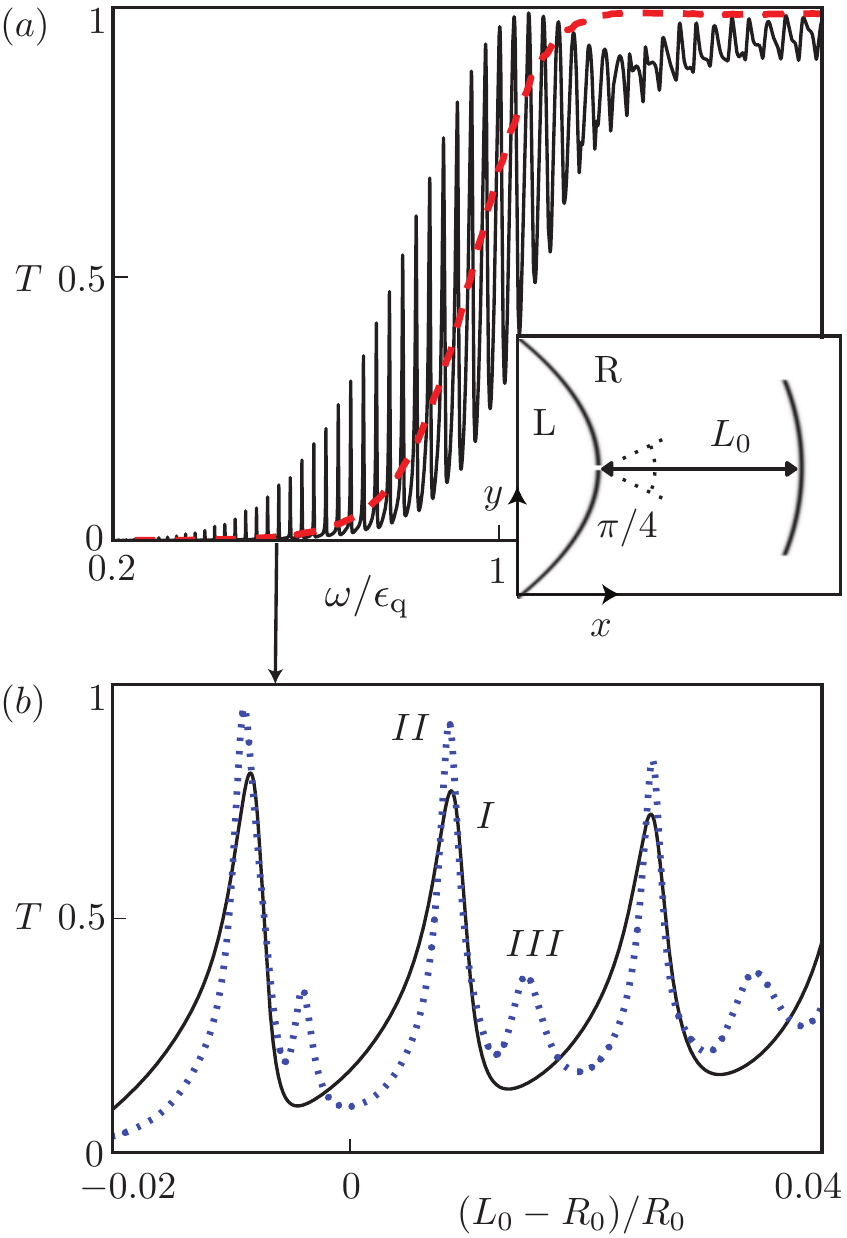} 
	\caption[]{\label{fig1}
		Transmission modulation of a QPC due to a structured lead. The
		inset shows the geometry of the QPC--cavity setup used in our
		numerical simulation: two parts of the 2DEG ($L$ and $R$) are
		separated by a potential barrier leaving only a narrow
		connecting channel that defines a QPC. At a distance $L_0$, we
		include a curved potential with radius $R_0$ and an opening
		angle $\pi/4$ (cavity mirror).  (a)  In the absence of the
		cavity mirror (dashed red line), the transmission \(T\) versus
		energy \(\omega\) shows the expected conductance quantization,
		with an arbitrary quantization energy $\epsilon_{\rm q}$. In
		the presence of the $\pi/4$ cavity with \(R_0 = L_0\) (solid
		line), the transmission is modulated by pronounced, well
		separated resonances.  (b) Transmission through the QPC at
		$\omega=0.54 \epsilon_\textrm{q}$ and fixed radius $R_0$ as a
		function of the cavity-mirror position \(L_0\). The results of
		two mirror opening-angles are presented, a narrow mirror with
		opening angle $\pi/4$ (solid line) and a wider mirror with
		opening angle $\pi/2$ (dotted blue line).  The former shows a
		regular pattern of isolated resonances, while the latter shows
		main resonances accompanied by additional side resonances. For
		a larger distance from the QPC, the resonances become broader
		and less transmitting due to additional scattering into the
		surrounding lead. I, II, and III denote peaks in transmission
		for which the scattering single-particle states are depicted
		in Figs.  \ref{fig2}(b)--(d), respectively.
	} %
\end{figure}
We consider a QPC connecting two extended leads, one of which is structured by
a mirror gate [see the inset of Fig.\ \ref{fig1}(a)]. Such a setup is known to
modulate the transmission through the QPC by forming ballistic resonator
modes\cite{katine_point_1997, hersch_diffractive_1999,roessler:2015}.  We
model this device using the numerical transport package KWANT\cite{kwant}.
This involves a discretization of the system using a fine square-lattice mesh.
To model the QPC, we separate the 2DEG into two parts along the \(y\)-axis,
denoted as left ($\mathrm{L}$) and right ($\mathrm{R}$), by a large on-site
potential leaving only a small channel connecting them. In the absence of the
mirror gate, the transport between the two separated 2DEG parts shows the
well-known conductance quantization steps~\cite{wees:88,wharam:88} as a
function of energy \(\omega\) [see the red dashed line in Fig.~\ref{fig1}(a)
following the first quantization step].  The curved mirror and QPC gates are
positioned symmetrically around the \(x\)-axis and define a half-circular
cavity for electrons which causes a pronounced oscillation in the transmission
through the QPC [see the solid line in Fig.~\ref{fig1}(a)].
In the experiment, a gate voltage was applied to the mirror gate that reduced
the length of the radial confinement of the cavity\cite{roessler:2015}.  We
model this by studying the transmission at a fixed energy $\omega$ as a function of
the distance of the mirror from the QPC [see Fig.~\ref{fig1}(b)]. We observe a
regular pattern of resonances, in accord with the experiment. Increasing the
opening angle from \(\pi/4\) to \(\pi/2\) shows additional resonances,
demonstrating that the optimal cavity has to be carefully tuned in order to
arrive at well-separated but still sharp transport resonances. To better
understand the properties of these resonances, we study the eigenstates in a
closed half-disk quantum box that has a shape similar to that of the cavity. Indeed, these
states are closely related to the resonances seen in an open microwave
billiard\cite{hersch_diffractive_1999}.
We start with the eigenstates of a circular box with radius $R_0$, and  work
in both Cartesian \(x\), \(y\) and circular \(r\), \(\varphi\) coordinates,
according to convenience. In circular coordinates, the Schr\"odinger equation
for free particles is
\begin{equation}
\frac{1}{r}\frac{\partial}{\partial r}
\left(r\frac{\partial \psi}{\partial r}\right)+
\frac{1}{r^2}\frac{\partial^2 \psi}{\partial \varphi^2}+k^2\psi=0.
\end{equation}
Adding a circular confining potential generates eigenstates of the form
$\psi_{n,\pm m}({\bf r})\propto J_{m}(k_{n}r)e^{\pm im\varphi}$ with the
cylindrical Bessel functions $J_{m}$ and $k_{n}$ the $n$-th solution of
$J_{m}(k R_0)=0$. The integers \(n\) and \(m\) are the radial and azimuthal
quantum numbers, respectively. The solutions with $\pm m$ are energetically
degenerate, such that arbitrary superpositions of them are solutions as well.
The wave functions must vanish at the QPC gates, which lie on the \(y\) axis.
For a given \(|m|>0\), we can create superpositions which fulfill this
criterion
\begin{equation}
\tilde \psi_{n,m}(\mathbf{r}) \propto J_m(k_nr)
\sin\left[m(\varphi+\pi/2)\right],
\label{eq:eigenstate}
\end{equation}
where we take \(\varphi=0\) to lie on the \(x\)-axis.
These specific superpositions are eigenstates of the half circular disk
defined on \(-\pi/2<\varphi<\pi/2\). A selection of such states is shown in
Fig.\ \ref{fig2}(a).
 \begin{figure}
 	\centering
 	\includegraphics[width=7.0cm]{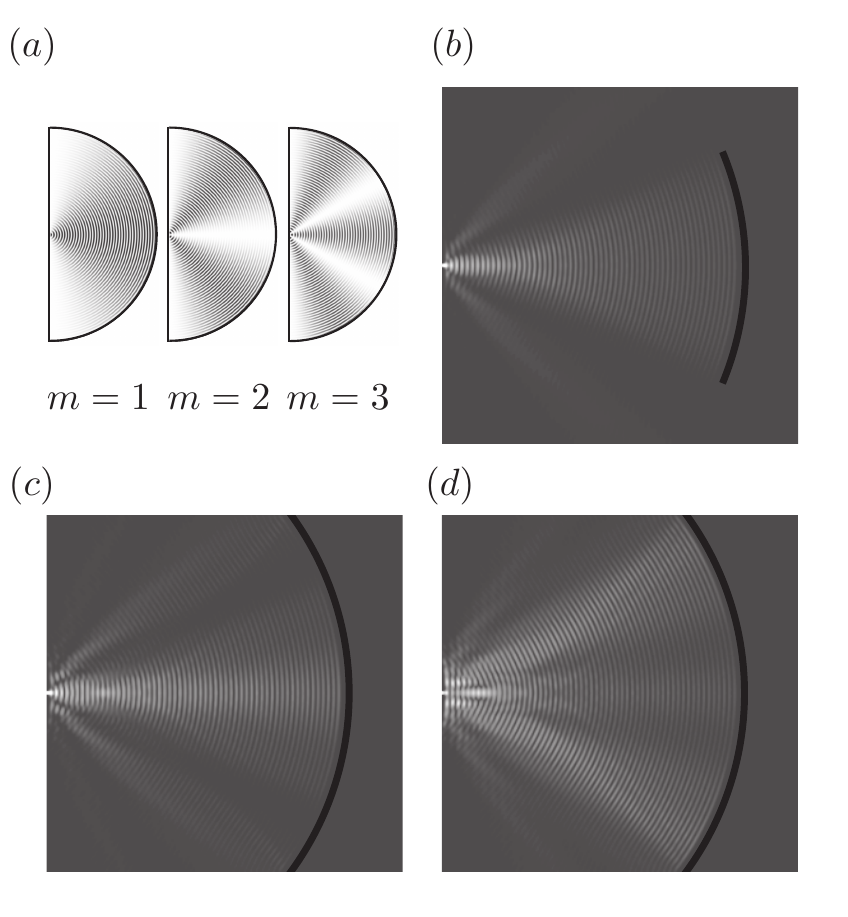}
	\caption[]{\label{fig2}
		Resonances of closed and open half-disks.
		(a) Eigenstates of a closed half-circular quantum box. Here
		the modes with \(n=40\) and \(m=1,2,3\) are plotted [see Eq.
		\eqref{eq:eigenstate}]. For clarity, we show the local density
		of states of the modes multiplied by the radius $r$,
		$r|\tilde{\psi}_{40,m}(\mathbf{r})|^2$.  (b)--(d) The local
		density of states of a plane wave scattering through the QPC
		into the structured lead as obtained using the numerical
		transport package KWANT\cite{kwant}. In (b) and (c), we show
		the modes that lead to the pronounced resonances I and II in
		Fig.~\ref{fig1}(b) of the narrow and wide mirrors,
		respectively. These modes both have angular quantum number
		\(m=1\) and correspond to the same state in a closed system.
		Their respective resonances differ in height and width because
		the narrow mirror is less effective in confining the mode. In
		(d), we depict the LDOS corresponding to the side resonance III
		in Fig.~\ref{fig1}(b) of the wide mirror with angular quantum
		number \(m=3\).
	}
 \end{figure}
We now consider how a wave incoming from the left of the QPC will tunnel into
cavity states on the right hand side. First, we establish a selection rule on
\(m\) due to the parity of the initial and final wave functions. The incoming
wave \(e^{ikx}\) is even under the sign change \(y \to -y\) and hence scatters
into states \eqref{eq:eigenstate} with odd \(m\). Second, due to the effective
potential of high angular-momentum modes, the local density of states (LDOS)
of the cavity eigenstates at the QPC is suppressed for large azimuthal quantum
numbers \(m\) and thus they couple less strongly to the incoming wave. Hence,
promising solutions that strongly couple across the QPC are modes with small
odd values of \(m\).
Inspiration for optimizing the cavity geometry to produce a strong and
coherent dot--cavity coupling can be drawn from quantum electrodynamical (QED)
setups\cite{Haroche2013}. We define the \(Q\) factor of our cavity to
be the ratio of the peak-to-peak distance and the full width at half maximum
(inverse lifetime) of the peaks. Reaching the strong coupling limit between a
QED cavity photon and an atom requires a high \(Q\)-factor. In the electronic
system, the cavity electron strongly interacts with the electrons on the
quantum dot (artificial atom) and a moderate \(Q\)-factor is sufficient.  We
can maximise the \(Q\)-factor by optimizing the cavity geometry and applying
the quantum engineering insights obtained above. 

Our tuning parameters for the \(Q\)-factor are then the lifetime of the states
in the cavity and their distance in energy space. The lifetime highly depends on how
well-confined the states are and hence on the opening angle of the cavity
mirror. If a large fraction of the weight of a state is located in a region
which is not confined by the mirror gate, it will leak out very fast and thus
the lifetime will be short, leading to broad resonances when the opening angle
of the cavity is small. On the other hand, a small cavity opening leads to the
disappearance of high-\(m\) modes and thus the peak-to-peak distance
increases.  These features are illustrated in Fig.\ \ref{fig1}(b), where we
see that the side peak with \(m=3\) broadens and disappears when the cavity
mirror is narrowed. Selected states corresponding to the peaks in transmission
in Fig.\ \ref{fig1}(b) are pictured in Figs.\ \ref{fig2}(b)--(d). 

Thus, decreasing the opening angle of the cavity has two competing implications
for \(Q\): (i) high-\(m\) modes are not confined and therefore side-peaks
vanish such that \(Q\) increases, and (ii) the main \(m=1\) mode broadens,
decreasing the lifetime and thus \(Q\). We find a cavity opening angle
\(\pi/4\) to be a good compromise between isolated and sharp resonances.  A
further analysis of the quality of the cavity depends on the exact shape of
the gates, their relative sizes, and its robustness to disorder and is beyond
the scope of this work. 

In this section, we investigated the single-particle properties of the cavity.
We found that when tuned correctly, the cavity acts as an effectively
one-dimensional box coupled to a reservoir. We also showed that we can use
analytic tools to provide design guidelines, and that KWANT~\cite{kwant} can
function as a low-cost test bed for the design of future devices. In
Sec.~\ref{sec:eff}, we will combine these results with the interacting quantum
dot to create effective models for the entire device.
%

\section{Effective models}\label{sec:eff}

%
Next, we set up an effective model for an interacting dot--cavity system,
where we replace the QPC by an interacting quantum dot including its spin
degree degree of freedom and include cavity states as discussed above.  Given
the large extent of the cavity, we assume that cavity charging effects are
screened and thus can be ignored. The cavity modes are then assumed to solely
affect the tunneling amplitude from the dot into a `structured' lead.
 \begin{figure}
 \centering
  \includegraphics{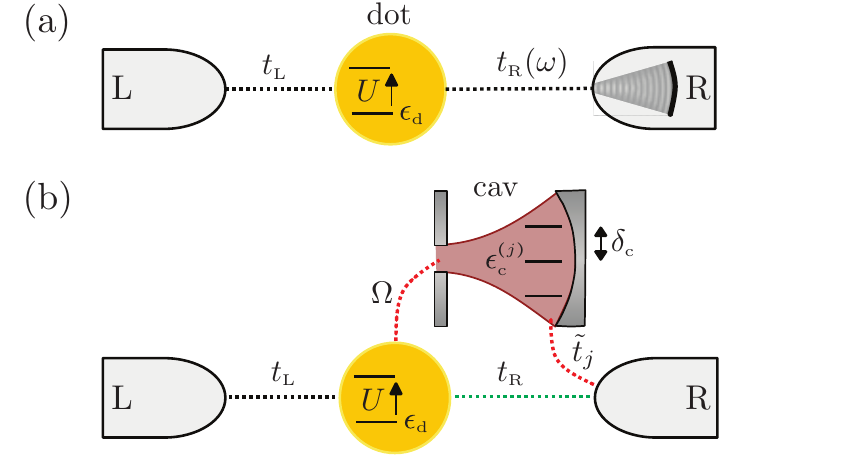}
 \caption[]{\label{models} Effective models describing the interacting
 dot--cavity system.  (a) Anderson-type model where the effect of the
 structured lead is accounted for by an energy-dependent transmission
 coefficient $t_{\rm \scriptscriptstyle R}(\omega)$ [see Eq.~\eqref{eq:deft}].
 (b) Dot--cavity model where the effect of the structured lead is accounted
 for by discrete cavity states separately coupled to the dot and to the right
 lead.  Closed paths connecting the dot via the cavity to the right lead (red
 dotted lines) and back to the dot (green dotted) might generate Fano
 interference, however, the summation over such trajectories (see Fig.\
 \ref{models2}) will suppress this effect.}
 \end{figure}

\textit{Our first model Hamiltonian} describes the quantum dot in terms of an
Anderson model with a standard (unstructured) coupling to the left lead and
includes the cavity in terms of a structured right lead with an
energy-dependent transmission coefficient [see Fig.\ \ref{models}(a)],
\begin{align}\label{eq:ham}
	H=H_{\mathrm{leads}}+H_{\mathrm{dot}}+H_{\mathrm{tun}},
\end{align}
where 
\begin{align}
H_{\mathrm{leads}}
=\sum_{k,\sigma} \epsilon_{{\rsL} k} c_{{\rsL} k\sigma}^\dagger 
c_{{\rsL} k\sigma}^\pdagger
+\sum_{k,\sigma} \epsilon_{{\rsR} k} c_{{\rsR} k\sigma}^\dagger 
c_{{\rsR} k\sigma}^\pdagger
\end{align}
describes the left and right leads with creation and annihilation operators of
lead states $c_{\mathrm{a} k\sigma}^\dagger$ and $c_{\mathrm{a}
k\sigma}^\pdagger$ with energy \(\epsilon_{\mathrm{a}k}\) and where
\(\sigma\) and $k$ denote spin and momenta of the states in the left
(\(\mathrm{a} = \mathrm{L}\)) and right (\(\mathrm{a} =\mathrm{R}\)) leads. We
consider a dot with a single spin-degenerate level as described by an
Anderson model~\cite{anderson:61}
\begin{align}
H_{\mathrm{dot}}
=\sum_{\sigma}\epsilon_{\D}\, d_{\sigma}^\dagger d_{\sigma}^\pdagger
+U n_{\uparrow}n_{\downarrow},
\label{Eq:dotH}
\end{align}
with creation and annihilation operators $d_\sigma^\dagger$ and
$d_\sigma^\pdagger$ of the dot level with energy $\epsilon_\D$ and spin
$\sigma$, $n_{\sigma}=d^\dagger_{\sigma}d^\pdagger_{\sigma}$, and the onsite  Coulomb
interaction is denoted by $U$.  The coupling between the leads and the dot is
described by the tunneling Hamiltonian
\begin{align}\label{eq:Htunor}
	H_{\mathrm{tun}}=\sum_{k,\sigma} t_{\rsL} d_{\sigma}^\dagger 
	c_{{\rsL} k\sigma}^\pdagger 
	+\sum_{k,\sigma} t_{\rsR}(\omega_k)\, d_{\sigma}^\dagger  
	c_{{\rsR} k\sigma}^\pdagger +\mathrm{H.c.},
\end{align}
where we assume a constant tunneling amplitude $t_{\rsL}$ between the left
lead and the dot level and an \emph{energy-dependent} tunneling amplitude
$t_{\rsR}(\omega)$ between the dot level and the right lead state with energy
$\omega$. The energies \(\omega_k\) of the lead states are related to their
momenta \(\hbar k\) through the density of states
\(\rho=\mathrm{d}k/\mathrm{d}\omega_k\). The energy-dependence of the
tunneling amplitude to the right lead is given by the specific shape of the
resonances in Fig.\ \ref{fig1}(a). This shape is well approximated by
separated Lorentzian peaks on top of a constant background,
\begin{align}
	t_{\rs R}(\omega)
	=t_{\rsR} 
	+ \sum_j \frac{\lambda_{j}}{\omega-\epsilon_{\C}^{\s (j)}
        +i\Gamma_j/2}.
	\label{eq:deft}
\end{align}
Here, the amplitude $t_{\rsR}$ describes the direct transmission from the dot
to the right lead and the sum over Lorentzians accounts for the transmission
into the right lead via the cavity states. The energy of the cavity resonances
is given by $\epsilon_{\C}^{\s (j)}$ while $\Gamma_{j}$ and $\lambda_{j}$
describe their width and coupling strength, respectively. Note that the analytic
expression in Eq.~\eqref{eq:deft} incorporates the effect of the cavity modes
as resonances in the transmission coefficient.  In doing so, we account for
the cavity coherence in the right lead via a (large) finite lifetime
$\hbar/\Gamma_j$, i.e., narrow resonances.

\textit{Second effective model.} On the other hand, we can go one step further
and describe these cavity states as discrete levels that are tunnel coupled to
the dot and are broadened by tunnel coupling to the background states of the
right lead. This setting is described by the Hamiltonian
\begin{align}
	\label{eq:effHam}
	\bar{H}=H_{\mathrm{leads}}+H_{\mathrm{dot}}+H_{\rm cav}
               +H_{\rm coupl}+\bar{H}_{\rm tun},
\end{align}
where the cavity Hamiltonian
\begin{align}
	H_{\mathrm{cav}}=\sum_{\sigma,j} \epsilon_{\C}^{\s (j)} 
			 f_{j\sigma}^\dagger f_{j\sigma}^\pdagger
			\label{Eq:cav}
\end{align}
describes the discrete cavity levels $\epsilon_{\C}^{\s (j)}$ with creation and 
annihilation operators $f_{j\sigma}^\dagger$ and $f_{j\sigma}^\pdagger$, the 
coupling Hamiltonian 
\begin{align}
	H_{\mathrm{coupl}}&=\sum_{j,\sigma} \Omega_j  f_{j\sigma}^\dagger 
						d_{\sigma}^\pdagger
			    +\mathrm{H.c.}
					\label{Eq:coupl}
\end{align}
accounts for the coupling between the dot and cavity states, and the modified 
tunneling Hamiltonian 
\begin{align}
	\bar{H}_{\mathrm{tun}}&= 
	H_{\mathrm{tun}}^{\s\D\rsL}+\bar{H}_{\mathrm{tun}}^{\s\D\rsR}
	+\bar{H}_{\mathrm{tun}}^{\s\C\rsR}
	\nonumber\\
	&=\sum_{k,\sigma} (t_{\rsL} d_{\sigma}^\dagger 
	c_{{\rsL} k\sigma}^\pdagger +\mathrm{H.c.})
       +\sum_{k,\sigma} (t_{\rsR} d_{\sigma}^\dagger c_{{\rsR} k\sigma}^\pdagger
			 +\mathrm{H.c.})
      	\nonumber\\
	&\quad+\sum_{j,k,\sigma} (t_j f_{j\sigma}^\dagger 
	c_{{\rsR} k\sigma}^\pdagger +\mathrm{H.c.})
				\label{eq:Htun}
\end{align}
describes the tunneling between both unstructured leads and the dot as well
as the coupling of the cavity to the unstructured right lead.  The tunneling
amplitudes $t_{j}$ that describe the coupling of the cavity levels to
the right lead are related to the cavity resonance widths
by~\cite{bruus-flensberg} $\Gamma_{j}=2\pi \rho_{\rsR} |t_{j}|^2$,
with $\rho_{\rsR}$ the density of states in the right lead.  The hybridization
amplitude $\Omega_j = \lambda_j/t_j$ between the dot and the cavity is
related to the strength $\lambda_j$ of the resonance and its
width\cite{bruus-flensberg}. Similarly, the amplitudes $t_{\rm a}$ that couple
the dot to the left ($\mathrm{a} = \mathrm{L}$) and right
($\mathrm{a}=\mathrm{R}$) leads give rise to rates
\begin{equation}
   \Gamma_{{\rm a}}=2\pi \rho_{\rm a} \left|t_{\rm a}\right|^2,
\label{eq:dotRate}
\end{equation}
that broaden the dot level. 

In the following, we will consider a fixed cavity level spacing $\delta_{\C}$,
i.e., $\epsilon_{\C}^{\s (j)}= \epsilon_{\C}+j \> \delta_{\C}$, and
additionally assume that all cavity levels are tunnel coupled to the dot and
leads with the same amplitudes $\Omega_j=\Omega$ and $t_j =
t_\C$, respectively.  Correspondingly, the rates coupling the
cavity levels to the right lead are identical, \(\Gamma_{j} = \Gamma_{{\C}}\)
with 
\begin{equation}
   \Gamma_{{\C}} = 2\pi \rho_{\rsR} |t_{\C}|^2.
\label{eq:cavRate}
\end{equation}
As a result, the dot--lead Hamiltonian corresponds to a standard
Anderson-model type description of an interacting dot with tunneling into
unstructured leads\cite{anderson:61} and an additional energy-dependent
channel due to tunneling via the cavity [see Fig.\ \ref{models}(b)].  Note that
obtaining the Hamiltonian $H$ from $\bar H$ corresponds to tracing out the
effect of the cavity. Here, we have performed the opposite procedure and
``gave birth'' to the coherent electronic cavity physics. 

\begin{figure}
        \centering
        \includegraphics[width=7.6cm]{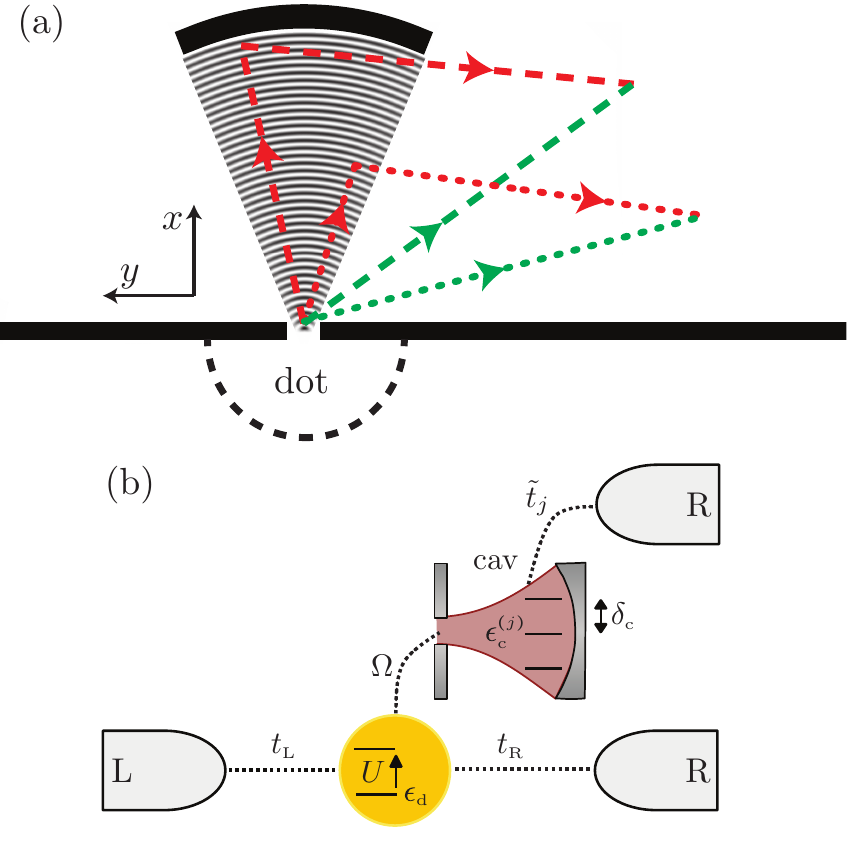}
	\caption[]{\label{models2}
		Suppression of Fano interference.  (a) Two examples of
		interfering paths from the dot to a structured lead.  Each set
		of paths (dashed and dotted) shows interferences between a
		trajectory going directly from the dot into the lead (green
		lines) and a trajectory connecting to the lead via the cavity
		(red lines).  All such trajectories starting and ending in the
		same points must be averaged over. This turns into a double
		averaging, first over the endpoint (red-green vertex)
		including the 2D extent of the lead and second over the
		(red-red) cavity vertices within the cavity area.  This
		summation causes the phase to average out and eliminates any
		Fano effect in the dot-cavity device, in agreement with the
		experiment\cite{roessler:2015} where no such interference
		effects could be observed.  (b) Effective 0D model (coupled to Fermi leads) accounting
		properly for the elimination of Fano-resonances by
		independently coupling the dot and cavity to the right lead.
		Note the differences to Fig.~\ref{models}(b) with the
		red--green loop connecting the dot, cavity, and reservoir; the
		latter is now replaced by two independent reservoirs.
	}
\end{figure}
\textit{No Fano-interference.} In general in model~\eqref{eq:effHam}, both
$t_{\rsR}$ and $\Omega_{j}$ may have an energy-dependent relative phase, which
could give rise to Fano-type interferences\cite{fano:61}. Recalling our 2D
geometry, we argue that the direct transmission into the right lead and the
transmission via the cavity into the right lead are phase-averaged [see
Fig.~\ref{models2}(a)]. As a result, we conclude that these two processes do
not interfere and that the Fano-effect is suppressed. We therefore will make
sure in the following to sum these terms incoherently by creating a second
copy of the right lead [see Fig.~\ref{models2}(b)] and thereby neglect
Fano-type processes, in contrast to previous Kondo-box analyses
\cite{dias_da_silva_zero-field_2006,dias_da_silva_spin-polarized_2013}.
In the following we will use the model in Fig.\ \ref{models2}(b) in Secs.
\ref{sec:ed} and \ref{master}. Within our discussion of the Kondo physics in
Sec. \ref{sec:eom}, the two models are equivalent (as we drop Fano-type
interference effects) and we will use the structured-lead formulation of
Eq.~\eqref{eq:Htunor} and Fig.~\ref{models}(a).
\begin{figure}[h]
        \centering
        \includegraphics{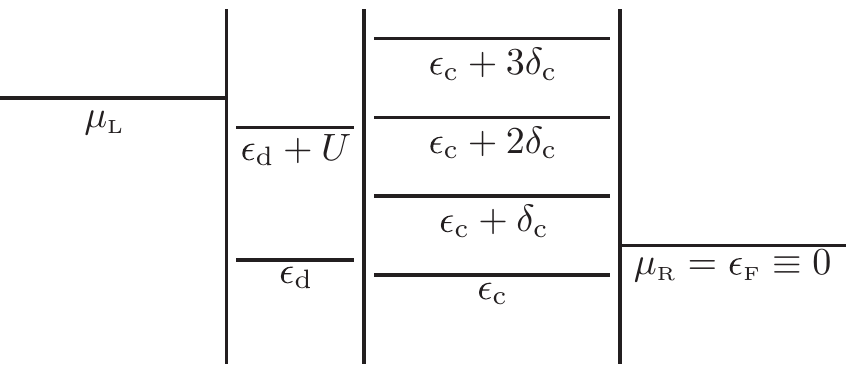}
	\caption[]{\label{fig:params} Schematic view of the different tuning
	parameters in the dot--cavity system. We will investigate the effects
	of tuning the left chemical potential \(\mu_\mathrm{\rsL}\), the dot
	level \(\epsilon_\D\), and the lowest cavity level \(\epsilon_\C\),
	keeping the chemical potential \(\mu_{\rsR}\) in the right lead fixed.
	We use \(\mu_{\rsR} =0 \) as our zero of energy.}
\end{figure}

\textit{Transport configurations and capacitive cross talk.} When analyzing
the many-body properties of the Hamiltonian $\bar{H}$ in Eq.\
(\ref{eq:effHam}), we will study typical configurations as sketched in Fig.\
\ref{fig:params}, with a fixed chemical potential \(\mu_{\rsR} =
\epsilon_{\scriptscriptstyle \mathrm{F}}\) on the right lead which we take as
our zero of energy, \(\mu_{\rsR} = 0\). Transport across the device then is
studied by changing the left chemical potential \(\mu_{\rsL}\) and tuning the
dot \(\epsilon_\D\) and cavity \(\epsilon_\C\) levels. In addition, we assume
a capacitive cross talk in the system between the dot and cavity such that
\begin{align} \label{eq:ed}
	\epsilon_\mathrm{d}
	&=\epsilon_\mathrm{d}'+\alpha_{\C \D}\epsilon_\mathrm{c}' 
	+\alpha_{{\rsL} \D}\mu_\mathrm{\scriptscriptstyle L},\\ 
	\label{eq:ec}
	\epsilon_\mathrm{c} 
	&=\epsilon_\mathrm{c}'+ \alpha_{\D \C}\epsilon_\mathrm{d}',
\end{align}
where primed quantities \(\epsilon_\mathrm{d}'\) and \(\epsilon_\mathrm{c}'\)
refer to applied gate voltages on the dot and cavity, respectively, producing
the dot and cavity levels \(\epsilon_\mathrm{d}\) and \(\epsilon_\mathrm{c}\)
in the Hamiltonian.  The cross talks modify the diagrams through a global tilt
and stretch; we seek only qualitatively correct values for the capacitive
couplings and therefore use \(\alpha_{\C \D} = 4\alpha_{\D \C} = 0.2\) and
\(\alpha_{{\rsL} \D} = 0.5\), while quantitative values can be easily
extracted from a proper calibration of the experiment.
%

\section{Artificial dot--cavity molecule -- an exact diagonalization treatment} \label{sec:ed}

We study the many-body spinful interacting dot--cavity system coupled to leads
as derived in Sec.~\ref{sec:eff}. Assuming that the coupling to the leads is
weak relative to the dot--cavity coupling, we first analyze the (isolated)
central region using exact diagonalization (ED) (see Sec.~\ref{ED}), and find
the emergent ``artificial molecule" ground-state map describing the isolated
dot--cavity system. We find how this map changes as a function of various
system parameters and compare it with a standard double-dot picture.  We then
couple the leads perturbatively to the dot--cavity artificial molecule and
determine the linear transport response across the device using a
master-equation approach (see Sec.~\ref{master}). Signatures of Kondo transport
are not captured by this formalism as we restrict the discussion to the
lowest-order sequential transport. We will investigate Kondo physics in
Sec.~\ref{sec:eom} using an equation-of-motion approach.
%

\subsection{Exact diagonalization: Ground state map}
\label{ED}

The subsystem consisting of the dot and cavity is described by the Hamiltonian 
\begin{align} \label{eq:molecule} 
   H_{\rm dc}=H_{\rm dot} +H_{\rm cav}+H_{\rm coupl} 
\end{align}
[see Eqs.~\eqref{Eq:dotH}, \eqref{Eq:cav}, and \eqref{Eq:coupl}]. Using exact
diagonalization, we determine the eigenstates and eigenenergies of this
isolated dot--cavity system. Specifically, for a fixed number $\Nup$ of
electrons with spin-$\uparrow$ and $\Ndo$ electrons with spin-$\downarrow$, we
obtain the eigenstates $\kett{\psi^{\alpha}_{\Nup,\Ndo}}$ with the
corresponding energies $\epsilon^{\alpha}_{\Nup,\Ndo}$, where $\alpha$ labels
all eigenstates of this specific Fock space. These eigenstates can be written
as a normalized superposition of the many-body occupation basis
\(\ket{n_{\D}^{\uparrow},n_{\D}^{\downarrow}, m_{0}^\uparrow,
m_{0}^\downarrow,m_{1}^\uparrow,m_{1}^\downarrow,\dots}\),
\begin{align}
\ket{\psi^{\alpha}_{\Nup,\Ndo}} 
&= \sum_{n_{\D}^{\sigma} + \sum_{j} m_{j}^{\sigma}=N^{\sigma}}
\mathcal{C}^{\alpha}_{n_{\D}^{\uparrow},n_{\D}^{\downarrow}, m_{0}^\uparrow,
m_{0}^\downarrow,m_{1}^\uparrow, m_{1}^\downarrow,\dots}\\
&\hspace{50pt}\times\ket{n_{\D}^{\uparrow},n_{\D}^{\downarrow}, m_{0}^\uparrow,
m_{0}^\downarrow,m_{1}^\uparrow, m_{1}^\downarrow, \dots},\nonumber
\end{align} 
where $n_{\D}^{\sigma}= 0,1$ and $m_{j}^{\sigma}=0,1$ are the occupation
numbers of the dot and $j^{\rm th}$ cavity states with spin
$\sigma=\uparrow,\downarrow$, respectively.  We denote the lowest energy
eigenstate of each Fock sector with $\alpha=1$, i.e., $|\psi^{\s
1}_{\Nup,\Ndo}\rangle$ is the state with $\epsilon^{\s 1}_{\Nup,\Ndo}=
\min_{\alpha}\epsilon^{\alpha}_{\Nup,\Ndo}$, while the remaining states
$\alpha > 1$ correspond to excited states.
When the system is coupled to a reservoir, the electron number is not fixed.
Assuming that the system is weakly tunnel coupled to leads,
$\Gamma_{\rsL},\Gamma_{\rsR},\Gamma_{\C}\ll \delta_{\C},U$, the leads populate
the system in the ground state with $\Nup$ and $\Ndo$ electrons. We can then
go over to the addition spectrum representation~\cite{alhassidreview} with
energies \(\tilde {\epsilon}^{\scriptscriptstyle
(j)}_\C=\epsilon^{\scriptscriptstyle (j)}_\C - \epsilon_\F \) and
\(\tilde{\epsilon}_\D=\epsilon_\D - \epsilon_\F \) defined with respect to the
chemical potential of the reservoir; with our choice \(\epsilon_\F =0 \), we
have \(\tilde{\epsilon}^{\scriptscriptstyle
(j)}_\C=\epsilon^{\scriptscriptstyle (j)}_\C  \) and \(\tilde{\epsilon}_\D =
\epsilon_\D \) and we drop the tilde in the following.  The ground state then
is given by the state with the lowest energy \(\epsilon^{\s 1}_{\Nup,\Ndo}\)
as a function of filling $\Nup$ and $\Ndo$.  At vanishing bias \( \mu_{\rsL} =
0\), the open system will conduct when two neighbouring Fock sectors have
degenerate ground states that are also the ground states of the entire system.
In such a configuration, an electron can be added to or removed from the
artificial molecule through an energy conserving process and hence the total
particle number on the artificial molecule remains undetermined.
 \begin{figure}[h]
        \centering
        \includegraphics[width=7cm]{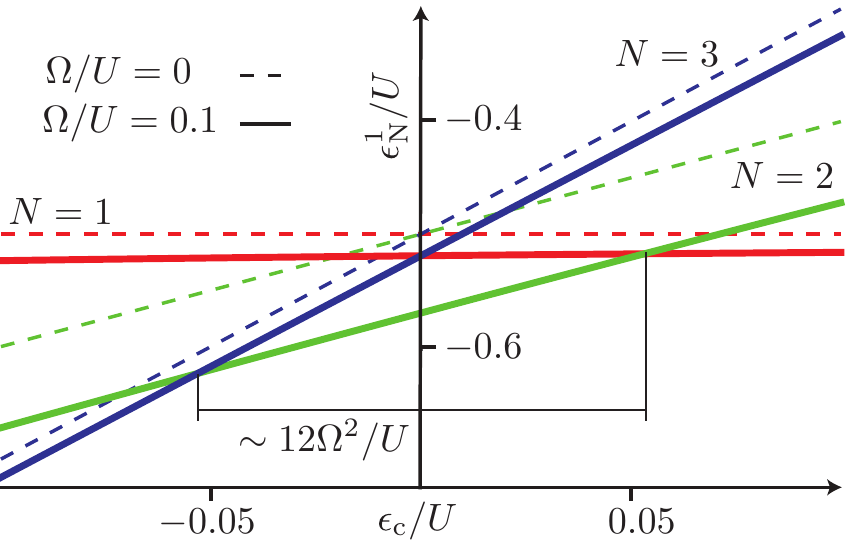}
	\caption[]{\label{addition}
		Addition spectrum calculated through exact diagonalization.
		The ground state energy of different Fock sectors (\(N=1\),
		red; \(N=2\), green; \(N=3\), blue) are plotted as a function
		of the cavity level while the dot is in Coulomb blockade,
		\(\epsilon_\D=-U/2\). All other Fock sectors are lifted
		further up in energy for this configuration. For vanishing
		temperature \(T\to 0\) and infinitesimal coupling to the
		leads, the energetically lowest Fock sector is the one which
		will be occupied.  Pushing the cavity level \(\epsilon_\C\)
		across the Fermi level \(\epsilon_\F\) from left to right
		entails its emptying \( N = 3 \to 2 \to 1 \). The dashed lines
		describe the situation for a vanishing dot--cavity hopping
		$\Omega$ and show that the cavity occupation changes by two
		particles (between red and blue) when tuning the cavity below
		the Fermi level, as illustrated by the crossing of red, green,
		and blue dashed lines in a single point. When the matrix
		element $\Omega$ is finite (solid lines) the ground state of
		the two-particle Fock sector is, to leading order in
		\(\Omega\), lowered by \(8\Omega^2/U\), while the odd sectors
		are lowered by \(2 \Omega^2/U\), leading to a splitting of
		\(12 \Omega^2/U\) of the two cavity levels. As a result, a
		$N=2$ dot--cavity ground-state singlet is formed at
		intermediate values \(|\epsilon_\C| \lesssim 0.05 \, U\)
		(lowest-energy green solid line) (see Appendix~\ref{app:ed})
		for more details.
	}
 \end{figure}
We can study the ground state energies of different Fock sectors as a function
of various system parameters: (i) the dot level $\epsilon_\D$, (ii) the on-site
interaction on the dot $U$, (iii) the lowest energy level of the cavity
$\epsilon_{\C}$, (iv) the cavity level spacing $\delta_{\C}$, and v) the
dot--cavity tunnel coupling $\Omega$. We focus on the experimental
situation~\cite{roessler:2015}, where the cavity level-spacing
$\delta_{\C}\sim U\gg \Omega$. In this regime, the impact of the cavity is
mostly due to a single one of its levels that is close to the chemical
potential. In Fig.~\ref{addition}, we explore the dot particle--hole symmetric
point \(\epsilon_\D= -U/2\) and find that the degeneracy of the cavity level
is lifted by a \(12\, \Omega^2 / U\) splitting (see Appendix~\ref{app:ed}).  In
this parameter range, both the dot and the cavity can be occupied by a single
electron each which will combine to form a singlet. 

\begin{figure*}
	\centering
	\includegraphics{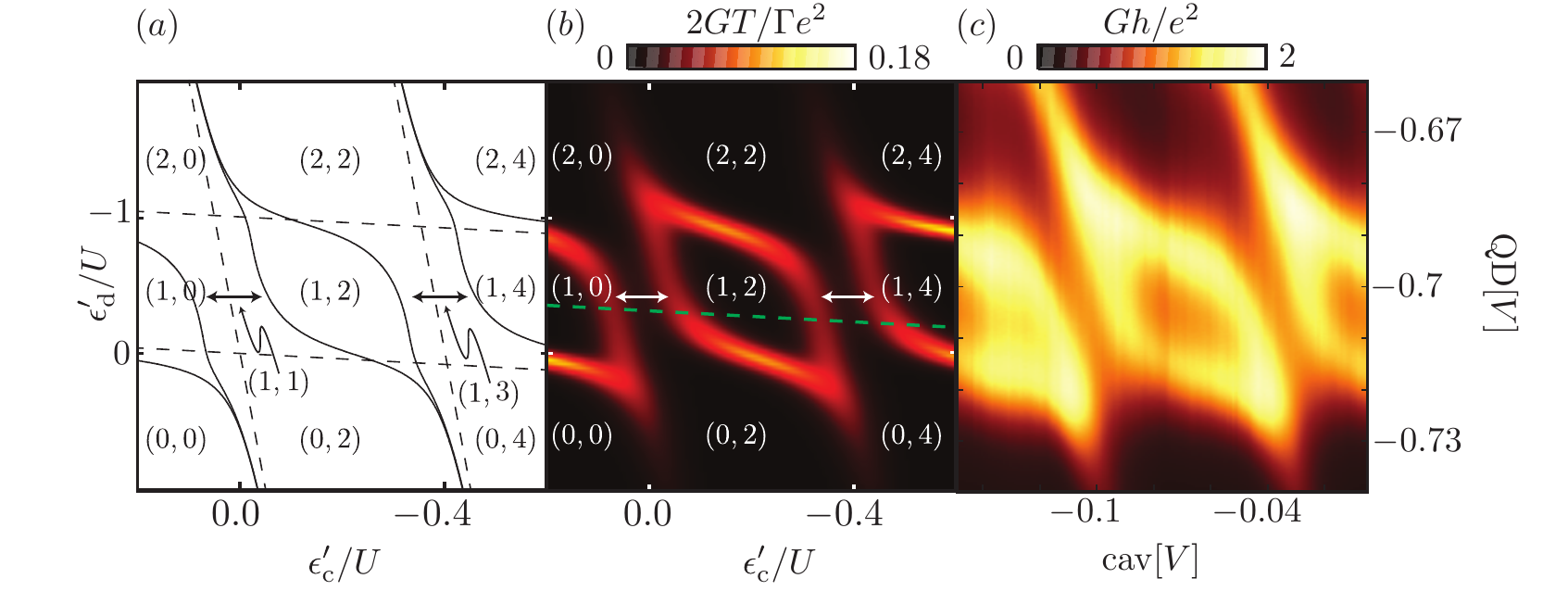}
	\caption[]{\label{groundMap}
		Ground state maps for a coupled dot--cavity system showing
		hybridization, (a) from exact diagonalization (see
		Sec.~\ref{sec:ed}),  (b) from a master-equation
		transport-analysis (see Sec.~\ref{master}), and (c) from
		transport experiments (see Ref.~[\onlinecite{roessler:2015}])
		[theoretical results in (a) and (b) assume a capacitive cross
		talk between the dot and cavity, see Eqs.~\eqref{eq:ed}
		and~\eqref{eq:ec}].  Note the different scales measuring
		\(\epsilon_\D'/U\) and \(\epsilon_\C'/U\), as the cavity levels are more dense than the dot’s Coulomb resonances.  (a) Ground state
		map versus dot and cavity energies as obtained from ED. Lines
		indicate parameter settings where different particle
		occupations become degenerate, thus allowing for transport
		across the artificial molecule.  Regimes with unique
		occupation are labeled with their dot, cavity occupation
		$(N_{\rm dot}, N_{\rm cav})$.  Dashed lines refer to the
		decoupled case with $\Omega/U=0$ and full lines correspond to
		$\Omega/U=0.12$.  A finite dot--cavity coupling $\Omega$ leads
		to a gap opening indicated by black arrows (of minimal width
		\(12\, \Omega^2/U\)) at \(\epsilon_\D'/U = -0.5\); the gap
		opens up a region of odd cavity-occupation with a
		molecular-singlet formation. (b) Differential conductance
		$G=\mathrm{d} I/\mathrm{d} V$ versus dot and cavity energies
		for the same parameters as in (a) and rates \(\Gamma_{\rsL} =
		\Gamma_{\rsR} = 5\Gamma_\C \equiv \Gamma / 2 \).  Gap openings
		are indicated by white arrows. The intensities of the
		conductance peaks depend on the tunneling rates
		\eqref{eq:rate1} and \eqref{eq:rate2}, while the temperature
		$T/U = 0.02$ determines their broadening. Kondo physics
		generates an enhanced conductance when the occupation of the
		artificial molecule is odd, i.e., inside the \((1,2n)\)
		regions with $n$ integer; such signatures are not captured by
		our master-equation description but show up in the experiment.
		In Sec.~\ref{sec:eom}, we take a cut along the green dashed
		line and address Kondo physics. Note that in (a) and (b), we
		truncate the Hilbert space at three spin-degenerate cavity
		levels. (c) Measured differential conductance through the
		experimental dot--cavity device as a function of bias on the
		dot plunger gate and cavity gate (see
		Ref.~[\onlinecite{roessler:2015}]).  The line shapes match the
		full black lines in (a) and the transport resonances in (b),
		thus confirming that the dot and cavity hybridize to form a
		coherent ``molecule".  The difference in intensities between
		occupancies \((1,2n)\) and \((0,2n)\), \((2,2n)\) is due to
		transport through a Kondo resonance.
	}
\end{figure*}

Figure \ref{groundMap} shows several ground state maps where both dot and
cavity levels are tuned through the Fermi energy \(\epsilon_\F\).  In
Fig.~\ref{groundMap}(a), we plot the dot--cavity ground state map derived from
ED as a function of applied voltages $\epsilon'_\D/U$ and $\epsilon'_{\C}/U$
[we include capacitive cross-talk to allow for better comparison with the
experimental result in \ref{groundMap}(c) [see Eqs.~\eqref{eq:ed}
and~\eqref{eq:ec}].  We label the different ground states by
$\left(N_{\D},N_{\C}\right)$, where $N_{\D}=\sum_{\sigma} n_{\D}^\sigma$
($N_{\C}=\sum_{j,\sigma} m_{j}^\sigma$) denote the total dot (cavity)
occupation. The boundaries between different regions mark those configurations
where the particle number on the dot--cavity system is undefined and the
system conducts.  For vanishingly small coupling $\Omega/U\rightarrow 0$
(dashed lines), the occupation of the cavity and the dot are independent and
the dot is empty for $\epsilon_{\D}> 0$, singly occupied for $\epsilon_{\D}<0
<\epsilon_{\D}+U$, and doubly occupied for $\epsilon_{\D} < - U$.  Similarly,
the cavity occupation changes by two electrons whenever a cavity level crosses
the Fermi energy because charging effects on the cavity are absent. At finite
dot--cavity coupling $\Omega$ (solid lines), the dot and cavity states
hybridize and form an intermediate singlet ground state (see also
Fig.~\ref{addition}). This results in a pronounced modification of the
ground-state map with additional regions of odd cavity-occupation states
separating regions with odd-dot--even-cavity regions.  The emerging
dot--cavity singlet extends over the entire hybrid system and thus defines an
artificial asymmetric dot--cavity molecule (note that these molecular states
are less prominent at even dot-fillings).  Given the large spatial extent of
the cavity, the existence of such a coherent state defines a lower bound on
the spin coherence length in GaAs heterostructures. It is this dot--cavity
molecular singlet that competes with the dot--lead Kondo singlet and that is
one of the most fascinating features characteristic of this device.

Next, we turn to the transport physics that renders the ground-state map
visible in an experiment. Indeed, since linear transport is restricted to the
degeneracy points where two ground states with different particle number
cross, we expect that a conductivity map $G = {\rm d}I/{\rm d}V$ will accurately
trace the lines of Fig.\ \ref{groundMap}(a). In the following, we use a master-equation
approach in a first attempt to map out the ground state diagram that can be
compared to experimental data [see Figs.\ \ref{groundMap}(b) and
\ref{groundMap}(c) respectively]. Furthermore, such an approach can be expanded to analyze
non-linear transport at large bias $\mu_{\rsL} \neq 0$. The inclusion
of Kondo physics requires a more sophisticated technique and we will discuss
this topic with the help of an equation-of-motion analysis in Sec.\
\ref{sec:eom} below.
%

%
\section{Master equation approach}
\label{master}
%

The master-equation successfully describes the transport through interacting
systems that are weakly coupled to leads~\cite{sakurai1985, beenakker:91,
Korotkov:1994, Koch:2006, thesis:Koch}, i.e., when $\Gamma_{\rm{a}},
\Gamma_{\C} \ll \max\left\{k_{\rs B}T, eV\right\}$, where
$eV=\mu_{\rsL}-\mu_{\rsR}$ is the bias-voltage between the left and right
leads, $T$ is the temperature, and $k_{\rs B}$ is the Boltzmann constant which
we set to unity, $k_{\rs B}=1$. The current
\(I\) across the artificial molecule is determined by the tunneling rates
\(W\) between the leads and the artificial molecule, as well as its occupation
probability \(P\) and the Fermi functions \(n_\mathrm{\scriptscriptstyle F}\)
of the leads.  Here, we describe transport through the dot--cavity system as
sequential, i.e., to lowest order in its couplings \(\Gamma_{\rm a}\) and
\(\Gamma_\C\) to the leads, but treat the dot--cavity coupling \(\Omega\) to
all orders by using our ED results for the molecular states. Thus, in a
picture where the coupling between the dot and the cavity is treated
perturbatively, our analysis captures the co-tunneling processes involving
the dot and the cavity, which we call cavity-assisted co-tunneling processes.

Given the molecular eigenstates $\kett{\psi^{\alpha}_\mathbf{N}}$ with
$\mathbf{N}=(\Nup,\Ndo)_\sigma$, we define the associated occupation
probabilities \(P^{\alpha}_{\mathbf{N}}\). Here, we distinguish the spin-tuple
\((\Nup,\Ndo)_\sigma\) describing the molecular spin occupation from the
dot--cavity occupation-tuple \((N_{\D},N_{\C})\).  The occupation dynamics
\(\dot{P}\) is determined by the master equation
\begin{align}\label{eq:rate}
\partial_{t} P^{\alpha}_{\mathbf{N}}=
\sum_{\mathbf{N}',\alpha'} (W^{\alpha,\alpha'}_{\mathbf{N},\mathbf{N}'} 
P^{\alpha'}_{\mathbf{N}'}
-W^{\alpha',\alpha}_{\mathbf{N}',\mathbf{N}} P^{\alpha}_{\mathbf{N}}),
\end{align}
where the rates $W^{\alpha,\alpha'}_{\mathbf{N},\mathbf{N}'}$ describe the
transitions from state $\kett{\psi_{\mathbf{N}'}^{\alpha'}}$ to state
$\kett{\psi_{\mathbf{N}}^{\alpha}}$.  Restricting the analysis to sequential
tunneling processes and considering only transitions between states that
differ by one electron, the rates \(W^{\alpha,\alpha'}_{\mathbf{N}
\pm\mathbf{e}_{\sigma} ,\mathbf{N}}\) take us between charge sectors
\(\mathbf{N}\) and \(\mathbf{N}\pm \mathbf{e}_{\sigma}\) with
$\mathbf{e}_{\uparrow}=(1,0)_\sigma$ and $\mathbf{e}_{\downarrow} =
(0,1)_\sigma$.  We couple the artificial molecule to the leads via the
tunneling Hamiltonian \(\bar{H}_\mathrm{tun}\) [see Eq.\
(\ref{eq:Htun})], with the latter contributing three processes that induce
changes in  the occupation number of the artificial molecule. These have
corresponding rates $W^{\rs L}$, $W^{\rs R}$, $W^\C$ adding up to the total
transition rate,
\begin{align}
W^{\alpha,\alpha'}_{\mathbf{N}\pm\mathbf{e}_{\sigma},\mathbf{N}} =
W^{{\rsL} \,\alpha,\alpha'}_{\mathbf{N},\pm\mathbf{e}_{\sigma},\mathbf{N}}
+W^{{\rsR}\, \alpha,\alpha'}_{\mathbf{N}\pm\mathbf{e}_{\sigma},\mathbf{N}}
+W^{{\C}\, \alpha,\alpha'}_{\mathbf{N}\pm\mathbf{e}_{\sigma},\mathbf{N}}.
\end{align}
Note that we sum the rates $W^{\rs R}$ and $W^\C$ incoherently, see our
discussion of Fano resonances in Sec.~\ref{sec:eff} and Fig.~\ref{models2}.
The individual rates are derived in Appendix~\ref{app:mastereq} and the result
is
\begin{align}
W^{\mathrm{a}\, \alpha,\alpha'}_{\mathbf{N}\pm\mathbf{e}_{\sigma},
\mathbf{N}} &=\frac{\Gamma_\mathrm{a}}{\hbar} | \langle 
\psi^{\alpha}_{\mathbf{N}\pm\mathbf{e}_{\sigma}} |
{d}_{\sigma}^\pm| \psi^{\alpha'}_{\mathbf{N}}\rangle|^2 
g^\mathrm{ a}_{\pm} (\epsilon_{\mathbf{N}\pm\mathbf{e}_{\sigma}}
-\epsilon_{\mathbf{N}}),\label{eq:rate1}\\
W^{\C\,\alpha,\alpha'}_{\mathbf{N}\pm\mathbf{e}_{\sigma},\mathbf{N}}
&=\frac{\Gamma_\C}{\hbar} |\langle \psi^{\alpha}_{\mathbf{N}\pm\mathbf{e}_{\sigma}}|
{f}_{\sigma}^\pm| \psi^{\alpha'}_{\mathbf{N}}\rangle|^2 
g^{\rs R}_{\pm}(\epsilon_{\mathbf{N}\pm\mathbf{e}_{\sigma}}
-\epsilon_{\mathbf{N}}),\label{eq:rate2}
\end{align}
where we have introduced the operator \({f}_\sigma = \sum_j
{f}_{j\sigma}\). The rates \(\Gamma_\mathrm{a}\) and \(\Gamma_\C\) are
given in Eqs.~\eqref{eq:dotRate} and \eqref{eq:cavRate}, $g^{\rm
a}_{+}(\epsilon)=n_{\rs F}(\epsilon-\mu_{\rm a})$ and $g^{\rm
a}_{-}(\epsilon)=1-n_{\rs F}(-\epsilon-\mu_{\rm a})$ derive from the
Fermi-Dirac distribution $n_{\rs F}(\epsilon)=1/(1+e^{\beta \epsilon})$, and
$\beta=1/T$. We use the operator notation $\mathcal{O}^{+}
=\mathcal{O}^\dagger$ and $\mathcal{O}^{-}=\mathcal{O}$.  To evaluate the
rates \eqref{eq:rate1} and \eqref{eq:rate2}, we use the eigenstates and
eigenenergies from our ED analysis (Sec.~\ref{sec:ed}).
The master equations \eqref{eq:rate} can be written in matrix form
\cite{Korotkov:1994,Koch:2006}
\begin{align}
   \partial_t {\bf P}= {\bf W} {\bf P},
\end{align}
with an occupation probability vector ${\bf P}$ and the rate matrix ${\bf W}$
that couples the different states. For the steady state, ${\bf W} {\bf P}=0$,
and we impose the normalization ${\bf P}\,\cdot\,{\bf e}=1$, where ${\bf e} =
(1,1,\dots)$.  Defining the square matrix ${\bf E}$ with all its rows given by
\({\bf e}\), we rewrite \(\mathbf{WP} + \mathbf{e} = \mathbf{e}\) in the form
\(\mathbf{WP}+\mathbf{EP} = \mathbf{e}\) and find that \(\mathbf{P}\) can be
written in the form
\begin{align}\label{eq:PW}
{\bf P}=({\bf W}+{\bf E})^{-1} {\bf e}.
\end{align}
Once the probability vector \(\mathbf{P}\) has been determined from Eq.\
\eqref{eq:PW}, we obtain the current \(I\) through the artificial molecule
from currents flowing between the dot (d) and the left lead (L)
\begin{align}
I=e \sum_{	
	\substack{
		\alpha,\alpha' \\
		\mathbf{N},\sigma
		}
	}
(W_{\mathbf{N}+\mathbf{e}_{\sigma},\mathbf{N}}^{{\D \rm L} \, \alpha,\alpha'}
-W_{\mathbf{N}-\mathbf{e}_{\sigma},\mathbf{N}}^{{\D\rm L}\, \alpha,\alpha'})
P_{\mathbf{N}}^{\alpha'},
\label{eq:master_current}
\end{align}
where the first and second term correspond to electrons entering and leaving
the dot, respectively. In the following, we will calculate and analyze the
differential conductance $G = \mathrm{d} I/\mathrm{d} V$, with $eV =
\mu_{\rsL}$ the applied voltage, in units of \(\Gamma e^2/T\). Note that the
factor of \(T\) appears when differentiating the Fermi-Dirac distribution in
the rates \(W\) with respect to the bias.

\subsection{Equilibrium linear transport} 

We first analyze the equilibrium transport at low temperatures \(T\) and small
bias \(eV = \mu_{\rsL}\) such that $\max\left\{k_{\rs B}T, eV\right\}$ remains
small compared to the level-spacing within each molecular Fock sector.  Under
these conditions, transport involves only the ground state configuration
\(\alpha = 1\) in each Fock sector \((N^\uparrow,N^\downarrow)_\sigma\). For
our spin-symmetric Hamiltonian, the ground state configurations are restricted
to $\mathbf{N}=(n,n)_\sigma$ and $\mathbf{N}=(n\pm 1,n)_\sigma$. We limit
ourselves to states with up to eight particles for computational reasons and thus
$n \leq 4$.  Calculating the linear response current, we can plot the
conductance $G_0=\lim_{V\to0}\mathrm{d}I/\mathrm{d}V$ as a function of the
system parameters [see Fig.~\ref{groundMap}(b)].  Linear-response transport
arises at the boundaries of the ground state map where two dot--cavity
molecular ground states are degenerate; in the following we refer to these
degeneracies as {\it molecular} resonances. While usual dot transport would
show Coulomb resonances at energies $\epsilon_{\D}\approx 0$ and
$\epsilon_{\D}+U\approx 0$, the molecular resonances give rise to additional
split resonances within the Coulomb blockade regime \([\epsilon_{\D},
\epsilon_\D+U]\) whenever a cavity level crosses the Fermi level, i.e.,
\(\epsilon^{\s (j)}_\C\approx 0\).  The shape of these transport resonances is
dictated by the formation of the molecular states and generates the
conductivity map of Fig.\ \ref{groundMap}(b) that is aligned with the ground
state map of \ref{groundMap}(a).  The intensities of the resonances encode the
overlap between eigenstates that determine the transition rates $W$ in Eqs.\
\eqref{eq:rate1} and \eqref{eq:rate2}, while the temperature leads to their
broadening.  The same signatures have been observed in the
experiment\cite{roessler:2015} [see Fig.~\ref{groundMap}(c)] that we take as
evidence for the formation of an extended dot--cavity molecular state (the
experimental data shown here is an unpublished result from the same device as
in Ref.\ [\onlinecite{roessler:2015}]).
\subsection{Non-equilibrium linear transport}

The above equilibrium analysis has provided us with some insights into the
ground state resonance structure of the dot--cavity hybrid. Going beyond this
equilibrium analysis, we now investigate the system when it is driven strongly
out of equilibrium. We apply a large bias \(\mu_{\rsL}=eV\) to the left lead
and tune the dot \(\epsilon'_\D\) and cavity \(\epsilon'_\C\) gates.  In the following,
we describe how the cavity modifies the out-of-equilibrium transport
signatures. We start with the dot's Coulomb diamond in the usual dot bias
$\epsilon'_\D$ versus source bias $\mu_{\rsL}$ plot. The presence of the
cavity then manifests itself through additional resonances within the Coulomb
diamond, similar to the inelastic co-tunneling features seen in a dot when
including excited
states\cite{proceedings,furusaki1995theory,zumbuhl_cotunneling_2004}. Next, we
take a cut through the diamond at fixed $\epsilon_\D$ and tune the cavity bias
$\epsilon'_\C$; the molecular-singlet formation shows up most prominently in
such a plot.
 \begin{figure*} 
 \centering 
 \includegraphics{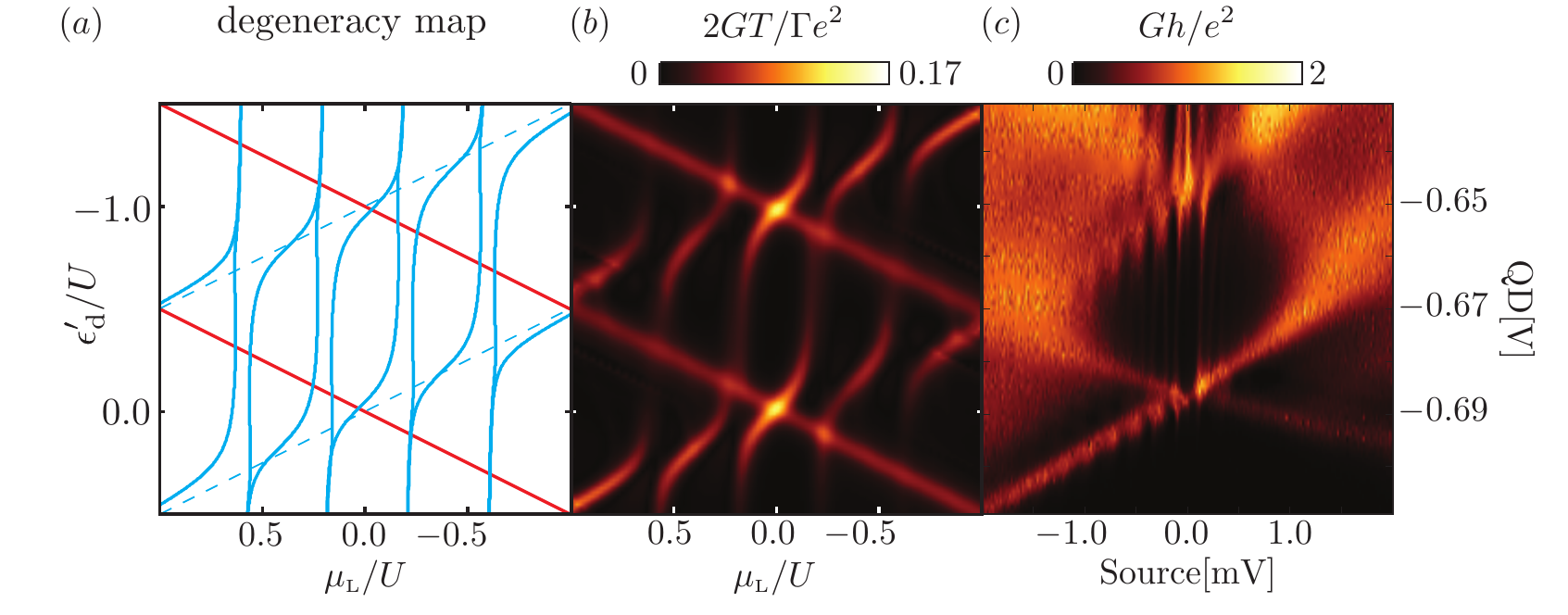}
 \caption{\label{diamond} Effect of the cavity on the dot Coulomb-diamond. (a)
 Ground state degeneracy maps for a fixed cavity level $\epsilon_\C$ as a
 function of dot gate-voltage \(\epsilon'_\D\) and chemical potential
 \(\mu_{\rsL} = eV\) of the left lead; shown are the calculated locations of
 ground state degeneracies in the addition spectra for the dot--cavity system
 with four cavity levels (\(\epsilon_\C-\epsilon_\F=-0.6U\) and
 \(\delta_\C=0.4U\)) when the artificial dot--cavity molecule is coupled to
 the left lead (blue) and the right lead (red).  The dashed lines are for
 vanishing dot--cavity coupling \(\Omega=0\) while the solid lines refer to a
 finite \(\Omega = 0.08U\). These degeneracy lines provide an outline for the
 out-of-equilibrium transport signatures in (b).  The blue lines are
 susceptible to both dot and cavity parameters \(\epsilon_\D - \mu_{\rsL}\)
 and \(\epsilon_\C-\mu_{\rsL}\) that are changed by the variation of
 \(\epsilon'_\D\) and \(\mu_{\rsL}\). The red lines, instead, depend on the
 parameters \(\epsilon_\D-\mu_{\rsR}\) and \(\epsilon_\C - \mu_{\rsR}\). The
 latter is independent of \(\epsilon'_\D\) and \(\mu_{\rsL}\) and hence the
 red lines do not show any features related to the cavity.  (b) Transport
 resonances as a function of dot gate voltage \(\epsilon'_\D\) and chemical
 potential \(\mu_{\rsL} = eV\) as derived from a master-equation approach
 assuming a capacitive cross talk between the dot gate and the chemical
 potential of the left lead as explained in Sec.\ \ref{sec:eff}
 (Eqs.~\eqref{eq:ed} and~\eqref{eq:ec}). The same dot--cavity system as in (a)
 is now simultaneously coupled to both leads at temperature \(T=0.02U\) with
 rates \(\Gamma_{\rsL} = \Gamma_{\rsR} = 5 \Gamma_\C=\Gamma/2\).  The
 transport resonances trace the degeneracies shown in (a): positively sloped
 Coulomb resonances (where the chemical potential \(\mu_{\rsL} = eV\) is
 aligned with the dot) are strongly modified by the cavity states, while the
 negatively sloped resonances remain largely unperturbed, as expected from the
 schematic view offered by the two spectra in (a). While the small splitting
 due to the molecular singlet formation shows up in the spectral map (a), this
 feature is not visible in the transport map shown in (b) due to the
 temperature broadening; it will show up in Fig.\ \ref{blocade}(c) at large
 coupling $\Omega$.  The vertical features showing up when the dot is in
 Coulomb blockade appear in a location where one expects co-tunneling
 features to manifest. In fact, our molecular description via ED captures
 processes which correspond to two-particle co-tunneling processes in a
 perturbative dot--cavity treatment (see Fig.\ \ref{blocade}) for details.  (c)
 Measured differential conductance through the dot--cavity device as a
 function of dot bias (vertical axis) and source-drain bias across the device
 (horizontal axis) (see Ref.~[\onlinecite{roessler:2015}]). The cavity levels
 generate avoided crossings on the positively slanted Coulomb resonances as
 predicted by our theoretical analysis, emphasizing the formation of a
 molecular singlet state.  In the experiment, the cavity levels are more
 densely spaced as compared to our theoretical modeling, where we have
 limited ourselves to four cavity levels due to computational reasons.
}
\end{figure*}
At finite bias \(eV\), excited states are populated and contribute to the
transport across the device. When calculating the rates in (\ref{eq:rate1})
and (\ref{eq:rate2}), all eigenstates and eigenenergies obtained by the exact
diagonalization have to be included.  In doing so, we account  for all spin
configurations \(\mathbf{N} = (n,m)_\sigma\) with \(n\) and \(m\) bounded by
the total number of single particle levels in the system.  We first
investigate the effect of the electronic cavity on the standard Coulomb
diamond by calculating the differential conductance $G=\mathrm{d} I/\mathrm{d}
V$ and plotting the result versus dot gate voltage \(\epsilon'_\D\) and
source-drain bias \(\mu_{\rsL} = eV\), while keeping the cavity level
\(\epsilon_\C\) fixed (see Fig.\ \ref{diamond}).

In the absence of the cavity, transport signatures appear when the dot levels
align with the chemical potential in either lead, \(\epsilon_\D \textrm{ or
}\epsilon_\D + U=\mu_{\rsL} \textrm{ or }\mu_{\rsR}\). Including a cavity with
levels at \(\epsilon^{\scriptscriptstyle (j)}_\C\), we expect signatures to
appear when the molecular (rather than the dot) levels align with either
\(\mu_{\rsL}\) or \(\mu_{\rsR}\). When aligning the molecular level with the
right lead the only available tuning parameter is the dot level;
correspondingly, we take a near vertical cut at fixed cavity energy
\(\epsilon_\C\) through the molecular ground state map in Fig.\
\ref{groundMap}, leaving its qualitative behaviour unchanged.  On the other
hand, the left lead can be aligned with the molecular level via two
parameters, the dot level \(\epsilon_\D\) and the chemical potential
\(\mu_{\rsL}\) in the left lead (which in turn changes the cavity levels
\(\epsilon^{\scriptscriptstyle (j)}_\C\)) and we thus expect to explore the
entire ground state map in Fig.\ \ref{groundMap}. The expected transport
signatures are shown in Fig.\ \ref{diamond}(a) where we plot the locations
where additional molecular (groundstate) degeneracies enter/leave the bias
window. At these locations, the degeneracies between molecular states and a
lead chemical potential add or remove a transport channel, producing a signal
in the non-linear conductivity $G$.
 \begin{figure*}
 \centering
 \includegraphics[width=15cm]{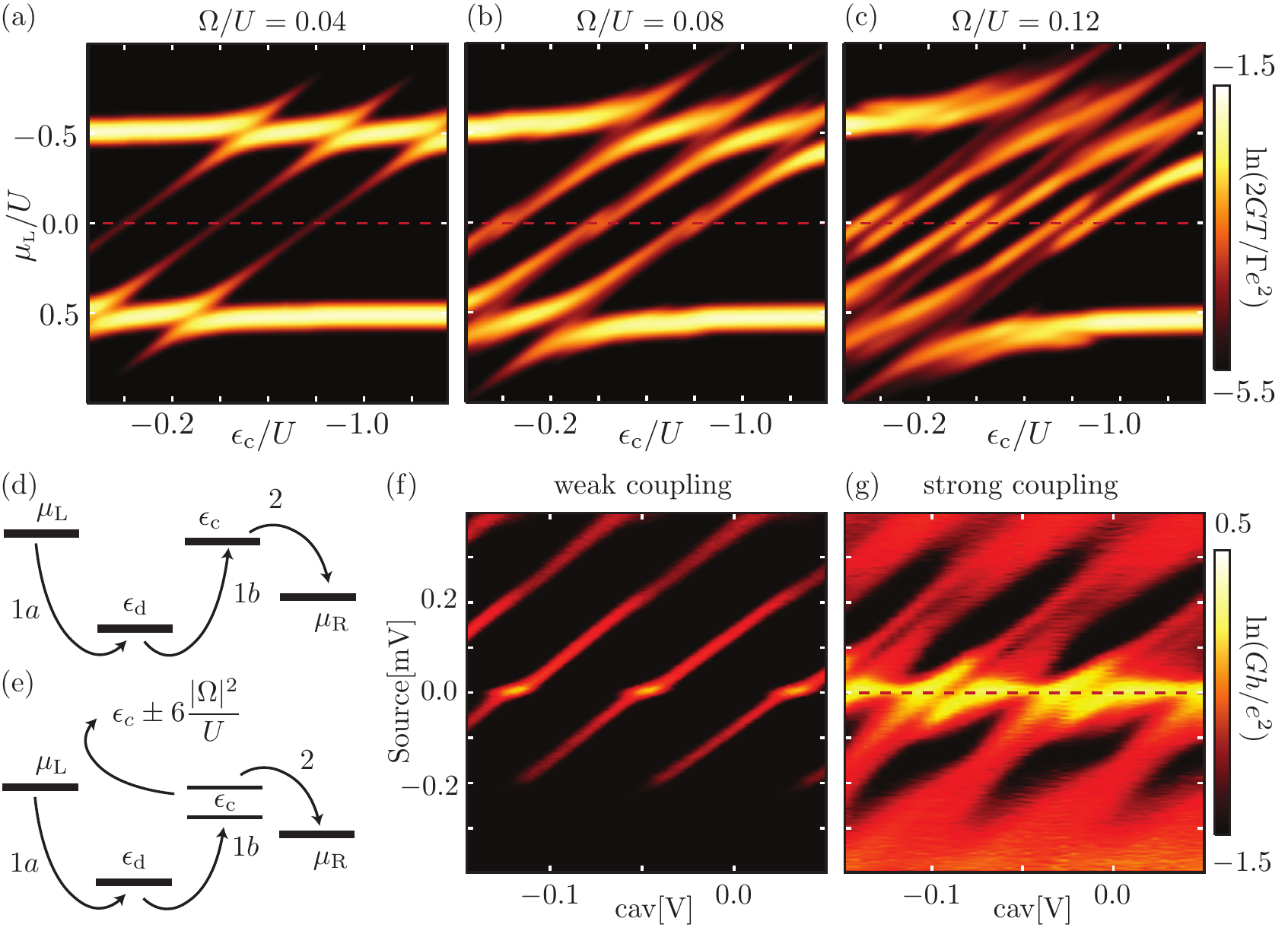}
 \caption{Differential conductance in the Coulomb blockade region at fixed dot
 level \(\epsilon_\D\) versus source bias \(\mu_{\rsL}\) and cavity level
 \(\epsilon_\C\) for different dot--cavity couplings \(\Omega\). A logarithmic
 scale is used in order to highlight the weak co-tunneling features.  The dot
 is placed at the particle--hole symmetric point within the blockaded region
 \(\epsilon_\D = -U/2\), the temperature is set to \(T=0.02\,U\) and the leads
 are coupled to the artificial molecule through the rates \(\Gamma_{\rsL} =
 \Gamma_{\rsR} = 5 \Gamma_\C=\Gamma/2\). The two horizontal lines mark the
 dot's Coulomb peaks and the array of weak diagonal lines is due to the
 molecular resonances (or cavity-assisted co-tunneling processes) appearing
 as the next cavity level \(\epsilon_\C^{\scriptscriptstyle (j)}\) is shifted
 across \(\mu_{\rsR}\).  These resonances are suppressed due to the indirect
 coupling of the cavity (via the dot) to the left lead and it is the
 hybridization with the dot that leads to the transport signature.  Increasing
 the coupling \(\Omega\), the cavity degeneracy is first slightly lifted, [see
 (b)], and then fully lifted with a splitting \(\sim 12\,\Omega^2/U\) in (c);
 figures (a)--(c) include 3 spin-degenerate cavity levels.  The red dotted
 line marks the location where Kondo transport would be present (not captured
 by the master-equation analysis discussed in this section).  (d) Illustration
 of a cavity-assisted co-tunneling process: the arrows represent hopping
 events with \(1a\) and \(1b\) describing a two electron co-tunneling event
 from the left lead into the cavity via the dot.  The arrow labeled by \(2\)
 marks a single electron tunneling from the cavity into the right lead. (e)
 The same process including the splitting of the cavity level due to the
 hybridization with the dot, corresponding to the split co-tunneling
 signatures in (b) and (c).  (f)--(g) Differential conductance measurements
 through the dot--cavity device\cite{roessler:2015}.  (f) Weak coupling
 \(\Omega\ll t_\mathrm{\scriptscriptstyle L}\) and (g) strong coupling
 \(\Omega\gg t_\mathrm{\scriptscriptstyle L}\). Although the splitting is
 accurately captured in the master-equation approach, [see (c)], the additional
 Kondo resonance along the zero source-drain bias line (\(\mathrm{source}=0\))
 is not captured by the result of the master-equation approach in (c).  }
 {\label{blocade}}
 \end{figure*}

The degeneracy map of Fig.\ \ref{diamond}(a) then has to be compared with the
transport map in Fig.\ \ref{diamond}(b) where we plot the differential
conductance calculated from Eq.\ \eqref{eq:master_current}.  As expected, the
boundaries of the Coulomb diamond are modulated by the presence of the cavity
with anti-crossings appearing whenever cut by $\epsilon^{\scriptscriptstyle
(j)}_\C$. These anti-crossings appear on the positively sloped Coulomb
resonance when the chemical potential in the left lead lines up with the
molecular degeneracies.  The negatively sloped lines originate from aligning
the dot level with the right chemical potential via the direct dot--lead
coupling. No signatures appear due to the cavity as $\epsilon^{
\scriptscriptstyle (j)}_\C$ remains fixed with respect to $\mu_{\rsR}$; in
Figs.\ \ref{diamond}(a) and (b), we set all cavity levels $\epsilon^{
\scriptscriptstyle (j)}_\C$ far away from the chemical potential in the right
lead $\mu_{\rsR}$.  As in the equilibrium situation, the intensities of the
high conductance lines is given by the wave-function overlaps in Eqs.\
\eqref{eq:rate1} and \eqref{eq:rate2}, while their broadening is due to the
temperature $k_{\rm\scriptscriptstyle B} T = 0.02U$.  Figure~\ref{diamond}(c)
shows the corresponding out-of-equilibrium transport data measured in the
dot--cavity device of Ref.~[\onlinecite{roessler:2015}].  We observe a good
qualitative agreement with our model predictions, noting, that we have assumed
a lower density of cavity levels in our theoretical analysis due to
computational reasons.

The appearance of additional resonances within the Coulomb diamonds is well
known from conventional dots: accounting for the dot's excited states,
higher-order transport channels open up (or close) when multi-particle
processes become allowed (or disallowed), thereby changing the total current
through the dot---such changes then generate (weaker) resonance structures in
the differential conductance $G$ and are known as inelastic co-tunneling
features. We can cast the appearance of our molecular resonance structure due
to the presence of the cavity in this language as well; rather than a complex
dot spectrum, it is the dot--cavity hybridization that generates these
features.  Formulating the transport within a perturbative picture in the
dot--cavity coupling $\Omega$, the additional resonances then are viewed as
cavity-assisted co-tunneling processes involving two-electron processes,
i.e., a coherent hopping of one electron from the left lead to the dot, while
the (second) dot electron moves out of the way by hopping to the cavity (see
Fig.~\ref{blocade}).  Note that, in making use of exact molecular states, our
description includes processes to lowest order in the coupling to the leads
and all orders in the coupling $\Omega$ between the dot and the cavity and
thus goes beyond the co-tunneling result.

Next, we focus on the transport signatures at a fixed dot level
\(\epsilon_\D\) when tuning the source (\(\mu_{\rsL}\)) and cavity level
\(\epsilon_\C\) (see Fig.\ \ref{blocade}). We place the dot level into the
blockaded position \(\epsilon_\D \approx -U/2\) at zero bias \(\mu_{\rsL} =
0\) with one electron on the dot.  Figure~\ref{blocade} then shows the Coulomb
blockaded region between two subsequent Coulomb peaks. As the dot couples to
the right lead directly via \(t_{\rsR}\), the dot-like molecular resonances
appear as bright horizontal features whenever a dot level is aligned with the
left lead at finite bias \(\mu_{\rsL}\).  The weak (diagonal) features are
associated with the cavity-like molecular levels and appear whenever a cavity
level enters/leaves the bias window between \(\mu_{\rsL}\) and \(\mu_{\rsR}\).
These molecular resonances involve little, \(\propto \Omega^2\), spectral
weight on the dot and hence are largely suppressed [see the weak transport
features in Fig.\ \ref{blocade}(a)].  Increasing the coupling \(\Omega\), the
singlet gap \(12 \Omega^2/U\) in the molecular spectrum and the spectral
weight on the dot increase; this manifests itself in the transport as an
increased intensity and splitting of the molecular features [see Figs.\
\ref{blocade}(b) and (c)].  All transport signatures, except for the Kondo
resonance at zero source bias, compare well with the experimental findings
shown in Figs.\ \ref{blocade}(f) and (g).  The observed splitting of the
cavity level confirms that the transport across the device involves the
spin-coherent dot--cavity singlet.

In the above discussion, we have treated the dot--cavity system as a single
artificial molecule and have analyzed the transport as sequential tunneling
from the left lead via the artificial molecule to the right lead.
Alternatively, we can formulate the transport via the dot--cavity system as a
co-tunneling process involving two electrons and the dot and cavity as two
separate entities.  The jump from the left lead to the cavity (and
subsequently to the right lead) then involves the dot in a virtual process
which contributes a factor $\propto \Omega^2$ [see Fig.\ \ref{blocade}(d)].
This process is turned on/off as the cavity level enters/leaves the bias
window. In Fig.\ \ref{blocade}(e), we show the same process at large coupling,
where the cavity level is split by the presence of the dot, resulting in two
closeby co-tunneling features separated by the singlet gap \(12 \Omega^2/U\).
We call this process a cavity-assisted co-tunneling process, where the
virtual hop through the dot is helped by the presence of the cavity by
providing a large final density of states. Note that: (i)  In
Ref.~[\onlinecite{proceedings}], we have shown that such cavity assisted
co-tunneling processes provide further spectroscopic information on the dot.
(ii) Within the co-tunneling picture, we have to artificially account for the
cavity level-splitting. On the contrary, the discussion based on the molecular
picture relies on the exact solution of the dot--cavity problem; this includes
all orders of tunneling (\(\Omega\)) between the dot and the cavity and thus
provides us with the proper level splitting.

The experimental data in Fig.\ \ref{blocade}(g) exhibits an additional
zero-bias peak due to the Kondo effect. This Kondo resonance is due to a
singlet formation between the dot and the leads and is broken up by the
dot--cavity molecular singlet when the cavity is tuned across the equilibrium
chemical potential.  The switching between a many-body Kondo- and a
dot--cavity molecular singlet is one of the outstanding results of Ref.\
[\onlinecite{roessler:2015}]; its theoretical understanding will be developed
in the next section.
%

\section{Equation of motion}\label{sec:eom}
%
To analyze the competition between Kondo physics and dot-cavity hybridization,
we make use of an equation-of-motion (EOM) technique~\cite{bruus-flensberg}
and focus on the equilibrium linear-response properties. The EOM method is
ubiquitous in the discussion of quantum-dot physics described by the Anderson
impurity model\cite{lacroix:81,lacroix:82,entin-wohlman:05,kashcheyevs:06,
Lavagna2010,thesis:Roermund} ; it provides us with the dot-Green's function
$G_\sigma(\omega)$, from which we obtain the conductance $G_0$ of the device
through a modified Meir-Wingreen formula\cite{Meir1992} as discussed in
Appendix \ref{app:Meir-Wingreen} for the limit of finite but small
temperatures \(T\ll\Gamma\). In this limit, we recover the standard
expression~\cite{Meir1992, thesis:Roermund}
\begin{align}\label{eq:G}
  G_0=- \frac{e^2}{h} \tilde{\Gamma}(\epsilon_\F) \sum_\sigma
  2 \im \, G_\sigma(\epsilon_{\rs F}),
\end{align}
with the rate \(\tilde{\Gamma}\) derived in Appendix \ref{app:Meir-Wingreen},
\begin{align}\label{eq:Gt}
\tilde{\Gamma}(\epsilon_{\rs F}) 
= \frac{\Gamma}{2} \frac{1+\beta}{2+\beta}, \quad
 \beta = 2\frac{\Gamma_\C }{\Gamma}\,
	\frac{\Omega^2}{\epsilon_\C^2 +\Gamma_\C^2/4}.
\end{align}
Here, we have considered a minimal situation with a single cavity level and
\(\Gamma_{\rsR} = \Gamma_{\rsL} =\Gamma/2\); the rate \eqref{eq:Gt} is then
bounded by \(\Gamma/4 < \tilde\Gamma < \Gamma/2 \). Aligning the cavity with
the Fermi level, we have $\epsilon_\C = 0$ and tuning $\Omega$ from weak to
strong coupling, the parameter $\beta$ changes from a small value to one above
unity.  The strong coupling is exemplified in Fig.\ \ref{blocade}(g), where
the competition between Kondo- and molecular- singlets is prominent.

The following discussion makes use of Hamiltonian \eqref{eq:ham} in the
compact notation
\begin{align}\label{eq:eomham}
H
=&\sum_{\sigma}\epsilon_{\D} d_{\sigma}^\dagger d_{\sigma}^\pdagger
+U n_{\uparrow}n_{\downarrow} \\ \nonumber
&+ \sum_{k,\sigma} \epsilon_{k} c_{ k\sigma}^\dagger
c_{k\sigma}^\pdagger
+\sum_{k,\sigma} \bigl(t_kd_{\sigma}^\dagger  c_{ k\sigma}^\pdagger
+\mathrm{H.c.}\bigr),
\end{align}
where \(t_k\) encodes all tunnelings to and from the dot; Fano-type
interference effects have to be excluded as discussed in Sec.~\ref{sec:eff}.
In Appendix~\ref{app:selfenergy}, we show that within our EOM analysis, the
Hamiltonian \eqref{eq:eomham} is equivalent to \(\bar{H}\) [see Eq.\
\eqref{eq:effHam} and Fig.\ \ref{models}(b)]. When using the Hamiltonian
\eqref{eq:eomham} the couplings \(t_{\rsL}\), \(\Omega\), and \(t_{\rsR}\) are
accounted for in the same order.

\subsection{Formalism and lowest-order approximation}

For completeness, we provide a brief overview of the equation-of-motion method
leading to the dot Green's function \(G_\sigma\) and highlight the physical
implications of the different truncation schemes (see Refs.\
[\onlinecite{entin-wohlman:05,kashcheyevs:06,thesis:Roermund,Lavagna2010}]
and Appendix \ref{app:eom} for more details). We start with the retarded
thermal Green's functions of the dot
\begin{align}
   G_\sigma= - i \lim_{\eta\to 0^+} \!\! \int\limits_{-\infty}^{\infty}\!\!dt\,
   \Theta(t) \langle \{d_\sigma^\pdagger(t),d_\sigma^\dagger(0)\}\rangle
   e^{i(\omega+i\eta)t},
\end{align}
where \(\langle \cdots \rangle\) denotes the finite-temperature
quantum-statistical average and \(\Theta\) is the Heaviside function. The
equation of motion for the (retarded) Green's function \(G_\sigma\) generates
an infinite hierarchy of additional equations for higher-order Green's
functions including lead electrons. It is convenient to introduce the Zubarev
notation\cite{Zubarev1960} (with \(A\) and \(B\) two fermionic operators)
\begin{align}
\langle\!\langle A;B \rangle\!\rangle_z \!=
\mp i \lim_{\eta\to 0^+} \!\! \int\limits_{-\infty}^{\infty}\!\!\!dt\,
\Theta(\pm t)
\langle \{A(t),B(0)\}\rangle
e^{i(\omega\pm i\eta)t},
 \end{align}
with the shorthand \(z=\omega\pm i\eta\) for the complex frequency; we will
suppress this subscript from now on. The EOMs for such Green's functions are
found by taking time derivatives and transforming to frequency space,
\begin{align} \label{eq:eom}
   z\langle\!\langle A;B \rangle\!\rangle &= \langle \{A,B\} \rangle 
   + \langle\!\langle [A,H];B \rangle\!\rangle \nonumber\\
   &=\langle \{A,B\} \rangle - \langle\!\langle A;[B,H] \rangle\!\rangle,
\end{align}
where (\(\{\cdot,\cdot\}\)) \([\cdot,\cdot]\) denote usual (anti-)commutators.
Applying these equations to the dot Green's function and the dot--lead
correlator  \(\langle\!\langle c^\pdagger_{k\sigma};
d_\sigma^\dagger\rangle\!\rangle\), and eliminating the latter, we
obtain\cite{bruus-flensberg} ($\bar\sigma$ and $\sigma$ denote complementary
spins)
\begin{align} \label{eq:greenself}
   G_{\sigma}(z) \equiv \langle\!\langle d^\pdagger_\sigma;d^\dagger_\sigma 
   \rangle \!\rangle = \frac{1 +U\langle\!\langle n_{\bar{\sigma}} 
   d^\pdagger_\sigma;d^\dagger_\sigma\rangle\!\rangle}{z-\epsilon_\D-\Sigma(z)},
\end{align}
where the dot--lead correlator gives rise to the network self-energy
(see Appendix~\ref{app:selfenergy})
\begin{align} \label{eq:selfen}
   \Sigma(\omega\pm i\eta) =
   \mp i \frac{\Gamma_{\rsL}+\Gamma_{\rsR}}{2}
   + \frac{|\Omega|^2}{\omega-\epsilon_{\C}\pm i\Gamma_c/2}.
\end{align}

The non-interacting limit \(U=0\) of Eq.~(\ref{eq:greenself}) provides us with
a Lorentzian spectrum centred around the (spin-degenerate) dot level
\(\epsilon_\D\).  At finite \(U\), we first truncate at the level of the
four-point Green's function in~\eqref{eq:greenself}; using a cumulant
expansion, we obtain
\begin{align}
   \langle\!\langle n_{\bar{\sigma}} d^\pdagger_\sigma;d^\dagger_\sigma 
   \rangle\!\rangle
   \approx \langle n_{\bar{\sigma}}\rangle \langle\!\langle 
   d_\sigma^\pdagger;d_\sigma^\dagger \rangle\!\rangle
\end{align}
(see Refs.\ [\onlinecite{kashcheyevs:06,thesis:Roermund,Lavagna2010,Kubo1962}]
and Appendix \ref{app:eom} for details).  This truncation is \(O(t^0)\) and
provides a shift \(U\langle n_{\bar{\sigma}}\rangle\) in the dot-level
depending on its filling. We must therefore replace $\epsilon_\D \to
\epsilon_\D + U\langle n_{\bar{\sigma}}\rangle$ in the ${U=0}$ Green's
function. To calculate the occupation on the dot, we can use the spectral
theorem
\begin{align} \label{spectral}
  \langle n_{\bar{\sigma}} \rangle =\frac{i}{2\pi}  \oint dz \, n_{\rs F}(z) 
  G_\sigma(z),
\end{align}
where the integration contour runs clockwise around the real axis with the
advanced Green's function (\(G^\mathrm{a}_\sigma = G_\sigma^*\)) in the lower
half of the complex plane. As we work at equilibrium with \(V = 0\), the same
Fermi-Dirac distribution \(n_{\rs F}(\omega)\) applies to all the leads.
Combining the results for $G_\sigma$ and $\langle n_{\bar{\sigma}} \rangle$
self-consistently, the conductance \(G_0(\epsilon_\D)\) then exhibits
Lorentzian peaks~\cite{bruus-flensberg} around $\epsilon_\D = 0$ and
$-U$.

\subsection{Lacroix and further truncation schemes}

Proceeding with the next order, the equation of motion for the four-point
Green's function in \eqref{eq:greenself} takes the form
\begin{align} \label{eq:eom2particle}
   (z-\epsilon_\D-U ) \langle\!\langle 
   n_{\bar{\sigma}}d^\pdagger_\sigma;d^\dagger_\sigma \rangle\!\rangle
   &\,= \langle n_{\bar{\sigma}} \rangle + \sum_{k}[
   t_k\langle\!\langle n_{\bar{\sigma}}c_{k\sigma};d^\dagger_\sigma 
   \rangle\!\rangle\\ \nonumber
   + \,t_k\langle\!\langle d_{\bar{\sigma}}^\dagger &
   c_{k\bar{\sigma}} d_\sigma^\pdagger;d^\dagger_\sigma \rangle\!\rangle
   -t_k^*\langle\!\langle c_{k\bar{\sigma}}^\dagger d_{\bar{\sigma}}^\pdagger 
   d_\sigma^\pdagger;d^\dagger_\sigma \rangle\!\rangle]\nonumber
\end{align}
and introduces three new correlators involving a lead state \(k\sigma\) or
\(k\bar\sigma\).  The second term \(t_k\langle\!\langle
d_{\bar{\sigma}}^\dagger c_{k\bar{\sigma}} d_\sigma^\pdagger;d^\dagger_\sigma
\rangle\!\rangle\) in the sum describes a spin exchange between the dot and a
lead electron involving a tunneling process and is the basic process
responsible for the Kondo resonance.  Truncating the EOM at this \(O(t)\)
level is not sufficient to capture the Kondo effect, as the last two terms
cancel (see Appendix \ref{app:eom}). Nonetheless, it is instructive to note
that the corresponding dot Green's function
\begin{align} \label{eq:Gtsquared}
   G_\sigma (z)=&\frac{1-\langle n_{\bar{\sigma}}\rangle}{z-\epsilon_\D-\Sigma(z)
   \left(1+\frac{U\langle n_{\bar{\sigma}} \rangle}{z-(\epsilon_\D+U)}\right)}\\
   &+ \frac{\langle n_{\bar{\sigma}}\rangle}{z-(\epsilon_\D+U)-\Sigma(z)
   \left(1-\frac{U(1-\langle n_{\bar{\sigma}} \rangle)}{z-\epsilon_\D}\right)}
\end{align}
now shows two split Lorentzians as a function of $z$, one centred around
\(\epsilon_\D\) and the other around \(\epsilon_\D+U\), respectively.
Proceeding with the equations of motion for the three four-point Green's
functions including a lead electron and truncating at order \(O(t^2)\), i.e., using the
Lacroix truncation~\cite{lacroix:81,lacroix:82,kashcheyevs:06}, one arrives at the relations
\begin{align} \label{eq:eomhigh1}
\sum_k t_k&  \langle\!\langle n_{\bar{\sigma}}c_{k\sigma};d^\dagger_\sigma 
\rangle\!\rangle = \langle\!\langle 
n_{\bar{\sigma}}d^\pdagger_\sigma;d^\dagger_\sigma \rangle\!\rangle\, 
\Sigma(z),\\\label{eq:eomhigh2}
\sum_k t_k& \langle\! \langle d_{\bar{\sigma}}^\dagger 
c_{k\bar{\sigma}} d_\sigma^\pdagger;d^\dagger_\sigma \rangle\!\rangle =
\langle\!\langle n_{\bar{\sigma}}d^\pdagger_\sigma;d^\dagger_\sigma 
\rangle\!\rangle\, \Sigma(z)\\ \nonumber
&+\left[1+\Sigma(z)G_\sigma(z)\right] P_{\bar{\sigma}}(z)-G_\sigma(z) 
Q_{\bar{\sigma}}(z),\\ \label{eq:eomhigh3}
-\sum_k t_k^*&\langle\! \langle c_{k\bar{\sigma}}^\dagger 
d_{\bar{\sigma}}^\pdagger d_\sigma^\pdagger;d^\dagger_\sigma \rangle\!\rangle = 
-\langle\!\langle n_{\bar{\sigma}}d^\pdagger_\sigma;d^\dagger_\sigma 
\rangle\!\rangle\, \Sigma(z_2)\\ \nonumber 
&+\left[1+\Sigma(z)G_\sigma(z)\right] P_{\bar{\sigma}}(z_2)+G_\sigma(z) 
Q_{\bar{\sigma}}(z_2),
\end{align}
with the shifted variable 
\begin{align}
  z_2=2\epsilon_{\rm d}+U-z.
\end{align}
Equations~\eqref{eq:eomhigh1}--\eqref{eq:eomhigh3} express the various dot--lead
correlators in terms of the four-point dot correlator \(\langle\!\langle
n_{\bar{\sigma}} d^\pdagger_\sigma; d^\dagger_\sigma \rangle\!\rangle\), the
dot Green's function $G_\sigma$ and self-energy $\Sigma$, as well as the two
new functions\cite{kashcheyevs:06}
\begin{align}
\label{eq:P}
P_\sigma(z) &\equiv \sum_k \frac{t_k \langle d_\sigma^\dagger
        c_{k\sigma}^\pdagger\rangle}{z-\epsilon_k} =\mathcal{F}_{\sigma z}[G],\\
\label{eq:Q}
Q_\sigma(z) &\equiv \sum_{kk'} \frac{t_{k'} t_k^*\langle c_{k\sigma}^\dagger
        c_{k'\sigma}^\pdagger\rangle}{z-\epsilon_k'} =
\mathcal{F}_{\sigma z}[1+\Sigma G].
\end{align}
The last equalities describe the functions \(P_\sigma(z)\) and \(Q_\sigma(z)\)
in terms of an integral over the dot Green's function \(G_\sigma\) (see 
Ref.~[\onlinecite{kashcheyevs:06}] and Appendix~\ref{app:eom}), e.g.,
\begin{align} \label{eq:functional}
   \mathcal{F}_{\sigma z}[G] \equiv \frac{i}{2\pi}\oint_{C}dz' n_\F (z')
   G_\sigma(z')\frac{\Sigma(z')-\Sigma(z)}{z-z'},
\end{align}
and similar for \(\mathcal{F}_{\sigma z}[1+\Sigma G]\).  Note that $\langle
d_\sigma^\dagger c_{k\sigma}^\pdagger\rangle$ and $\langle c_{k\sigma}^\dagger
c_{k'\sigma}^\pdagger\rangle$ relate to $\langle\!\langle
d_\sigma\pdagger;d_\sigma^\dagger\rangle\!\rangle= G_\sigma$ via the {spectral
theorem} and the equations of motion.  Combining the equations
\eqref{eq:greenself}, \eqref{eq:eom2particle}, and
\eqref{eq:eomhigh1}--\eqref{eq:eomhigh3}, the final expression for
the dot Green's function is~\cite{kashcheyevs:06}
\begin{widetext}
	\begin{align} \label{eq:green}
	G_{\sigma}(z) = \frac{g^{-1}_2(z)+\langle 
		n_{\sigma}\rangle\, U+[P_{\bar\sigma}(z)+P_{\bar\sigma}(z_2)]\,U}
	{g^{-1}_2(z)\, g^{-1}(z) 
                -[P_{\bar\sigma}(z)+P_{\bar\sigma}(z_2)]\,U\,\Sigma(z)
                + [Q_{\bar\sigma}(z)-Q_{\bar\sigma}(z_2)]\,U},
	\end{align}
\end{widetext}
where we have introduced the characteristic functions
\begin{align}\label{eq:g}
  g^{-1}(z) &= [z-\epsilon_\D - \Sigma(z)], \\ \label{eq:g2}
  g^{-1}_2(z) &= [z-(\epsilon_\D + U) - 2\Sigma(z) + \Sigma(z_2)],
\end{align}
that define the poles associated with the two dot levels separated by $U$.  The set
of equations \eqref{eq:P}, \eqref{eq:Q}, and \eqref{eq:green} can be solved
numerically for a self-consistent solution providing the dot's Green's function
\(G_\sigma\), from which the conductance at the Fermi energy follows through
the Meir-Wingreen\cite{Meir1992} formula \eqref{eq:G}.
\begin{figure}
	\centering
	\includegraphics{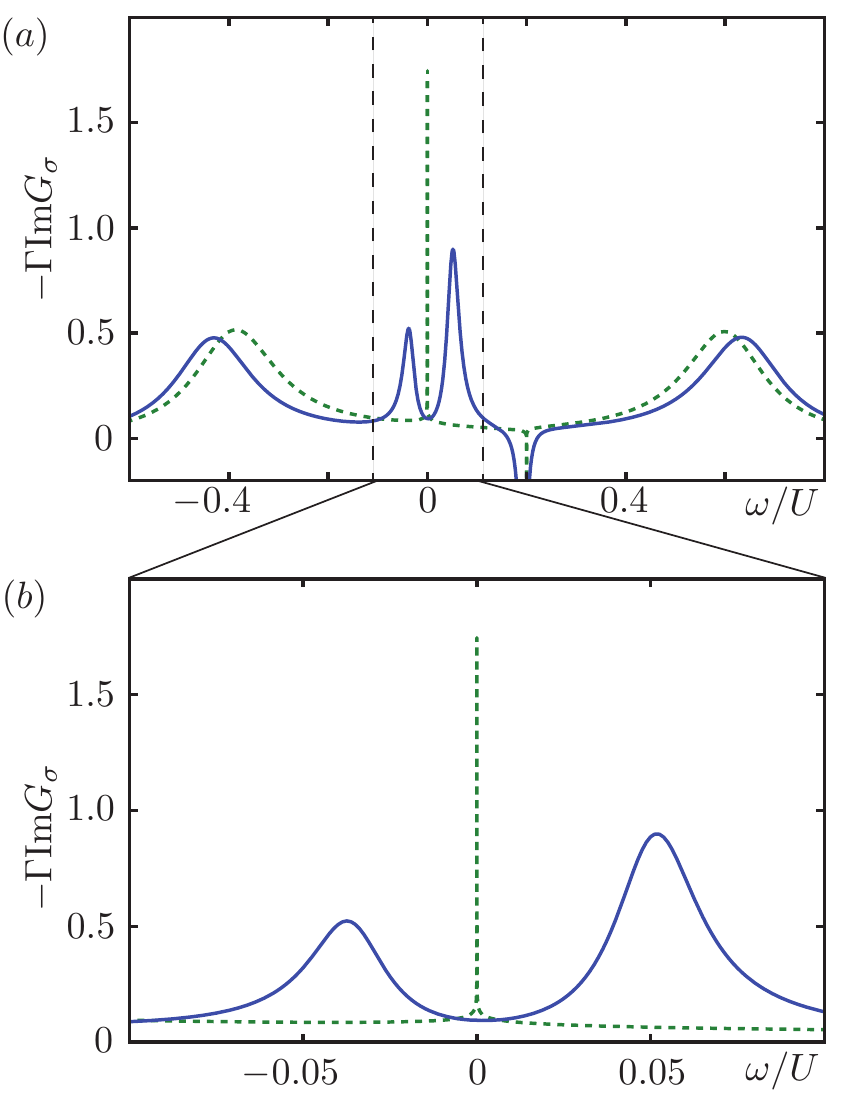}
	\caption[]{\label{fig:KondoDOS}
		Dot density of states \(\mathrm{Im}G_\sigma\) obtained through
		the EOM method versus energy \(\omega/U\). The dashed green
		line shows the density of states in the dot without a cavity
		as a function of energy for \(2\Gamma_{\rsL} = 2\Gamma_{\rsR}
		= 5\Gamma_\C = \Gamma = U/10\) and in the Coulomb blockade
		regime with \(\epsilon_\D=-2U/5\) as calculated
		self-consistently using \eqref{eq:green} at a temperature \(T
		\ll T_{\rs K}\).  We find the expected Coulomb resonances at
		\(\omega/U=-2/5, 3/5\) in (a) and a sharp Kondo resonance at
		the Fermi energy \(\omega=0\) The negative peak at
		\(\omega=2\epsilon_\D+U=U/5\) is an artifact of the Lacroix
		truncation.  The solid blue line shows the same dot
		configuration with the dot coupled to a cavity  with a
		strength \(\Omega = \Gamma = 0.1U\) and the single cavity
		level $\epsilon_\C = 0$ tuned to the Fermi energy.  Outside
		the region marked by the thin vertical dashed lines (a) the
		features remain essentially unchanged.  The lower plot (b)
		shows an enlargement of the area around the Fermi energy. A
		finite coupling to the cavity suppresses the Kondo peak and
		establishes a pair of molecular resonances separated by
		\(12\alpha \Omega^2/U\) [see Eq.~\eqref{eq:alpha}].
	}
\end{figure}

Before discussing the result, we analyze the various features in the Green's
function \eqref{eq:green} at the positions \(z =
\epsilon_\D,~0,~2\epsilon_\D+U,~ \epsilon_\D+U\) (see the dashed green line in
Fig.~\ref{fig:KondoDOS}).  The functions \(g^{-1}(z)\) and \(g^{-1}_2(z)\)
account for the Lorentzian shaped dot levels at \(\epsilon_\D\) and
\(\epsilon_\D+U\). The functions \(P(z)\) and \(Q(z)\) contribute additional
divergences (at $T=0$) at the Fermi level \(z= \epsilon_\F = 0\)
and thus give rise to the Kondo unit conductance. On the other hand, at
\(z=2\epsilon_\D+U\) (\(z_2=0\)) it is \(P(z_2)\) and \(Q(z_2)\) that diverge
and we obtain an unphysical negative conductance --- a well known artifact of
the Lacroix truncation scheme\cite{kashcheyevs:06}.  Away from the
particle--hole symmetric point these two divergences appear in separate points
($z_2 \neq -z$) as illustrated in Fig.\ \ref{fig:KondoDOS}. Tuning towards the
particle--hole symmetric point the two divergences interfere and the Kondo
peak vanishes.  

There are several approaches that allow to cure the anomaly appearing at $z_2
= 0$, among them higher truncation schemes
\cite{Goldberg2005,Monreal2005,Lavagna2010} or numerical methods
\cite{NRGReview1970,NRGReview2008,White1992, DMRGReview2005}. One possibility is to replace the
self-consistent determination of $Q_\sigma$ in Eq.\ \eqref{eq:Q} by a
predetermined function of $\Sigma$\cite{Goldberg2005} within a generalized
Lacroix scheme.  A step beyond the Lacroix scheme that avoids the anomaly at
$z_2=0$ while keeping the Kondo peak at $z=0$ is adopting a so-called
\(O(t^4)\) decoupling method \cite{Monreal2005,Lavagna2010}; this provides
expressions for the self-energies associated with the four-point functions
\(\langle\!  \langle d_{\bar{\sigma}}^\dagger c_{k\bar{\sigma}}
d_\sigma^\pdagger;d^\dagger_\sigma \rangle\! \rangle\) (Kondo) and \(\langle
\!\langle c_{k\bar{\sigma}}^\dagger d_{\bar{\sigma}}^\pdagger
d_\sigma^\pdagger;d^\dagger_\sigma \rangle\!  \rangle\) (anomaly). One then
finds \cite{Lavagna2010,thesis:Roermund} that the self-energy for the
spin-flip excitations responsible for the Kondo effect vanishes in the Coulomb
blockade region; the resulting excitations are long-lived and contribute to
the transport. On the contrary, the excitations at \(z_2=0\) have finite
self-energies; they are short-lived and do not contribute to transport.  A
precise numerical approach makes use of the renormalization group and has been
used in various embedded impurity systems \cite{NRGReview1970,NRGReview2008},
in the Kondo-box problem \cite{thimm_kondo_1999}, and most recently for the
present experiment \cite{Ulloa2017} using a model Hamiltonian different from
ours, i.e., without removing the Fano interference term (see Fig.\
\ref{models2}) and a sharp Kondo peak surviving at all values of cavity
energies, at variance with the experimental findings and our results presented
below.  Here, we choose to keep the Lacroix truncation scheme and stay away
from the dangerous particle-hole symmetric point. Furthermore, we invoke
Fermi-liquid arguments \cite{Langreth1966} to confirm the validity of our
findings.

\subsection{Results}

The above EOM method describes Kondo physics and can be applied directly to
our dot--cavity--leads system as the cavity appears only within the definition
of the network self-energy $\Sigma$ (see Appendix~\ref{app:selfenergy}). To
obtain the transport features, we must solve Eqs.~\eqref{eq:P}, \eqref{eq:Q},
and \eqref{eq:green} self-consistently.  This can be done numerically or, in
principle, in the zero temperature limit analytically. Here, we shortly
comment on the $T=0$ analytic approach that provides us with a useful upper
bound on the conductance that turns exact in the limits $\epsilon_\C \to
0,\infty$ for $\epsilon_\D = -U/2$. However, this method is unsuitable in
describing a situation with a non-trivial self-energy $\Sigma(\omega)$ as is
the case for our dot--cavity setup.  We therefore resort to a self-consistent
numerical analysis and find that the results properly explain the changeover
from Kondo- to molecular singlet, thus providing insights beyond the master
equation method in Sec.~\ref{master}.

In the zero temperature limit, following Ref.\ [\onlinecite{kashcheyevs:06}],
we can isolate the divergent terms in the numerator and denominator of
$G_\sigma$ close to the Fermi energy $z = \epsilon_\F$ and obtain
\begin{align} 
G_\sigma(0)=\lim\limits_{z\to 0}
\frac{1}{Q_{\bar{\sigma}}(z)/P_{\bar{\sigma}}(z)
	-\Sigma(z)}.
\end{align}
Expanding the functional~\eqref{eq:functional} around \(\omega = 0\) at
\(T=0\) and canceling diverging contributions, we find that
\begin{align}
G_\sigma(0)=\frac{1}{1/G^*_\sigma(0) - \Sigma + \Sigma^*},
\end{align}
from which follows the relation $\mathrm{Im} G_\sigma^{-1} (0) = -\mathrm{Im}
\Sigma(0)$, i.e., $\mathrm{Im} G_\sigma^{-1} (0)$ is given by the network
self-energy. This is the first of a set of Fermi-liquid relations
\cite{Langreth1966} and is respected by the EOM in the Lacroix truncation
scheme. However, using this scheme, we have no direct handle on $\mathrm{Re}
G_\sigma^{-1} (0)$ at zero temperature. Commonly, one constrains the real part
of the inverse Green's function using Friedel's sum rule~\cite{Langreth1966}
\begin{align}\label{Eq:FriedelSumRule}
   \mathrm{Re}G^{-1}_\sigma(0) &= [\mathrm{Im} \Sigma(0)] \cot(\pi\tilde{n}_\sigma),\\
   \tilde{n}_\sigma - \langle n_\sigma \rangle &\equiv \gamma 
   = \frac{1}{\pi}\mathrm{Im} \!\int\!\! \mathrm{d}\omega \, n_\F(\omega) 
   \frac{\partial \Sigma(\omega)}{\partial \omega} G_\sigma(\omega).
\end{align} 
Equation \eqref{Eq:FriedelSumRule} is the second of two Fermi liquid relations
and is violated in the Lacroix truncation of the EOM method.

For unstructured leads with a constant self-energy, one finds that
\(\gamma=0\); furthermore, \(\langle n_\sigma\rangle=1/2\) for a half-filled
dot in Coulomb blockade and hence the real part $\mathrm{Re} G_\sigma^{-1} (0)
= 0$ vanishes. As a result, the conductance of the system at zero temperature
assumes its maximal value~\cite{kashcheyevs:06}, \(G_0 = 2e^2/\hbar\). For the
structured lead, the parameter \(\gamma\) must be evaluated using the
(self-consistent) solution for the Green's function \(G_\sigma(\omega)\).
Furthermore, the occupation of the dot \(\langle n_\sigma \rangle \) may
deviate from \(1/2\) when the network self-energy \(\Sigma\) is not symmetric
around the Fermi energy and must be calculated as well.  However, the
conductance obtained by setting \(\tilde{n}=1/2\) defines an upper bound on
the conductance that can be shown~\cite{Ferguson2018} to become a very good
approximation to the exact result around \(\epsilon_\C = 0\). In the
following, we make use of a numerical self-consistent solution of
Eqs.~\eqref{eq:P}-\eqref{eq:green} and use the Fermi liquid relations at
\(\epsilon_\C = 0 \) to substantiate our numerical EOM results.

\begin{figure}
	\centering
	\includegraphics{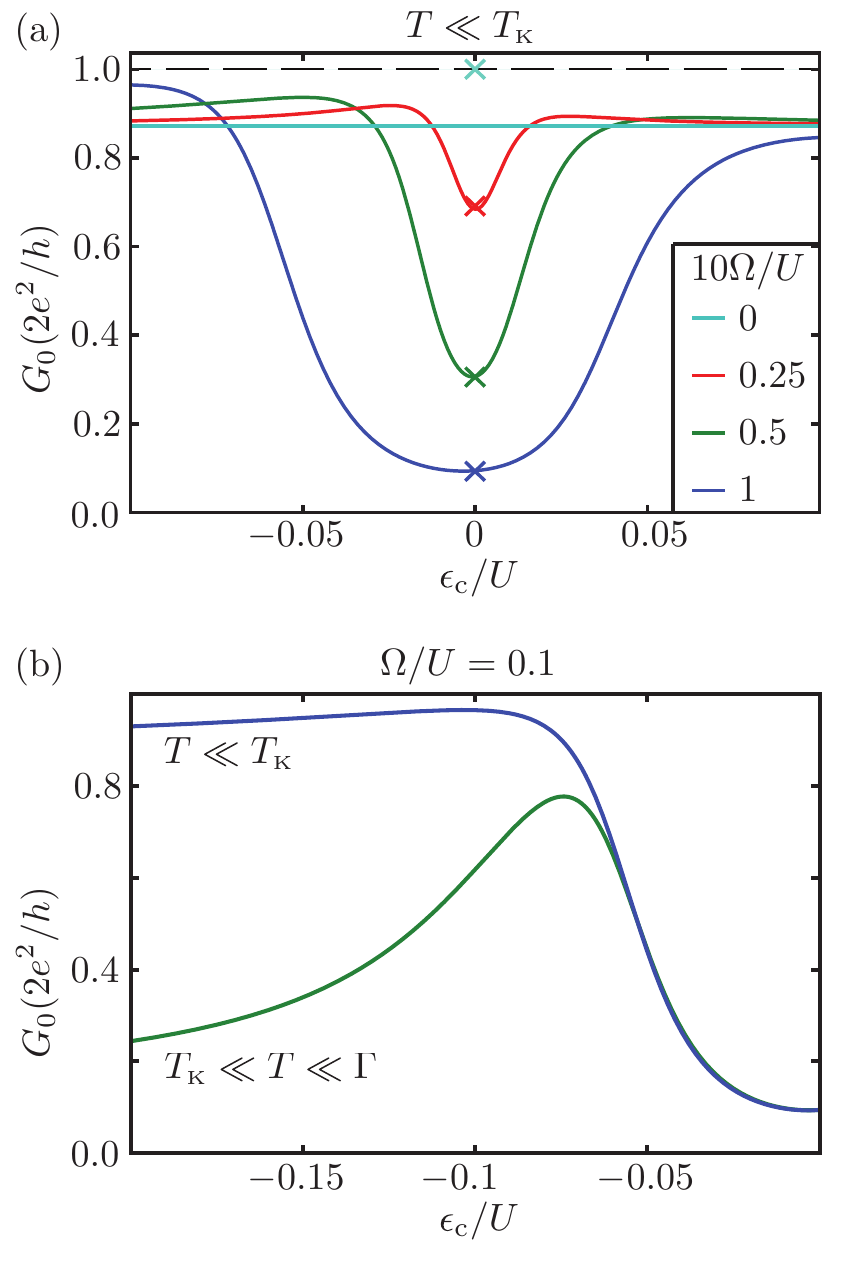}
	\caption[]{\label{results}
		Equilibrium conductance \(G_0\) as a function of cavity-level
		position \(\epsilon_\C/U\), corresponding to the green dashed
		line in figure \ref{groundMap}(b) but for different couplings
		\(\Omega/U\) and temperatures \(T/T_{\rs K}\), with
		\(2\Gamma_L = 2\Gamma_R= 5\Gamma_\C= \Gamma = U/10\) and
		\(\epsilon_d/U=-2/5\). In (a), we show the conductance for
		very low temperatures \(T \ll T_{\rs K}\) and different
		coupling strengths \(\Omega\). The cavity suppresses the Kondo
		dot--lead singlet as the dot--cavity coupling is increased.
		The crosses correspond to the upper bounds given by the Fermi
		liquid theory~\cite{Langreth1966} and become exact in
		the limit where \(\epsilon_\D=-U/2\)~\cite{Ferguson2018}. In
		(b), we plot the strong-coupling $\Omega = 0.1 \,U$,
		low-temperature transport [dark blue line in (a) and (b)] and
		the transport for the same coupling but at a temperature much
		larger than \(T_{\rs K}\) (green line).  We observe the
		disappearance of the Kondo conductance for a detuned cavity
		level $\epsilon_\C \neq 0$.  This unveils the cavity-assisted
		co-tunneling features as the split cavity levels  pass the
		Fermi level at \(\epsilon_\C\pm 6 \alpha \Omega^2/U=0\) [see
		Eq.~\eqref{eq:alpha}] (only one level shown here).
	}
\end{figure}

The functions \(P_\sigma(z)\) and \(Q_\sigma(z)\) in \eqref{eq:green} depend
themselves on the dot's Green's function \(G_\sigma(z)\), such that we have to
solve \eqref{eq:green} via an iterative approach. In our numerical solution,
we take account of one cavity mode and make use of the self-energy
\(\Sigma(z)\) as given by \eqref{eq:selfen}.  We initiate the iteration with a
Green's function that is peaked around the two Coulomb resonances
\(\epsilon_\D <0\) and \(\epsilon_\D+U>0\) and proceed until convergence is
achieved. The discretization of energies around \(z=0\) and \(z_2=0\) must be
sufficiently fine on the level of the Kondo temperature~\cite{Haldane1978} \(T_{\rs K}\), which
sets the width of the Kondo peak at \(z = 0\).  The green dashed line in Fig.\
\ref{fig:KondoDOS} shows the standard dot--lead result (i.e., $\Omega = 0$)
for the dot spectral function \(-\textrm{Im}G_\sigma(z)\) away from the dot
particle--hole symmetric point, i.e., \(\epsilon_\D \neq -U/2\). In this
regime, the Lacroix truncation works well, and the Kondo peak at $\omega = 0$
is framed by the Coulomb resonances at \(\omega = \epsilon_\D\) and \(\omega =
\epsilon_\D + U\) with \(\epsilon_\D  =- 2U/5\) in Fig.\ \ref{fig:KondoDOS}.
The width of the Coulomb resonances is set by \(\Gamma/2 =U/20\) (we choose
\(\Gamma / 2 = \Gamma_{\rsL} = \Gamma_{\rsR}\)), while in the present
approximation the width of the Kondo peak is given by the Kondo
temperature\cite{kashcheyevs:06,Lavagna2010,thesis:Roermund}
\begin{align}
  T_{\rs K} =(2\epsilon_\D + U)\exp\left[\frac{2\pi 
	\epsilon_\D(\epsilon_\D+U)}{\Gamma U}\right],
\end{align}
which underestimates the true width~\footnote{The expression for the exact
Kondo temperature \(\hspace{20pt} T_{\rs K}^{\rm ex}
=(U\Gamma/4)^{1/2}\exp\left[\pi \epsilon_\D(\epsilon_\D+U)/(\Gamma U)\right]
\) can be found in Ref.~[\onlinecite{Haldane1978}] along with its derivation.}
(note that \(\epsilon_\D < 0 < \epsilon_\D+U \)). 

The blue solid line (Fig.~\ref{fig:KondoDOS}) shows the same result in the
presence of a cavity (i.e., $\Omega \neq 0$) at \(\epsilon_\C = 0\): the Kondo
peak has vanished, giving way to two molecular states separated by \(12 \alpha
\Omega^2 /U\), where the enhancement
\begin{align} \label{eq:alpha}
   \alpha = \frac{1}{1 - (1+2\epsilon_\D/U)^2},
\end{align}
accounts for deviation from the dot's particle--hole symmetric point
(\(\epsilon_\D\neq-U/2\)) (see Appendix \ref{app:ed} for details).  

Such a changeover from a dot--lead (Kondo) to a dot--cavity (molecular)
singlet has been observed in the experiment [see Fig.\ \ref{blocade}(g)] (but
fails to show up in Figs.\ \ref{blocade}(a)--(c) derived within a master
equation analysis).  In Fig.\ \ref{results}, we study this competition
systematically using the results of our EOM analysis. In
Fig.~\ref{results}(a), we show the conductance at vanishing temperature $T \ll
T_{\rs K}$ as a function of the cavity level $\epsilon_\C$, corresponding to
the green dashed line in Fig.\ \ref{groundMap}(b). As the cavity level
$\epsilon_\C$ approaches $\epsilon_\F$, the formation of the molecular state
suppresses the Kondo resonance peak; the latter reappears when the split
cavity levels have crossed $\epsilon_\F$, resulting in a depression of the
Kondo peak over a distance $\sim \Omega^2/U$. As the coupling $\Omega$ is
decreased, the region where the Kondo peak is suppressed shrinks and vanishes
in the limit $\Omega \to 0$.

As discussed above, Fermi-liquid theory provides us with an upper bound on the
conductance that approximates well the true result at $\epsilon_\C = 0$, i.e.,
at maximal depression. The crosses in Fig.\ \ref{results} marking these
Fermi-liquid bounds nearly coincide with the numerical results away from small
couplings $\Omega/U > 0.01$, indicating that corrections by the leads become
perturbative when the cavity becomes resonant and the dot--cavity singlet is
formed.  At small coupling $\Omega/U$ or at large cavity detuning
$\epsilon_\C$, the violation of the Friedel sum rule by the EOM method becomes
apparent as the conductance does not reach the unitary value \(2e^2/\hbar\).

In Fig.~\ref{results}(b), we keep the large dot--cavity coupling $\Omega$ from
Fig.~\ref{results}(a) and increase the temperature $T$ instead. Again, the
Kondo resonance is suppressed, while the singlet gap stays constant. As a
result, the split cavity levels manifest themselves as pronounced
co-tunneling resonances.  Choosing a configuration away from the
particle--hole symmetric point results in an asymmetry between the two
conductance peaks.  The result of our equation-of-motion analysis is then in
good agreement with the experimental data measured along the red dashed line
in Fig.\ \ref{blocade}(g). Note the different widths of the
conductance peaks along the source bias direction, the narrow Kondo peak
of width $T_{\rs K}$ versus molecular singlet peaks extending over a region
$\Gamma_{\C}$.

The combination of exact-diagonalization-, master-equation-,
equation-of-motion-, and Fermi-liquid-theory- analysis therefore provides us
with a complete and consistent understanding of transport across the
dot--cavity system that is in agreement with the experimental findings. As a
result, we conclude that the electrons in the dot--cavity hybrid indeed form
an extended molecular singlet state that competes with the many-body Kondo
singlet.
%


\section{Summary and Outlook}
\label{sec:con}

Novel types of mesoscopic setups motivated, among other, by the prospects of
quantum computing, combine various geometrical structures, including quantum
channels, (multiple) dots, corrals, and cavities. Such structures define
elementary building blocks of quantum engineering that can be combined into
complex devices with specific functionalities. The coherent operation and
coupling of these elements is a mandatory requirement for the operation of
such devices. With the experiment in Ref.~[\onlinecite{roessler:2015}], an
important step has been made in demonstrating spin coherent operation of an
electronic device involving a quantum dot coupled to an extended cavity of
micrometer scale for the first time. This paper contributes a careful
theoretical analysis of this experiment.

In our study, we first translated the experimental setup into a theoretical
model. We have analyzed the single-particle transport properties of a cavity
with electronic injection through a quantum point contact using the numerical
package KWANT\cite{kwant} (see Fig.~\ref{fig1}). Combining our numerical
findings with analytic results for the eigenstates of semicircular disks
(Fig.~\ref{fig2}) has lead us to specific design principles for a cavity with
separated sharp levels that are strongly coupled to the quantum point contact
(see Fig.~\ref{models}). The same methodology can serve as a testbed for a rapid
and inexpensive optimization of new cavity designs, e.g., cavities connecting
several quantum dots, thereby serving as a bus for the information transfer
between qubits. In a subsequent step, we have constructed a model Hamiltonian
that combines the single particle properties of the cavity with the
interacting physics of a quantum dot, thus defining an asymmetric artificial
molecule, that is further coupled to leads in order to reproduce the transport
geometry of the experimental setup [see Fig.~\ref{models}(b)].  In constructing
this model Hamiltonian, care has to be taken to avoid the appearance of Fano
interference terms\cite{fano:61} that are not present in the experiment;
within our formulation, the absence of such resonances is due to a phase
averaging over the two-dimensional extent of the cavity (see
Fig~\ref{models2}).

Next, we have analyzed this model in three stages of increasing complexity,
exact diagonalization (ED), master equation (ME), and equation of motion (EOM)
approaches.  Isolating the artificial molecule from the lead, the size of the
involved Hilbert space became tractable and we could perform an ED study of
the isolated dot--cavity system. We thus obtained ground-state- and degeneracy
maps which already match well the shape of the spectroscopic data provided by
the experiment, confirming the validity of our dot--cavity molecular setup,
(see Figs.~\ref{groundMap} and~\ref{diamond}). In a second step, we have made
use of the ED results in a master equation approach\cite{sakurai1985,
beenakker:91, Korotkov:1994, Koch:2006, thesis:Koch}. We then obtained
transport signatures that reproduce all the main features of the experiment,
including the equilibrium and non-equilibrium transport data, ground state
maps, modulated Coulomb diamonds, and cavity assisted co-tunneling features
in the blockaded region (see Figs.~\ref{groundMap}--\ref{blocade}), with the
exception of the zero-bias Kondo resonance.  While the ME combined with the
results from ED could capture the many-body physics of the dot--cavity system,
it could not simultaneously deal with the many-body dot--lead physics
responsible for the Kondo effect. We addressed this deficiency by applying an
equation of motion analysis, using the method developed in
Refs.~[\onlinecite{kashcheyevs:06,Lavagna2010,thesis:Roermund}] enhanced by
the presence of the cavity which enters the formalism through its contribution
to the network self-energy (see Fig.~\ref{fig:KondoDOS}). Once the dot Green's
function has been found, the transport features could be calculated with the
help of the Meir-Wingreen formula\cite{Meir1992}, finally allowing us to
analyze the changeover from the Kondo dot--lead to the molecular dot--cavity
singlet (see Fig.~\ref{results}), and thus theoretically substantiating the
main experimental claim of Ref.~[\onlinecite{roessler:2015}], the observation
of spin-coherent transport across an extended quantum engineered system.

The setup discussed in this paper and further extensions thereof
provide great opportunities for future research. Examples within the present
dot--cavity system are the dependence of the molecular- to Kondo-singlet
transition on disorder, magnetic field or the level spacing in the cavity. On
the theory side, it would be interesting to apply further, systematic
methods to this problem such as the renormalization group (NRG and DMRG)
\cite{NRGReview1970,NRGReview2008,DMRGReview2005,White1992} and the Bethe
ansatz~\cite{Andrei1980,Wiegmann1983,Tsvelick1983}. Most importantly, the
ability to combine different functional elements opens the door for new
designs and experiments.  For example, it seems possible to use an electronic
cavity as a `quantum bus' which connects distant qubits, allowing them to
exchange quantum information through fully coherent operation.  Another
proposal is the study of two distant dots fused into an artificial molecule
via a cavity. Such a device would give access to interesting Kondo physics
involving a competition between several dots and leads, in particular, a
superposition of two (macroscopic) Kondo clouds that may define a Kondo cat
state.


\begin{acknowledgements}
	We thank E.\ van Nieuwenburg, T.\ Wolf, M.\ Goldstein, and R.\ Chitra for illuminating
	discussions and acknowledge financial support from the Swiss National Science Foundation, Division 2, and through the National Centre of Competence in Research ''QSIT - Quantum Science and Technology''.
	
	M.S.F. and D.O. contributed equally to this work.
\end{acknowledgements}

%
\appendix

\section{Exact diagonalization}\label{app:ed}

To perform exact diagonalization for our dot--cavity system, we choose for each 
Fock space with particle numbers $\mathbf{N}=(\Nup,\Ndo)_\sigma$ a basis 
\begin{align}
\Biggl\{
\ket{n_{\D}^{\uparrow},n_{\D}^{\downarrow}, m_{0}^\uparrow,
	m_{0}^\downarrow,m_{1}^\uparrow,
	m_{1}^\downarrow,\dots}
\,\,\Bigg|\,\,
n_{\D}^{\sigma}+\sum_j m_{j}^{\sigma}=N^{\sigma}
\Biggr\}.
\end{align}
Expressing the Hamiltonian \(H_{\D\C}\) \eqref{eq:molecule} in this basis, we
obtain a matrix $H_{\rm dc}^{\mathbf{N}}$.  The diagonalization of this matrix
provides us with the eigenenergies $\epsilon_{\Nup,\Ndo}^{\alpha}$ and the
eigenstates
\begin{align}
\ket{\psi^{\alpha}_{\Nup,\Ndo}}
&= \sum_{n_{\D}^{\sigma} + \sum_{j} m_{j}^{\sigma}=N^{\sigma}}
\mathcal{C}^{\alpha}_{n_{\D}^{\uparrow},n_{\D}^{\downarrow}, m_{0}^\uparrow,
	m_{0}^\downarrow,m_{1}^\uparrow,
	m_{1}^\downarrow,\dots}\\
&\hspace{55pt}\times\ket{n_{\D}^{\uparrow},n_{\D}^{\downarrow}, m_{0}^\uparrow,
	m_{0}^\downarrow,m_{1}^\uparrow,
	m_{1}^\downarrow,\dots}.\nonumber
\end{align}
As an example, we consider the case of a dot with a single level at
\(\epsilon_\D\) and onsite interaction \(U\), coupled via \(\Omega\) to a
single cavity mode at \(\epsilon_\C\). This case includes up to four particles
and the basis vectors for the different Fock sectors are
\begin{align}
\{\ket{0,0,0,0}\},\\
\{\ket{1,0,0,0},\ket{0,0,1,0},\ket{0,1,0,0},\ket{0,0,0,1}\},\\
\Bigg\{\ket{1,1,0,0},\frac{1}{\sqrt{2}}(\ket{1,0,0,1}+\ket{0,1,1,0}),\ket{0,0,1,1},
\nonumber \\
\frac{1}{\sqrt{2}}(\ket{1,0,0,1}-\ket{0,1,1,0}),
\ket{1,0,1,0},\ket{0,1,0,1}\Bigg\},\\
\{\ket{1,1,1,0},\ket{1,0,1,1},\ket{1,1,0,1},\ket{0,1,1,1}\},\\
\{\ket{1,1,1,1}\},
\end{align}
where we introduced a rotation in the two particle Fock sector which
simplifies the corresponding Hamiltonian. In this basis the Hamiltonians become
\begin{align}
H^{0} &= (0),\\
H^{1}&=
\begin{pmatrix}
\epsilon_\D & \Omega\\
\Omega^* & \epsilon_\C
\end{pmatrix}
\otimes
\begin{pmatrix}
1 & 0\\
0 & 1
\end{pmatrix},
\\
H^{2} &= 
\begin{pmatrix}
2\epsilon_\D+U & \sqrt{2}\Omega & 0 & 0 \\
\sqrt{2} \Omega^* &\epsilon_\D +\epsilon_\C &\sqrt{2}\Omega & 0\\
0 & \sqrt{2}\Omega^* & 2\epsilon_\C & 0\\
0 & 0 & 0 & (\epsilon_\C+\epsilon_\D) \mathds{1}_3
\end{pmatrix},
\\
H^{3} &= 
\begin{pmatrix}
2\epsilon_\D + U+\epsilon_\C & \Omega 
\\
\Omega^* & \epsilon_\D + 2 \epsilon_\C
\end{pmatrix}
\otimes
\begin{pmatrix}
1 & 0\\
0 & 1
\end{pmatrix},
\\
H^{4} &= (2\epsilon_\D+2\epsilon_\C+U),
\end{align}
where we have maintained the ordering from the basis states above and
\(\mathds{1}_3\) is the identity matrix in three dimensions such that \(H^2\)
spans the six dimensional basis of the two-particle Fock sector. The gap opens
in the Coulomb blockade region where the two-particle Fock sector ground state
has a lower energy than the ground state of the one- and three-particle Fock
sectors (see Fig.~\ref{addition}). In the region of interest \(|\epsilon_\C|
\ll U\), \(\epsilon_\D\approx-U/2\), we can expand the ground state energies
of the one-, two- and three-particle Fock sectors in \(\Omega\) and
\(\epsilon_\C\) to obtain
\begin{align}
\epsilon^1_1&\approx  
\epsilon_\D -2 \frac{|\Omega|^2}{U}\frac{1}{1-\delta_{\mathrm ph}}
\\
\epsilon^1_2&\approx\epsilon_\D - 8\frac{|\Omega|^2}{U}
\frac{1}{1-\delta_{\mathrm ph}^2} +\epsilon_\C,\\
\epsilon^1_3&\approx\epsilon_\D - 2\frac{|\Omega|^2}{U} \frac{1}{1+\delta_{\mathrm ph}}+2\epsilon_\C,
\end{align}
where \(\delta_{\mathrm ph}=1+2\epsilon_\D/U\) is the parameter which quantifies how far away from the particle hole symmetric point \(\delta_{\mathrm ph}=0\) the dot is set.
We then solve for the cavity level position where the different Fock sectors
are degenerate and obtain the conditions
\begin{align}
\epsilon_\C^\pm = \pm \frac{6 
	|\Omega|^2}{U}\left(\frac{1\pm\delta_{\mathrm ph}/3}{1-\delta_{\mathrm ph}^2}\right)
\end{align}
for the one- and two-particle, and the two- and three-particle Fock sectors
being degenerate, respectively. This leads to the minimal hybridization gap
\(12|\Omega|^2/U\) in the ground state map. This gap diverges around the
points \(\epsilon_\D=0,U\) where the coupling \(\Omega\) is no longer small
compared to the energy difference between the basis states and the above
non-degenerate perturbative approach is no longer valid. In this
configuration, one obtains a gap of order \(\Omega\), the usual result of
degenerate perturbation theory.  We use the procedure outlined above to create
the Hamiltonian matrices for larger systems and then diagonalize them
numerically. We thus produce the eigenstates that we use to produce ground
state maps (Sec.~\ref{sec:ed}) and that enter in the master equation
(Sec.~\ref{master}).

\section{Master equation approach} \label{app:mastereq}

The crucial ingredient of the master-equation approach are the rates
\eqref{eq:rate1} and \eqref{eq:rate2} which we will derive in the following.
The sequential tunneling transition rates
$W^{\alpha,\alpha'}_{\mathbf{N},\mathbf{N}'}$ from a state
$\kett{\psi_{\mathbf{N}'}^{\alpha'}}$ to state
$\kett{\psi_{\mathbf{N}}^{\alpha}}$ are given by Fermi's golden rule
\cite{bruus-flensberg}
\begin{align}
W^{\alpha,\alpha'}_{\mathbf{N},\mathbf{N}'}
=
\frac{2\pi}{\hbar} \sum_{f,i}
\bigl| \bigl\langle
\psi_{\mathbf{N}}^{\alpha},\varphi_f | \bar{H}_{\rm tun} | 
\psi_{\mathbf{N}'}^{\alpha'},\varphi_i
\bigr\rangle\bigr|^2 W_{i} \delta(E_f-E_i),
\label{eq:fermigolden}
\end{align}
with the notation $| \psi,\varphi \rangle=|\psi\rangle\otimes
|\varphi\rangle$, where $|\varphi_{i}\rangle$ and $|\varphi_{f}\rangle$
represent the initial and final lead states, $W_i$ is the probability for the
lead to be in state $|\varphi_i\rangle$, and $E_{i}$ and $E_{f}$ correspond to
the initial and final energies of the lead and molecular states. The
tunneling Hamiltonian \eqref{eq:Htun} consists of three parts, tunneling
between the dot and the left lead, between the dot and the right lead, and
between the cavity to the right lead. All of these processes have to be
treated incoherently as discussed in Sec.~\ref{sec:eff}, thus giving rise to
the rates
\begin{align}
W_{\mathbf{N},\mathbf{N}'}^{\alpha,\alpha'}=
W^{{\s \D \rsL}{\alpha,\alpha'}}_{\mathbf{N},\mathbf{N}'}
+
W^{{\s \D \rsR}{\alpha,\alpha'}}_{\mathbf{N},\mathbf{N}'}
+
W^{{\s \C \rsR}{\alpha,\alpha'}}_{\mathbf{N},\mathbf{N}'},
\end{align}
where each term is given by an expression of the form of
Eq.~\eqref{eq:fermigolden} with the corresponding part of the tunneling
Hamiltonian (see Eq.~\eqref{eq:Htun}). The sequential tunneling rates only
change the molecular occupation by one electron, i.e., $W_{\mathbf{N} \pm
\mathbf{e}_{\sigma},\mathbf{N}}^{\alpha,\alpha'}\neq 0$. For the rates, where
the molecular occupation is increased from the left lead, the initial and
final lead states differ by one electron in the left lead, i.e.,
$|\varphi_{f}\rangle=\sum_{l} c_{l\sigma}|\varphi_i\rangle$. Summing over all
initial lead states, we obtain
\begin{align}
W^{{\rs \D \rs L }{\alpha,\alpha'}}_{\mathbf{N}+\mathbf{e}_{\sigma},\mathbf{N}}
&=
\frac{2\pi}{\hbar} \sum_{l}
\bigl| \bigl\langle
\psi_{\mathbf{N}+\mathbf{e}_{\sigma}}^{\alpha}
| t_{\rs L} d_{\sigma}^\dagger | \psi_{\mathbf{N}}^{\alpha'}
\bigr\rangle\bigr|^2  
\label{eq:WdL+}\\
&\qquad \times n_{\rs F}(\epsilon_l-\mu_{\rs L})
\delta(\epsilon_{\mathbf{N}+\mathbf{e}_{\sigma}}^{\alpha}-
\epsilon_l-\epsilon_{\mathbf{N}}^{\alpha'})
\nonumber\\
&
=\frac{\Gamma_{\rs L}}{\hbar}
\bigl| \bigl\langle
\psi_{\mathbf{N}+\mathbf{e}_{\sigma}}^{\alpha}| d_{\sigma}^\dagger | 
\psi_{\mathbf{N}}^{\alpha'}
\bigr\rangle\bigr|^2
n_{\rs F}(
\epsilon_{\mathbf{N}+\mathbf{e}_{\sigma}}^{\alpha} 
-\epsilon_{\mathbf{N}}^{\alpha'}
-\mu_{\rs L})\nonumber,
\end{align}
where we used $\sum_i \langle \varphi_i
|c^\dagger_{l\sigma}c^\pdagger_{l\sigma}|\varphi_i\rangle W_{i}=n_{\rs
F}(\epsilon_l-\mu_{\rs L})$, replaced $\sum_l \rightarrow \int d\epsilon_l
\rho_{\rs L}$, and made use of the definition of the rate $\Gamma_{\rs L}=
2\pi \rho_{\rs L} \left|t_{\rm L}\right|^2$. The rate $W^{{\rs \D \rs R}
{\alpha,\alpha'}}_{\mathbf{N}+\mathbf{e}_{\sigma},\mathbf{N}}$ follows
analogously by replacing ${\rm L} \rightarrow {\rm R}$ in the expression
above.  The derivation of the rate $W^{{\s \C \rsR}
{\alpha,\alpha'}}_{\mathbf{N} + \mathbf{e}_{\sigma},\mathbf{N}}$ follows the
same arguments and the result is given by
\begin{align}
W^{{\s \C \rsR}{\alpha,\alpha'}}_{\mathbf{N}+\mathbf{e}_{\sigma},\mathbf{N}}
&=
\frac{\Gamma_\C}{\hbar}
\bigl| \bigl\langle
\psi_{\mathbf{N}+\mathbf{e}_{\sigma}}^{\alpha}
| \sum_j f_{j \sigma}^\dagger | \psi_{\mathbf{N}}^{\alpha'}
\bigr\rangle\bigr|^2
n_{\rs F}(
\epsilon_{\mathbf{N}+\mathbf{e}_{\sigma}}^{\alpha}- 
\epsilon_{\mathbf{N}}^{\alpha'}
-\mu_{\rs R})\nonumber\\
&\approx
\frac{\Gamma_\C}{\hbar}
\sum_j \bigl| \bigl\langle
\psi_{\mathbf{N}+\mathbf{e}_{\sigma}}^{\alpha}
| f_{j \sigma}^\dagger | \psi_{\mathbf{N}}^{\alpha'}
\bigr\rangle\bigr|^2
n_{\rs F}(
\epsilon_{\mathbf{N}+\mathbf{e}_{\sigma}}^{\alpha} 
-\epsilon_{\mathbf{N}}^{\alpha'}
-\mu_{\rs R}).
\label{eq:WcR+}
\end{align}
In the last step, we used the fact that the cavity level spacing is large on
the scale of temperature \(T\); in this situation, interference terms
involving different cavity levels are suppressed and we sum these processes
incoherently.  To calculate the rates which decrease the number of electrons
on the artificial molecule we make use of the relation $\sum_i \langle
\varphi_i |c_{l \sigma}c^\dagger_{l \sigma}|\varphi_i\rangle W_{i}=1-n_{\rs
F}(\epsilon_l-\mu_{\rs L})$ and obtain
\begin{align}
W^{\rs \D \rs a }_{\mathbf{N}-\mathbf{e}_{\sigma},\mathbf{N}}
&=\frac{\Gamma_{\rm a}}{\hbar}
\bigl| \bigl\langle
\psi_{\mathbf{N}-\mathbf{e}_{\sigma}}^{\alpha}| d_{\sigma} | 
\psi_{\mathbf{N}}^{\alpha'}
\bigr\rangle\bigr|^2
\nonumber\\
&\qquad\times
\bigl[1-n_{\rs F}(\epsilon_{\mathbf{N}}^{\alpha'} 
-\epsilon_{\mathbf{N}-\mathbf{e}_{\sigma}}^{\alpha}
-\mu_{\rs a})\bigr],
\label{eq:Wda-}\\
W^{\rs \C \rs R }_{\mathbf{N}-\mathbf{e}_{\sigma},\mathbf{N}}
&=\frac{\Gamma_\C }{\hbar}\sum_j
\bigl| \bigl\langle
\psi_{\mathbf{N}-\mathbf{e}_{\sigma}}^{\alpha}| f_{j\sigma} | 
\psi_{\mathbf{N}}^{\alpha'}
\bigr\rangle\bigr|^2
\nonumber\\
&\qquad\times
\bigl[1-n_{\rs F}(\epsilon_{\mathbf{N}}^{\alpha'} 
-\epsilon_{\mathbf{N}-\mathbf{e}_{\sigma}}^{\alpha}
-\mu_{\rs R})\bigr].
\label{eq:WcR-}
\end{align}
The rates \eqref{eq:WdL+} -- \eqref{eq:WcR-} constitute the main building
blocks of our master-equation calculation.  In determining these rates, we can
apply the exact diagonalization results by using the dot--cavity molecular
spectrum $\epsilon_{\mathbf{N}}^{\alpha}$ as well as the associated
eigenstates in the calculation of matrix elements of the type $\bigl\langle
\psi_{\mathbf{N} + \mathbf{e}_{\sigma}}^{\alpha} | d_{\sigma}^\dagger |
\psi_{\mathbf{N}}^{\alpha'} \bigr\rangle$ between different Fock states. We
can thus produce the conductance maps in Figs.~\ref{groundMap}--\ref{blocade}.

\section{Meir-Wingreen formula}\label{app:Meir-Wingreen}

The equation of motion method will provide us with the retarded Green's
functions \(G_\sigma\) of the dot. We will show here that in equilibrium and
at low temperatures \(T\ll\Gamma\) this quantity is sufficient to calculate
the conductance of the device. For a spin-independent tunneling, the
Meir-Wingreen formula for the current from the dot to a lead
\(\mathrm{a}=\mathrm{L,R}\) is given by\cite{Meir1992}
\begin{align*}
  I_{\rm a}  = \frac{ie}{h} \sum_\sigma \! \int \!\! 
  \mathrm{d} \omega \, \Gamma_{\rm a}(\omega)
 [i\,n_{\F}(\omega - \mu_{\rm a})  \mathrm{Im}G_\sigma (\omega)
 \!+ \!G_\sigma^<(\omega)],
\end{align*}
with the lesser Green's function \(G^<_\sigma\) to be determined with the help
of the Keldysh formalism\cite{Keldysh1964}. We consider only a single cavity
level at \(\epsilon_\C\) and drop the (artificial) Fano contribution to obtain
the energy dependent rates
\begin{align}
\Gamma_{\rsL}(\omega) &= \Gamma_{\rsL},\\
\Gamma_{\rsR}(\omega) &= \Gamma_{\rsR} 
+ \Gamma_\C \frac{|\Omega|^2}{(\epsilon_\C-\omega)^2+\Gamma_\C^2/4}.
\end{align}
Employing the symmetrization procedure of Ref.~[\onlinecite{Meir1992}], we can
rewrite the current through the dot in the form (we use that \(I_{\rsL} =  -
I_{\rsR}\))
\begin{align}
   I &= \frac{I_{\rsR}\Gamma_{\rsL} - I_{\rsL}\Gamma_{\rsR}(\epsilon_\F)}
   {\Gamma_{\rsL}+\Gamma_{\rsR}(\epsilon_\F)}.
\end{align}
The contribution to the current $I$ is limited to an energy
window of extension $T$ around the Fermi energy $\epsilon_\F$.
Limiting ourselves to temperatures smaller than the tunneling rates divided be the derivative of the tunneling rates with respect to the energy
\(T\partial_\omega \Gamma_{\rsR}\ll\Gamma_{\rsR}\), the tunneling rate to the right lead is effectively constant,
\(\Gamma_{\rsR}(\omega)\approx\Gamma_{\rsR}(\epsilon_\F=0)\), and the
symmetrization procedure of Ref.~[\onlinecite{Meir1992}] can be carried
through. As a result, we obtain an expression for the current that does no
longer depends on $G^<$,
\begin{align}
I \approx
\frac{2e}{h}\tilde{\Gamma}\sum_\sigma \!\!\int\!\!\! \mathrm{d} \omega 
[n_{\F}(\omega\!-\! \mu_{\rsL}) - n_{\F}(\omega\!-\! \mu_{\rsR})]
\mathrm{Im} G_\sigma(\omega), 
\end{align}
with the rate 
\begin{align}\label{eq:Ga}
\tilde{\Gamma} = \frac{\Gamma_{\rsL} \Gamma_{\rsR} (\epsilon_\F)}
{\Gamma_{\rsL} +\Gamma_{\rsR} (\epsilon_\F)}.
\end{align}
Differentiating with respect to the bias voltage \(V = -\mu_{\rsL}/e\) and
taking the equilibrium limit, we arrive at a formula which relates the
conductance to the retarded Green's function of the dot,
\begin{align}
G_0 = \lim_{V\to0}\frac{\mathrm{d}I}{\mathrm{d}V} \approx 
-\frac{2 e^2}{h}\tilde{\Gamma} \sum_\sigma \int \mathrm{d} \omega \, 
\frac{\beta \mathrm{Im} G_\sigma ( \omega ) }
{4 \cosh^2 \frac{\beta\omega}{2}}.
\end{align}
Combining this result with the equation of motion method described in
Sec.~\ref{sec:eom}, we can determine the conductance of the device including
the Kondo effect.

\section{Equation of motion method}\label{app:eom}

In this Appendix, we discuss some more technical aspects of the
equation-of-motion method as found in the
literature~\cite{kashcheyevs:06,Lavagna2010,thesis:Roermund}. We start with
the equations of motion for the required correlators and their truncation and
then add some explanations on the derivation of the integral representations
of the \(P\) and \(Q\) functions [Eqs.~\eqref{eq:P} and~\eqref{eq:Q}]. 

Taking time derivatives on successive Green's functions, we obtain the
following sets of equations of motion (after transformation to frequency
space); the dot Green's function $\langle\! \langle
d^\pdagger_\sigma;d^\dagger_\sigma \rangle\! \rangle$,
\begin{align}
(z-e_\D)
\langle\! \langle d^\pdagger_\sigma;d^\dagger_\sigma \rangle\! \rangle
&= 
1 +\sum_k t_k \langle\! \langle c_{k\sigma}^\pdagger; d_\D^\dagger \rangle 
\!\rangle \\ \nonumber
&\qquad\quad +U\langle\! \langle n^\pdagger_{\bar{\sigma}} 
d^\pdagger_\sigma;d^\dagger_\sigma 
\rangle\!\rangle
\end{align}
couples to the two-point correlator involving a lead electron
\begin{align}
(z-\epsilon_k) 
\langle\! \langle c_{k\sigma}^\pdagger; d_\D^\dagger \rangle\! \rangle 
&=
t_k^*\langle\! \langle d^\pdagger_\sigma;d^\dagger_\sigma \rangle\!  \rangle,
\end{align}
and to the four-point correlator with an additional particle on the dot
\begin{widetext}
	\begin{align}
	(z-\epsilon_\D-U )
	\langle\! \langle 
	n_{\bar{\sigma}}d^\pdagger_\sigma;d^\dagger_\sigma \rangle\!\rangle
	&=
	\langle n_{\bar{\sigma}} \rangle + \sum_{k}[
	t_k\langle\! \langle n_{\bar{\sigma}}c_{k\sigma};d^\dagger_\sigma \rangle 
	\!\rangle
	+t_k\langle\! \langle d_{\bar{\sigma}}^\dagger
	c_{k\bar{\sigma}} d_\sigma^\pdagger;d^\dagger_\sigma \rangle\! \rangle
	-t_k^*\langle\! \langle c_{k\bar{\sigma}}^\dagger d_{\bar{\sigma}}^\pdagger 
	d_\sigma^\pdagger;d^\dagger_\sigma \rangle\! \rangle].
	\end{align}
	The latter couples to further four-point correlators involving lead electrons,
	\begin{align}
	\label{eq:eom2}
	(z-\epsilon_k)
	\langle\! \langle n_{\bar{\sigma}}c_{k\sigma};d^\dagger_\sigma \rangle 
	\!\rangle &=
	t_k^* \langle\! \langle 
	n_{\bar{\sigma}}d^\pdagger_\sigma;d^\dagger_\sigma \rangle\!\rangle
	+ \sum_{k'}^{}\left[
	t_{k'} \langle\! \langle
	d_{\bar{\sigma}}^\dagger c_{k'{\bar{\sigma}}}^\pdagger c_{k\sigma} ; 
	d^\dagger_\sigma
	\rangle\! \rangle
	-t_{k'}^* \langle\! \langle
	c_{k'\bar{\sigma}}^\dagger d_{{\bar{\sigma}}}^\pdagger c_{k\sigma} ; 
	d^\dagger_\sigma
	\rangle\! \rangle
	\right], \\
	\label{eq:KondoEOM2}
	(z-\epsilon_k)
	\langle\! \langle d_{\bar{\sigma}}^\dagger
	c_{k\bar{\sigma}} d_\sigma^\pdagger;d^\dagger_\sigma \rangle\! \rangle &=
	\langle d^\dagger_{\bar{\sigma}} c_{k\bar{\sigma}}^\pdagger \rangle +
	t_k^* \langle\! \langle 
	n_{\bar{\sigma}}d^\pdagger_\sigma;d^\dagger_\sigma \rangle\!\rangle
	+ \sum_{k'}^{}\left[
	t_{k'} \langle\! \langle
	d_{\bar{\sigma}}^\dagger c_{k{\bar{\sigma}}}^\pdagger c_{k'\sigma}^\pdagger 
	; 
	d^\dagger_\sigma
	\rangle\! \rangle
	-t_{k'}^* \langle\! \langle
	c_{k'\bar{\sigma}}^\dagger c_{k{\bar{\sigma}}}^\pdagger d_{\sigma}^\pdagger 
	; 
	d^\dagger_\sigma
	\rangle\! \rangle
	\right],\\
	\label{eq:KondoEOM3}
	(z-2\epsilon_\D-U+\epsilon_k)
	\langle\! \langle c_{k\bar{\sigma}}^\dagger d_{\bar{\sigma}}^\pdagger 
	d_\sigma^\pdagger;d^\dagger_\sigma \rangle\! \rangle &=
	\langle c_{k\bar{\sigma}}^\dagger  d^\pdagger_{\bar{\sigma}} \rangle -
	t_k \langle\! \langle 
	n_{\bar{\sigma}}d^\pdagger_\sigma;d^\dagger_\sigma \rangle\!\rangle
	+ \sum_{k'}^{}\left[
	t_{k'} \langle\! \langle
	c_{k\bar{\sigma}}^\dagger d_{{\bar{\sigma}}}^\pdagger c_{k'\sigma}^\pdagger 
	; 
	d^\dagger_\sigma
	\rangle\! \rangle
	+t_{k'} \langle\! \langle
	c_{k\bar{\sigma}}^\dagger c_{k'{\bar{\sigma}}}^\pdagger d_{\sigma}^\pdagger 
	; 
	d^\dagger_\sigma
	\rangle\! \rangle
	\right].
	\end{align}
\end{widetext}
We truncate the sequence at the lowest level that includes the Kondo physics
we are interested in.  In decoupling the four-point correlators, we
consistently decouple expressions with the same number of lead operators. This
provides us with three truncation schemes, the mean-field zeroth-order
\(O(t^0)\) truncation
\begin{align}
\langle\! \langle n^\pdagger_{\bar{\sigma}} 
d^\pdagger_\sigma;d^\dagger_\sigma 
\rangle\!\rangle \to \langle n_{\bar{\sigma}}^\pdagger \rangle \langle\! \langle 
d^\pdagger_\sigma;d^\dagger_\sigma \rangle\! \rangle,
\end{align}
the \(O(t)\) truncation,
\begin{align}
\langle\! \langle n_{\bar{\sigma}}c_{k\sigma};d^\dagger_\sigma \rangle\! \rangle
&\to \langle  n_{\bar{\sigma}}\rangle 
\langle\! \langle c_{k\sigma};d^\dagger_\sigma \rangle\! \rangle,\\
\langle\! \langle d_{\bar{\sigma}}^\dagger
c_{k\bar{\sigma}} d_\sigma^\pdagger;d^\dagger_\sigma \rangle\! \rangle
&\to \langle   d_{\bar{\sigma}}^\dagger
c_{k\bar{\sigma}}\rangle\langle\! \langle d_\sigma^\pdagger;d^\dagger_\sigma 
\rangle\! \rangle, \\
\langle\! \langle c_{k\bar{\sigma}}^\dagger d_{\bar{\sigma}}^\pdagger 
d_\sigma^\pdagger;d^\dagger_\sigma \rangle\! \rangle
&\to
\langle c_{k\bar{\sigma}}^\dagger d_{\bar{\sigma}}^\pdagger 
\rangle 
\langle\! \langle d_\sigma^\pdagger;d^\dagger_\sigma \rangle\! \rangle,
\end{align}
and finally the Lacroix \(O(t^2)\) truncation,
\begin{align}
\langle\! \langle
d_{\bar{\sigma}}^\dagger c_{k{\bar{\sigma}}}^\pdagger c_{k'\sigma}^\pdagger ; 
d^\dagger_\sigma
\rangle\! \rangle
&\to \langle d_{\bar{\sigma}}^\dagger c_{k{\bar{\sigma}}}^\pdagger \rangle
\langle\! \langle c_{k'\sigma}^\pdagger ; d^\dagger_\sigma \rangle\! \rangle,\\
\langle\! \langle
c_{k\bar{\sigma}}^\dagger d_{{\bar{\sigma}}}^\pdagger c_{k'\sigma}^\pdagger ; 
d^\dagger_\sigma
\rangle\! \rangle
&\to \langle c_{k\bar{\sigma}}^\dagger d_{{\bar{\sigma}}}^\pdagger \rangle
\langle\! \langle c_{k'\sigma}^\pdagger ; d^\dagger_\sigma \rangle\! \rangle,\\
\langle\! \langle
c_{k\bar{\sigma}}^\dagger c_{k'{\bar{\sigma}}}^\pdagger d_{\sigma}^\pdagger ; 
d^\dagger_\sigma
\rangle\! \rangle
&\to
\langle c_{k\bar{\sigma}}^\dagger c_{k'{\bar{\sigma}}}^\pdagger \rangle
\langle\! \langle d_{\sigma}^\pdagger ; d^\dagger_\sigma \rangle\! \rangle.
\end{align}
Note that all other decoupling terms vanish because the system
Hamiltonian~\eqref{eq:effHam} is particle and spin conserving.  We use the
spectral theorem [see \eqref{spectral}], for general fermionic operators \(A\)
and \(B\) to find the expectation values
\begin{align}
\label{eq:spectral}
\langle BA \rangle = \frac{i}{2\pi}  \oint dz \, n_{\rs F}(z) 
\langle\! \langle A;B \rangle\! \rangle.
\end{align} 
Combining the spectral theorem with the equations of motion~\eqref{eq:eom}, we
immediately find that
\begin{align}
t_k   \langle d_{\bar{\sigma}}^\dagger  c_{k\bar{\sigma}}^\pdagger  \rangle =
t_k^* \langle c_{k\bar{\sigma}}^\dagger d_{{\bar{\sigma}}}^\pdagger \rangle,
\end{align}
which greatly simplifies the \(O(t)\)  truncation.

The functions \(P\) and \(Q\) [Eqs.\ \eqref{eq:P} and \eqref{eq:Q}], can be
brought into an integral form by using the spectral
theorem~\eqref{eq:spectral}, the equation of motion~\eqref{eq:eom}, and some
algebra,
\begin{align} \nonumber
P_\sigma(z) &\equiv \sum_k \frac{t_k \langle d_\sigma^\dagger 
	c^\pdagger_{k\sigma}\rangle}{z-\epsilon_k} 
\\ \nonumber
&=\frac{i}{2\pi} \oint dz' n_{\rs F}(z') 
\sum_k \frac{t_k 
	\langle\! \langle c^\pdagger_{k\sigma} ; d_\sigma^\dagger\rangle 
	\!\rangle_{z'}}{z-\epsilon_k} 
\\ \nonumber
&=\frac{i}{2\pi}  \oint dz' n_{\rs F}(z') G_\sigma(z')
\sum_k \frac{|t_k|^2 
}{(z-\epsilon_k)(z'-\epsilon_k)}\\
&=\frac{i}{2\pi}  \oint dz' n_{\rs F}(z') G_\sigma(z')
\frac{\Sigma(z')-\Sigma(z)}{z-z'}.
\end{align}
We find the analogous expression for $Q_\sigma(z)$
\begin{align}
Q_\sigma (z)& \equiv \sum_{kk'} \frac{t_{k'} t_k^*\langle c_{k\sigma}^\dagger
	c_{k'\sigma}^\pdagger\rangle}{z-\epsilon_{k'}}\\
&= \frac{i}{2\pi}  \oint dz' n_{\rs F}(z') [1+\Sigma(z')G_\sigma(z')]
\frac{\Sigma(z')-\Sigma(z)}{z-z'}, \nonumber
\end{align}
by successively applying the two versions of the EOM~\eqref{eq:eom} to the
lead--lead correlator
\begin{align}
\langle\! \langle c_{k'\sigma}^\pdagger;c_{k\sigma}^\dagger\rangle\! \rangle
= \frac{\delta_{kk'}}{z-\epsilon_{k'}}+
\frac{t_kt_{k'}^*}{(z-\epsilon_k)(z-\epsilon_{k'})}G_\sigma.
\end{align}  
When attempting to remove the spurious peak in the density of states in
Lacroix truncation scheme one has to include a self-energy to the equation of
motion \eqref{eq:eom2}. This can be done rigorously as a further expansion in
\(t^2\), the so called \(O(t^4)\) truncation, yielding the self-energies to
equations~\eqref{eq:eom2}--\eqref{eq:KondoEOM3}; details can be found in
Refs.~[\onlinecite{Lavagna2010}] and~[\onlinecite{thesis:Roermund}]. In
Sec.~\ref{sec:eom} we stay away from the dot particle--hole symmetric point
(\(\epsilon_\D\neq-U/2\)) and thus avoid the anomaly.

\section{Network self-energy}\label{app:selfenergy}

In any equation-of-motion method accounting for the lead states, the latter
will appear explicitly only in the expression for the network self-energy,
defined as the self-energy of the full Hamiltonian in the absence of the dot.
Here, we show that the Hamiltonian with an energy-dependent tunneling to the
right lead~\eqref{eq:ham} and the Hamiltonian with discrete cavity
levels~\eqref{eq:effHam} give rise to the same self-energy expression and are
therefore equivalent in any truncation scheme that treats lead states on the
same footing.  The derivation of the self-energy is given in
Ref.~[\onlinecite{kashcheyevs:06}].  We focus first on our cavity model given
by the Hamiltonian \eqref{eq:effHam} with discrete cavity levels.  We consider
three independent contributions $\Sigma(z)=\Sigma_{\rsL}(z) + \Sigma_{\rsR}(z)
+ \Sigma_{\rm cav}(z)$, originating from the coupling to the source and drain
lead, $\Sigma_{\rsL}(z)$ and $\Sigma_{\rsR}(z)$, and to the cavity,
$\Sigma_{\rm cav}(z)$. The left and right leads are automatically independent
while the right lead and cavity contributions are treated such as not to
include artificial Fano resonances (see Sec.~\ref{sec:eff}). With all
microscopic quantities taken to be spin-independent, so are the self-energies.
The lead self-energies are given by $\Sigma_\alpha(z)= \sum_{k} |t_{\alpha
	k}|^2/(z-\epsilon_{k})$, resulting in
\begin{align}
\Sigma_{\alpha}(\omega\pm i\eta)&=\int d\omega' 
\frac{\rho_{\alpha}|t_{\alpha}|^2}{\omega-\omega'\pm i\eta}=\mp i\pi 
\rho_{\alpha}|t_{\alpha}|^2\nonumber\\
&\equiv\mp i\Gamma_\alpha/2,
\end{align}
and giving rise to an effective width $(\Gamma_{\rsL} +\Gamma_{\rsR})/2$ to
the level.  The cavity contribution is given by
\begin{align} \label{eq:cavSelf}
\Sigma_{\rm cav}(z)
=\sum_{j,l} \Omega^\ast_j[\mathcal{M}^{-1}(z)]_{jl} \Omega_l,
\end{align}
with $\mathcal{M}_{jl}(z)=(z -
\epsilon_{\C}^{(j)})\delta_{jl}-\tilde{\Sigma}_{jl}(z)$ the matrix which
diagonalizes the right lead (excluding its coupling to the dot) (see the
Appendix of Ref.~[\onlinecite{entin-wohlman:05}]).  Here, $\tilde{\Sigma}(z)$
is the self-energy resulting from the coupling of the cavity to the drain, in
particular, the diagonal elements $\tilde{\Sigma}_{jj}(\omega\pm i\eta)=\mp
i\Gamma_j/2 = \mp i \Gamma_\C/2$, where $\Gamma_\C$ is the rate defined
in~\eqref{eq:cavRate}, give rise to a finite width of the cavity levels, while
the off-diagonal elements $\tilde{\Sigma}_{jl}$ couple the different levels.
Neglecting the coupling between the levels, $\tilde{\Sigma}_{jl} = 0$, a
good approximation for $\delta_\C\gg \Gamma_{j}$, the inversion of
\(\mathcal{M}\) in~\eqref{eq:cavSelf} becomes trivial and we obtain
\begin{align}
\Sigma_{\rm cav}(\omega\pm 
i\eta)\approx\sum_{j}\frac{|\Omega_j|^2}{\omega-\epsilon_{\C}^{(j)}\pm 
	i\Gamma_j/2}.
\end{align}
Summing up all contributions, we obtain the network self-energy
\begin{align}
\Sigma(\omega\pm i \eta)&\approx
\mp i(\Gamma_{\rsL}+\Gamma_{\rsR})/2
+\sum_j
\frac{|\Omega_j|^2}{\omega-\epsilon_{c}^{(j)}\pm i\Gamma_j/2}.
\label{eq:selfenergyII}
\end{align}

Let us shortly consider the self-energy originating from the original model
\eqref{eq:ham}, where it consists of two contributions, $\Sigma(z) =
\Sigma_{\rsL}(z)+\Sigma_{\rsR}(z)$.  with $\Sigma_\alpha(z)= \sum_{k}
|t_{\alpha k}|^2/(z-\epsilon_{k})$. The contribution from the source lead is
unchanged, i.e., $\Sigma_{\rsL}(\omega\pm i\eta)=\mp i\Gamma_{\rsL}/2$.  For
the right lead, using the energy-dependent $t_{\rsR}(\omega)$ given
in Eq.~\eqref{eq:deft}, the self-energy is given by the expression
\begin{align}
\Sigma_{\rsR}(z)&=\int d \omega' \frac{\rho_{\rsR}}{z-\omega'}
\biggl| t_{\rsR} + \sum_j 
\frac{\lambda_j}{\omega'-\epsilon_{\C}^{(j)}+i\Gamma_j/2} \biggr|^2.
\end{align}
Note that the interference of the first and second terms leads to Fano
resonances, an artifact of the effective model (0D coupled to Fermi leads). We can easily avoid
such terms by considering instead
\begin{align}
\Sigma_{\rsR}(z)&=\int d\omega' \frac{\rho_{\rsR}}{z-\omega'} | t_{\rsR}|^2 
\nonumber\\
&\qquad+ \int d\omega' \frac{\rho_{\rsR}}{z-\omega'} 
\biggl|\sum_j \frac{\lambda_{j}}{\omega'-\epsilon_{\C}^{(j)}+i\Gamma_j/2} 
\biggr|^2,
\end{align}
where the first integral describes the unstructured lead, while the second one
originates from the cavity.  We consider the situation, where the cavity-level
separation is much larger than their width, i.e., $\delta_\C\gg \Gamma_{j}$,
leading to the approximate expression
\begin{align}
\Sigma_{\rsR}(\omega\pm i\eta)&\approx\int d\omega'
\frac{\rho_{\rsR}}{\omega-\omega'\pm i\eta} | t_{\rsR}|^2 
\nonumber\\
&\qquad+ \sum_j\int d\omega' \frac{\rho_{\rsR}}{\omega-\omega'\pm i\eta} 
\biggl| \frac{\lambda_{j}}{\omega'-\epsilon_{\C}^{(j)}+i\Gamma_j/2} \biggr|^2
\nonumber\\
&=
\mp i\Gamma_{\rsR}/2
+\sum_j
\frac{2\pi \rho_{\rsR} |\lambda_{j}|^2/\Gamma_j}{\omega-\epsilon_{\C}^{(j)}\pm 
	i \Gamma_j/2}.
\label{eq:selfenergyI}
\end{align}
Comparing to the result in Eq.~\eqref{eq:selfenergyII}, we see that the two
expressions coincide if we relate the two models via $\lambda_{j}=\Omega_j
t_j$.  Hence, any EOM approach treating the cavity on the same level
as the lead states will produce the same results for both our models. We use
the self-energy derived in this appendix along with the EOM method from
Sec.~\ref{sec:eom} to solve for the Green's function of the dot
self-consistently [see Eq.~\eqref{eq:green}]. The imaginary part \({\rm Im} G\)
of these Green's functions is shown in Fig.~\ref{fig:KondoDOS} and appears
indirectly through the Meir-Wingreen formula (see
Appendix~\ref{app:Meir-Wingreen}) in Fig.\ \ref{results}.


\begin{thebibliography}{67}%
	\makeatletter
	\providecommand \@ifxundefined [1]{%
		\@ifx{#1\undefined}
	}%
	\providecommand \@ifnum [1]{%
		\ifnum #1\expandafter \@firstoftwo
		\else \expandafter \@secondoftwo
		\fi
	}%
	\providecommand \@ifx [1]{%
		\ifx #1\expandafter \@firstoftwo
		\else \expandafter \@secondoftwo
		\fi
	}%
	\providecommand \natexlab [1]{#1}%
	\providecommand \enquote  [1]{``#1''}%
	\providecommand \bibnamefont  [1]{#1}%
	\providecommand \bibfnamefont [1]{#1}%
	\providecommand \citenamefont [1]{#1}%
	\providecommand \href@noop [0]{\@secondoftwo}%
	\providecommand \href [0]{\begingroup \@sanitize@url \@href}%
	\providecommand \@href[1]{\@@startlink{#1}\@@href}%
	\providecommand \@@href[1]{\endgroup#1\@@endlink}%
	\providecommand \@sanitize@url [0]{\catcode `\\12\catcode `\$12\catcode
		`\&12\catcode `\#12\catcode `\^12\catcode `\_12\catcode `\%12\relax}%
	\providecommand \@@startlink[1]{}%
	\providecommand \@@endlink[0]{}%
	\providecommand \url  [0]{\begingroup\@sanitize@url \@url }%
	\providecommand \@url [1]{\endgroup\@href {#1}{\urlprefix }}%
	\providecommand \urlprefix  [0]{URL }%
	\providecommand \Eprint [0]{\href }%
	\providecommand \doibase [0]{http://dx.doi.org/}%
	\providecommand \selectlanguage [0]{\@gobble}%
	\providecommand \bibinfo  [0]{\@secondoftwo}%
	\providecommand \bibfield  [0]{\@secondoftwo}%
	\providecommand \translation [1]{[#1]}%
	\providecommand \BibitemOpen [0]{}%
	\providecommand \bibitemStop [0]{}%
	\providecommand \bibitemNoStop [0]{.\EOS\space}%
	\providecommand \EOS [0]{\spacefactor3000\relax}%
	\providecommand \BibitemShut  [1]{\csname bibitem#1\endcsname}%
	\let\auto@bib@innerbib\@empty
	\bibitem [{\citenamefont {Ihn}(2010)}]{ihn2010semiconductor}%
	\BibitemOpen
	\bibfield  {author} {\bibinfo {author} {\bibfnamefont {T.}~\bibnamefont
			{Ihn}},\ }\href@noop {} {\emph {\bibinfo {title} {{Semiconductor
					Nanostructures: Quantum states and electronic transport}}}}\ (\bibinfo
	{publisher} {Oxford University Press},\ \bibinfo {year} {2010})\BibitemShut
	{NoStop}%
	\bibitem [{\citenamefont {Kouwenhoven}\ \emph {et~al.}(2001)\citenamefont
		{Kouwenhoven}, \citenamefont {Austing},\ and\ \citenamefont
		{Tarucha}}]{kouwenhoven_few-electron_2001}%
	\BibitemOpen
	\bibfield  {author} {\bibinfo {author} {\bibfnamefont {L.~P.}\ \bibnamefont
			{Kouwenhoven}}, \bibinfo {author} {\bibfnamefont {D.~G.}\ \bibnamefont
			{Austing}}, \ and\ \bibinfo {author} {\bibfnamefont {S.}~\bibnamefont
			{Tarucha}},\ }\href {\doibase 10.1088/0034-4885/64/6/201} {\bibfield
		{journal} {\bibinfo  {journal} {Reports on Progress in Physics}\ }\textbf
		{\bibinfo {volume} {64}},\ \bibinfo {pages} {701} (\bibinfo {year}
		{2001})}\BibitemShut {NoStop}%
	\bibitem [{\citenamefont {Goldhaber-Gordon}\ \emph {et~al.}(1998)\citenamefont
		{Goldhaber-Gordon}, \citenamefont {Shtrikman}, \citenamefont {Mahalu},
		\citenamefont {Abusch-Magder}, \citenamefont {Meirav},\ and\ \citenamefont
		{Kastner}}]{goldhaber-gordon_kondo_1998}%
	\BibitemOpen
	\bibfield  {author} {\bibinfo {author} {\bibfnamefont {D.}~\bibnamefont
			{Goldhaber-Gordon}}, \bibinfo {author} {\bibfnamefont {H.}~\bibnamefont
			{Shtrikman}}, \bibinfo {author} {\bibfnamefont {D.}~\bibnamefont {Mahalu}},
		\bibinfo {author} {\bibfnamefont {D.}~\bibnamefont {Abusch-Magder}}, \bibinfo
		{author} {\bibfnamefont {U.}~\bibnamefont {Meirav}}, \ and\ \bibinfo {author}
		{\bibfnamefont {M.~A.}\ \bibnamefont {Kastner}},\ }\href {\doibase
		10.1038/34373} {\bibfield  {journal} {\bibinfo  {journal} {Nature}\ }\textbf
		{\bibinfo {volume} {391}},\ \bibinfo {pages} {156} (\bibinfo {year}
		{1998})}\BibitemShut {NoStop}%
	\bibitem [{\citenamefont {Kouwenhoven}\ and\ \citenamefont
		{Glazman}(2001)}]{kouwenhoven_revival_2001}%
	\BibitemOpen
	\bibfield  {author} {\bibinfo {author} {\bibfnamefont {L.~P.}\ \bibnamefont
			{Kouwenhoven}}\ and\ \bibinfo {author} {\bibfnamefont {L.}~\bibnamefont
			{Glazman}},\ }\href@noop {} {\bibfield  {journal} {\bibinfo  {journal}
			{Physics World}\ }\textbf {\bibinfo {volume} {14}},\ \bibinfo {pages} {33}
		(\bibinfo {year} {2001})}\BibitemShut {NoStop}%
	\bibitem [{\citenamefont {Thomas}\ \emph {et~al.}(1996)\citenamefont {Thomas},
		\citenamefont {Nicholls}, \citenamefont {Simmons}, \citenamefont {Pepper},
		\citenamefont {Mace},\ and\ \citenamefont {Ritchie}}]{thomas:96}%
	\BibitemOpen
	\bibfield  {author} {\bibinfo {author} {\bibfnamefont {K.~J.}\ \bibnamefont
			{Thomas}}, \bibinfo {author} {\bibfnamefont {J.~T.}\ \bibnamefont
			{Nicholls}}, \bibinfo {author} {\bibfnamefont {M.~Y.}\ \bibnamefont
			{Simmons}}, \bibinfo {author} {\bibfnamefont {M.}~\bibnamefont {Pepper}},
		\bibinfo {author} {\bibfnamefont {D.~R.}\ \bibnamefont {Mace}}, \ and\
		\bibinfo {author} {\bibfnamefont {D.~A.}\ \bibnamefont {Ritchie}},\ }\href
	{\doibase 10.1103/PhysRevLett.77.135} {\bibfield  {journal} {\bibinfo
			{journal} {Phys. Rev. Lett.}\ }\textbf {\bibinfo {volume} {77}},\ \bibinfo
		{pages} {135} (\bibinfo {year} {1996})}\BibitemShut {NoStop}%
	\bibitem [{\citenamefont {Cronenwett}\ \emph {et~al.}(2002)\citenamefont
		{Cronenwett}, \citenamefont {Lynch}, \citenamefont {Goldhaber-Gordon},
		\citenamefont {Kouwenhoven}, \citenamefont {Marcus}, \citenamefont {Hirose},
		\citenamefont {Wingreen},\ and\ \citenamefont
		{Umansky}}]{cronenwett_low-temperature_2002}%
	\BibitemOpen
	\bibfield  {author} {\bibinfo {author} {\bibfnamefont {S.~M.}\ \bibnamefont
			{Cronenwett}}, \bibinfo {author} {\bibfnamefont {H.~J.}\ \bibnamefont
			{Lynch}}, \bibinfo {author} {\bibfnamefont {D.}~\bibnamefont
			{Goldhaber-Gordon}}, \bibinfo {author} {\bibfnamefont {L.~P.}\ \bibnamefont
			{Kouwenhoven}}, \bibinfo {author} {\bibfnamefont {C.~M.}\ \bibnamefont
			{Marcus}}, \bibinfo {author} {\bibfnamefont {K.}~\bibnamefont {Hirose}},
		\bibinfo {author} {\bibfnamefont {N.~S.}\ \bibnamefont {Wingreen}}, \ and\
		\bibinfo {author} {\bibfnamefont {V.}~\bibnamefont {Umansky}},\ }\href
	{\doibase 10.1103/PhysRevLett.88.226805} {\bibfield  {journal} {\bibinfo
			{journal} {Physical Review Letters}\ }\textbf {\bibinfo {volume} {88}},\
		\bibinfo {pages} {226805} (\bibinfo {year} {2002})}\BibitemShut {NoStop}%
	\bibitem [{\citenamefont {Hirose}\ \emph {et~al.}(2003)\citenamefont {Hirose},
		\citenamefont {Meir},\ and\ \citenamefont {Wingreen}}]{hirose:03}%
	\BibitemOpen
	\bibfield  {author} {\bibinfo {author} {\bibfnamefont {K.}~\bibnamefont
			{Hirose}}, \bibinfo {author} {\bibfnamefont {Y.}~\bibnamefont {Meir}}, \ and\
		\bibinfo {author} {\bibfnamefont {N.~S.}\ \bibnamefont {Wingreen}},\ }\href
	{\doibase 10.1103/PhysRevLett.90.026804} {\bibfield  {journal} {\bibinfo
			{journal} {Phys. Rev. Lett.}\ }\textbf {\bibinfo {volume} {90}},\ \bibinfo
		{pages} {026804} (\bibinfo {year} {2003})}\BibitemShut {NoStop}%
	\bibitem [{\citenamefont {Rejec}\ and\ \citenamefont {Meir}(2006)}]{meir-07}%
	\BibitemOpen
	\bibfield  {author} {\bibinfo {author} {\bibfnamefont {T.}~\bibnamefont
			{Rejec}}\ and\ \bibinfo {author} {\bibfnamefont {Y.}~\bibnamefont {Meir}},\
	}\href {http://dx.doi.org/10.1038/nature05054} {\bibfield  {journal}
	{\bibinfo  {journal} {Nature}\ }\textbf {\bibinfo {volume} {442}},\ \bibinfo
	{pages} {900} (\bibinfo {year} {2006})}\BibitemShut {NoStop}%
\bibitem [{\citenamefont {de~Picciotto}\ \emph {et~al.}(1997)\citenamefont
	{de~Picciotto}, \citenamefont {Reznikov}, \citenamefont {Heiblum},
	\citenamefont {Umansky}, \citenamefont {Bunin},\ and\ \citenamefont
	{Mahalu}}]{picciotto:97}%
\BibitemOpen
\bibfield  {author} {\bibinfo {author} {\bibfnamefont {R.}~\bibnamefont
		{de~Picciotto}}, \bibinfo {author} {\bibfnamefont {M.}~\bibnamefont
		{Reznikov}}, \bibinfo {author} {\bibfnamefont {M.}~\bibnamefont {Heiblum}},
	\bibinfo {author} {\bibfnamefont {V.}~\bibnamefont {Umansky}}, \bibinfo
	{author} {\bibfnamefont {G.}~\bibnamefont {Bunin}}, \ and\ \bibinfo {author}
	{\bibfnamefont {D.}~\bibnamefont {Mahalu}},\ }\href
{http://dx.doi.org/10.1038/38241} {\bibfield  {journal} {\bibinfo  {journal}
		{Nature}\ }\textbf {\bibinfo {volume} {389}},\ \bibinfo {pages} {162}
	(\bibinfo {year} {1997})}\BibitemShut {NoStop}%
\bibitem [{\citenamefont {Saminadayar}\ \emph {et~al.}(1997)\citenamefont
	{Saminadayar}, \citenamefont {Glattli}, \citenamefont {Jin},\ and\
	\citenamefont {Etienne}}]{saminadayar:97}%
\BibitemOpen
\bibfield  {author} {\bibinfo {author} {\bibfnamefont {L.}~\bibnamefont
		{Saminadayar}}, \bibinfo {author} {\bibfnamefont {D.~C.}\ \bibnamefont
		{Glattli}}, \bibinfo {author} {\bibfnamefont {Y.}~\bibnamefont {Jin}}, \ and\
	\bibinfo {author} {\bibfnamefont {B.}~\bibnamefont {Etienne}},\ }\href
{\doibase 10.1103/PhysRevLett.79.2526} {\bibfield  {journal} {\bibinfo
		{journal} {Phys. Rev. Lett.}\ }\textbf {\bibinfo {volume} {79}},\ \bibinfo
	{pages} {2526} (\bibinfo {year} {1997})}\BibitemShut {NoStop}%
\bibitem [{\citenamefont {Dolev}\ \emph {et~al.}(2008)\citenamefont {Dolev},
	\citenamefont {Heiblum}, \citenamefont {Umansky}, \citenamefont {Stern},\
	and\ \citenamefont {Mahalu}}]{dolev_observation_2008}%
\BibitemOpen
\bibfield  {author} {\bibinfo {author} {\bibfnamefont {M.}~\bibnamefont
		{Dolev}}, \bibinfo {author} {\bibfnamefont {M.}~\bibnamefont {Heiblum}},
	\bibinfo {author} {\bibfnamefont {V.}~\bibnamefont {Umansky}}, \bibinfo
	{author} {\bibfnamefont {A.}~\bibnamefont {Stern}}, \ and\ \bibinfo {author}
	{\bibfnamefont {D.}~\bibnamefont {Mahalu}},\ }\href {\doibase
	10.1038/nature06855} {\bibfield  {journal} {\bibinfo  {journal} {Nature}\
	}\textbf {\bibinfo {volume} {452}},\ \bibinfo {pages} {829} (\bibinfo {year}
	{2008})}\BibitemShut {NoStop}%
\bibitem [{\citenamefont {Katine}\ \emph {et~al.}(1997)\citenamefont {Katine},
	\citenamefont {Eriksson}, \citenamefont {Adourian}, \citenamefont
	{Westervelt}, \citenamefont {Edwards}, \citenamefont {Lupu-Sax},
	\citenamefont {Heller}, \citenamefont {Campman},\ and\ \citenamefont
	{Gossard}}]{katine_point_1997}%
\BibitemOpen
\bibfield  {author} {\bibinfo {author} {\bibfnamefont {J.~A.}\ \bibnamefont
		{Katine}}, \bibinfo {author} {\bibfnamefont {M.~A.}\ \bibnamefont
		{Eriksson}}, \bibinfo {author} {\bibfnamefont {A.~S.}\ \bibnamefont
		{Adourian}}, \bibinfo {author} {\bibfnamefont {R.~M.}\ \bibnamefont
		{Westervelt}}, \bibinfo {author} {\bibfnamefont {J.~D.}\ \bibnamefont
		{Edwards}}, \bibinfo {author} {\bibfnamefont {A.}~\bibnamefont {Lupu-Sax}},
	\bibinfo {author} {\bibfnamefont {E.~J.}\ \bibnamefont {Heller}}, \bibinfo
	{author} {\bibfnamefont {K.~L.}\ \bibnamefont {Campman}}, \ and\ \bibinfo
	{author} {\bibfnamefont {A.~C.}\ \bibnamefont {Gossard}},\ }\href {\doibase
	10.1103/PhysRevLett.79.4806} {\bibfield  {journal} {\bibinfo  {journal}
		{Physical Review Letters}\ }\textbf {\bibinfo {volume} {79}},\ \bibinfo
	{pages} {4806} (\bibinfo {year} {1997})}\BibitemShut {NoStop}%
\bibitem [{\citenamefont {Hersch}\ \emph {et~al.}(1999)\citenamefont {Hersch},
	\citenamefont {Haggerty},\ and\ \citenamefont
	{Heller}}]{hersch_diffractive_1999}%
\BibitemOpen
\bibfield  {author} {\bibinfo {author} {\bibfnamefont {J.~S.}\ \bibnamefont
		{Hersch}}, \bibinfo {author} {\bibfnamefont {M.~R.}\ \bibnamefont
		{Haggerty}}, \ and\ \bibinfo {author} {\bibfnamefont {E.~J.}\ \bibnamefont
		{Heller}},\ }\href {\doibase 10.1103/PhysRevLett.83.5342} {\bibfield
	{journal} {\bibinfo  {journal} {Physical Review Letters}\ }\textbf {\bibinfo
		{volume} {83}},\ \bibinfo {pages} {5342} (\bibinfo {year}
	{1999})}\BibitemShut {NoStop}%
\bibitem [{\citenamefont {Crommie}\ \emph {et~al.}(1993)\citenamefont
	{Crommie}, \citenamefont {Lutz},\ and\ \citenamefont
	{Eigler}}]{crommie_imaging_1993}%
\BibitemOpen
\bibfield  {author} {\bibinfo {author} {\bibfnamefont {M.~F.}\ \bibnamefont
		{Crommie}}, \bibinfo {author} {\bibfnamefont {C.~P.}\ \bibnamefont {Lutz}}, \
	and\ \bibinfo {author} {\bibfnamefont {D.~M.}\ \bibnamefont {Eigler}},\
}\href {\doibase 10.1038/363524a0} {\bibfield  {journal} {\bibinfo  {journal}
	{Nature}\ }\textbf {\bibinfo {volume} {363}},\ \bibinfo {pages} {524}
(\bibinfo {year} {1993})}\BibitemShut {NoStop}%
\bibitem [{\citenamefont {Manoharan}\ \emph {et~al.}(2000)\citenamefont
	{Manoharan}, \citenamefont {Lutz},\ and\ \citenamefont
	{Eigler}}]{manoharan_quantum_2000}%
\BibitemOpen
\bibfield  {author} {\bibinfo {author} {\bibfnamefont {H.~C.}\ \bibnamefont
		{Manoharan}}, \bibinfo {author} {\bibfnamefont {C.~P.}\ \bibnamefont {Lutz}},
	\ and\ \bibinfo {author} {\bibfnamefont {D.~M.}\ \bibnamefont {Eigler}},\
}\href {\doibase 10.1038/35000508} {\bibfield  {journal} {\bibinfo  {journal}
	{Nature}\ }\textbf {\bibinfo {volume} {403}},\ \bibinfo {pages} {512}
(\bibinfo {year} {2000})}\BibitemShut {NoStop}%
\bibitem [{\citenamefont {Agam}\ and\ \citenamefont
	{Schiller}(2001)}]{agam:01}%
\BibitemOpen
\bibfield  {author} {\bibinfo {author} {\bibfnamefont {O.}~\bibnamefont
		{Agam}}\ and\ \bibinfo {author} {\bibfnamefont {A.}~\bibnamefont
		{Schiller}},\ }\href {\doibase 10.1103/PhysRevLett.86.484} {\bibfield
	{journal} {\bibinfo  {journal} {Phys. Rev. Lett.}\ }\textbf {\bibinfo
		{volume} {86}},\ \bibinfo {pages} {484} (\bibinfo {year} {2001})}\BibitemShut
{NoStop}%
\bibitem [{\citenamefont {Fiete}\ and\ \citenamefont
	{Heller}(2003)}]{fiete:03}%
\BibitemOpen
\bibfield  {author} {\bibinfo {author} {\bibfnamefont {G.~A.}\ \bibnamefont
		{Fiete}}\ and\ \bibinfo {author} {\bibfnamefont {E.~J.}\ \bibnamefont
		{Heller}},\ }\href {\doibase 10.1103/RevModPhys.75.933} {\bibfield  {journal}
	{\bibinfo  {journal} {Rev. Mod. Phys.}\ }\textbf {\bibinfo {volume} {75}},\
	\bibinfo {pages} {933} (\bibinfo {year} {2003})}\BibitemShut {NoStop}%
\bibitem [{\citenamefont {Craig}\ \emph {et~al.}(2004)\citenamefont {Craig},
	\citenamefont {Taylor}, \citenamefont {Lester}, \citenamefont {Marcus},
	\citenamefont {Hanson},\ and\ \citenamefont {Gossard}}]{craig_tunable_2004}%
\BibitemOpen
\bibfield  {author} {\bibinfo {author} {\bibfnamefont {N.~J.}\ \bibnamefont
		{Craig}}, \bibinfo {author} {\bibfnamefont {J.~M.}\ \bibnamefont {Taylor}},
	\bibinfo {author} {\bibfnamefont {E.~A.}\ \bibnamefont {Lester}}, \bibinfo
	{author} {\bibfnamefont {C.~M.}\ \bibnamefont {Marcus}}, \bibinfo {author}
	{\bibfnamefont {M.~P.}\ \bibnamefont {Hanson}}, \ and\ \bibinfo {author}
	{\bibfnamefont {A.~C.}\ \bibnamefont {Gossard}},\ }\href {\doibase
	10.1126/science.1095452} {\bibfield  {journal} {\bibinfo  {journal}
		{Science}\ }\textbf {\bibinfo {volume} {304}},\ \bibinfo {pages} {565}
	(\bibinfo {year} {2004})}\BibitemShut {NoStop}%
\bibitem [{\citenamefont {Potok}\ \emph {et~al.}(2007)\citenamefont {Potok},
	\citenamefont {Rau}, \citenamefont {Shtrikman}, \citenamefont {Oreg},\ and\
	\citenamefont {Goldhaber-Gordon}}]{potok_observation_2007}%
\BibitemOpen
\bibfield  {author} {\bibinfo {author} {\bibfnamefont {R.~M.}\ \bibnamefont
		{Potok}}, \bibinfo {author} {\bibfnamefont {I.~G.}\ \bibnamefont {Rau}},
	\bibinfo {author} {\bibfnamefont {H.}~\bibnamefont {Shtrikman}}, \bibinfo
	{author} {\bibfnamefont {Y.}~\bibnamefont {Oreg}}, \ and\ \bibinfo {author}
	{\bibfnamefont {D.}~\bibnamefont {Goldhaber-Gordon}},\ }\href {\doibase
	10.1038/nature05556} {\bibfield  {journal} {\bibinfo  {journal} {Nature}\
	}\textbf {\bibinfo {volume} {446}},\ \bibinfo {pages} {167} (\bibinfo {year}
	{2007})}\BibitemShut {NoStop}%
\bibitem [{\citenamefont {Roch}\ \emph {et~al.}(2008)\citenamefont {Roch},
	\citenamefont {Florens}, \citenamefont {Bouchiat}, \citenamefont
	{Wernsdorfer},\ and\ \citenamefont {Balestro}}]{roch_quantum_2008}%
\BibitemOpen
\bibfield  {author} {\bibinfo {author} {\bibfnamefont {N.}~\bibnamefont
		{Roch}}, \bibinfo {author} {\bibfnamefont {S.}~\bibnamefont {Florens}},
	\bibinfo {author} {\bibfnamefont {V.}~\bibnamefont {Bouchiat}}, \bibinfo
	{author} {\bibfnamefont {W.}~\bibnamefont {Wernsdorfer}}, \ and\ \bibinfo
	{author} {\bibfnamefont {F.}~\bibnamefont {Balestro}},\ }\href {\doibase
	10.1038/nature06930} {\bibfield  {journal} {\bibinfo  {journal} {Nature}\
	}\textbf {\bibinfo {volume} {453}},\ \bibinfo {pages} {633} (\bibinfo {year}
	{2008})}\BibitemShut {NoStop}%
\bibitem [{\citenamefont {Hayashi}\ \emph {et~al.}(2003)\citenamefont
	{Hayashi}, \citenamefont {Fujisawa}, \citenamefont {Cheong}, \citenamefont
	{Jeong},\ and\ \citenamefont {Hirayama}}]{hayashi_coherent_2003}%
\BibitemOpen
\bibfield  {author} {\bibinfo {author} {\bibfnamefont {T.}~\bibnamefont
		{Hayashi}}, \bibinfo {author} {\bibfnamefont {T.}~\bibnamefont {Fujisawa}},
	\bibinfo {author} {\bibfnamefont {H.~D.}\ \bibnamefont {Cheong}}, \bibinfo
	{author} {\bibfnamefont {Y.~H.}\ \bibnamefont {Jeong}}, \ and\ \bibinfo
	{author} {\bibfnamefont {Y.}~\bibnamefont {Hirayama}},\ }\href {\doibase
	10.1103/PhysRevLett.91.226804} {\bibfield  {journal} {\bibinfo  {journal}
		{Physical Review Letters}\ }\textbf {\bibinfo {volume} {91}},\ \bibinfo
	{pages} {226804} (\bibinfo {year} {2003})}\BibitemShut {NoStop}%
\bibitem [{\citenamefont {Petta}\ \emph {et~al.}(2005)\citenamefont {Petta},
	\citenamefont {Johnson}, \citenamefont {Taylor}, \citenamefont {Laird},
	\citenamefont {Yacoby}, \citenamefont {Lukin}, \citenamefont {Marcus},
	\citenamefont {Hanson},\ and\ \citenamefont {Gossard}}]{petta_coherent_2005}%
\BibitemOpen
\bibfield  {author} {\bibinfo {author} {\bibfnamefont {J.~R.}\ \bibnamefont
		{Petta}}, \bibinfo {author} {\bibfnamefont {A.~C.}\ \bibnamefont {Johnson}},
	\bibinfo {author} {\bibfnamefont {J.~M.}\ \bibnamefont {Taylor}}, \bibinfo
	{author} {\bibfnamefont {E.~A.}\ \bibnamefont {Laird}}, \bibinfo {author}
	{\bibfnamefont {A.}~\bibnamefont {Yacoby}}, \bibinfo {author} {\bibfnamefont
		{M.~D.}\ \bibnamefont {Lukin}}, \bibinfo {author} {\bibfnamefont {C.~M.}\
		\bibnamefont {Marcus}}, \bibinfo {author} {\bibfnamefont {M.~P.}\
		\bibnamefont {Hanson}}, \ and\ \bibinfo {author} {\bibfnamefont {A.~C.}\
		\bibnamefont {Gossard}},\ }\href {\doibase 10.1126/science.1116955}
{\bibfield  {journal} {\bibinfo  {journal} {Science}\ }\textbf {\bibinfo
		{volume} {309}},\ \bibinfo {pages} {2180} (\bibinfo {year}
	{2005})}\BibitemShut {NoStop}%
\bibitem [{\citenamefont {R\"ossler}\ \emph {et~al.}(2015)\citenamefont
	{R\"ossler}, \citenamefont {Oehri}, \citenamefont {Zilberberg}, \citenamefont
	{Blatter}, \citenamefont {Karalic}, \citenamefont {Pijnenburg}, \citenamefont
	{Hofmann}, \citenamefont {Ihn}, \citenamefont {Ensslin}, \citenamefont
	{Reichl},\ and\ \citenamefont {Wegscheider}}]{roessler:2015}%
\BibitemOpen
\bibfield  {author} {\bibinfo {author} {\bibfnamefont {C.}~\bibnamefont
		{R\"ossler}}, \bibinfo {author} {\bibfnamefont {D.}~\bibnamefont {Oehri}},
	\bibinfo {author} {\bibfnamefont {O.}~\bibnamefont {Zilberberg}}, \bibinfo
	{author} {\bibfnamefont {G.}~\bibnamefont {Blatter}}, \bibinfo {author}
	{\bibfnamefont {M.}~\bibnamefont {Karalic}}, \bibinfo {author} {\bibfnamefont
		{J.}~\bibnamefont {Pijnenburg}}, \bibinfo {author} {\bibfnamefont
		{A.}~\bibnamefont {Hofmann}}, \bibinfo {author} {\bibfnamefont
		{T.}~\bibnamefont {Ihn}}, \bibinfo {author} {\bibfnamefont {K.}~\bibnamefont
		{Ensslin}}, \bibinfo {author} {\bibfnamefont {C.}~\bibnamefont {Reichl}}, \
	and\ \bibinfo {author} {\bibfnamefont {W.}~\bibnamefont {Wegscheider}},\
}\href {\doibase 10.1103/PhysRevLett.115.166603} {\bibfield  {journal}
{\bibinfo  {journal} {Phys. Rev. Lett.}\ }\textbf {\bibinfo {volume} {115}},\
\bibinfo {pages} {166603} (\bibinfo {year} {2015})}\BibitemShut {NoStop}%
\bibitem [{\citenamefont {Haroche}(2013)}]{Haroche2013}%
\BibitemOpen
\bibfield  {author} {\bibinfo {author} {\bibfnamefont {S.}~\bibnamefont
		{Haroche}},\ }\href {\doibase 10.1103/RevModPhys.85.1083} {\bibfield
	{journal} {\bibinfo  {journal} {Reviews of Modern Physics}\ }\textbf
	{\bibinfo {volume} {85}},\ \bibinfo {pages} {1083} (\bibinfo {year}
	{2013})}\BibitemShut {NoStop}%
\bibitem [{\citenamefont {Thimm}\ \emph {et~al.}(1999)\citenamefont {Thimm},
	\citenamefont {Kroha},\ and\ \citenamefont {von Delft}}]{thimm_kondo_1999}%
\BibitemOpen
\bibfield  {author} {\bibinfo {author} {\bibfnamefont {W.~B.}\ \bibnamefont
		{Thimm}}, \bibinfo {author} {\bibfnamefont {J.}~\bibnamefont {Kroha}}, \ and\
	\bibinfo {author} {\bibfnamefont {J.}~\bibnamefont {von Delft}},\ }\href
{\doibase 10.1103/PhysRevLett.82.2143} {\bibfield  {journal} {\bibinfo
		{journal} {Physical Review Letters}\ }\textbf {\bibinfo {volume} {82}},\
	\bibinfo {pages} {2143} (\bibinfo {year} {1999})}\BibitemShut {NoStop}%
\bibitem [{\citenamefont {Cornaglia}\ and\ \citenamefont
	{Balseiro}(2002)}]{Cornaglia2002}%
\BibitemOpen
\bibfield  {author} {\bibinfo {author} {\bibfnamefont {P.~S.}\ \bibnamefont
		{Cornaglia}}\ and\ \bibinfo {author} {\bibfnamefont {C.~A.}\ \bibnamefont
		{Balseiro}},\ }\href {\doibase 10.1103/PhysRevB.66.174404} {\bibfield
	{journal} {\bibinfo  {journal} {Phys. Rev. B}\ }\textbf {\bibinfo {volume}
		{66}},\ \bibinfo {pages} {174404} (\bibinfo {year} {2002})}\BibitemShut
{NoStop}%
\bibitem [{\citenamefont {Dias~da Silva}\ \emph {et~al.}(2006)\citenamefont
	{Dias~da Silva}, \citenamefont {Sandler}, \citenamefont {Ingersent},\ and\
	\citenamefont {Ulloa}}]{dias_da_silva_zero-field_2006}%
\BibitemOpen
\bibfield  {author} {\bibinfo {author} {\bibfnamefont {L.~G. G.~V.}\
		\bibnamefont {Dias~da Silva}}, \bibinfo {author} {\bibfnamefont {N.~P.}\
		\bibnamefont {Sandler}}, \bibinfo {author} {\bibfnamefont {K.}~\bibnamefont
		{Ingersent}}, \ and\ \bibinfo {author} {\bibfnamefont {S.~E.}\ \bibnamefont
		{Ulloa}},\ }\href {\doibase 10.1103/PhysRevLett.97.096603} {\bibfield
	{journal} {\bibinfo  {journal} {Physical Review Letters}\ }\textbf {\bibinfo
		{volume} {97}},\ \bibinfo {pages} {096603} (\bibinfo {year}
	{2006})}\BibitemShut {NoStop}%
\bibitem [{\citenamefont {Dias~da Silva}\ \emph {et~al.}(2013)\citenamefont
	{Dias~da Silva}, \citenamefont {Vernek}, \citenamefont {Ingersent},
	\citenamefont {Sandler},\ and\ \citenamefont
	{Ulloa}}]{dias_da_silva_spin-polarized_2013}%
\BibitemOpen
\bibfield  {author} {\bibinfo {author} {\bibfnamefont {L.~G. G.~V.}\
		\bibnamefont {Dias~da Silva}}, \bibinfo {author} {\bibfnamefont
		{E.}~\bibnamefont {Vernek}}, \bibinfo {author} {\bibfnamefont
		{K.}~\bibnamefont {Ingersent}}, \bibinfo {author} {\bibfnamefont
		{N.}~\bibnamefont {Sandler}}, \ and\ \bibinfo {author} {\bibfnamefont
		{S.~E.}\ \bibnamefont {Ulloa}},\ }\href {\doibase 10.1103/PhysRevB.87.205313}
{\bibfield  {journal} {\bibinfo  {journal} {Physical Review B}\ }\textbf
	{\bibinfo {volume} {87}},\ \bibinfo {pages} {205313} (\bibinfo {year}
	{2013})}\BibitemShut {NoStop}%
\bibitem [{\citenamefont {Groth}\ \emph {et~al.}(2014)\citenamefont {Groth},
	\citenamefont {Wimmer}, \citenamefont {Akhmerov},\ and\ \citenamefont
	{Waintal}}]{kwant}%
\BibitemOpen
\bibfield  {author} {\bibinfo {author} {\bibfnamefont {C.~W.}\ \bibnamefont
		{Groth}}, \bibinfo {author} {\bibfnamefont {M.}~\bibnamefont {Wimmer}},
	\bibinfo {author} {\bibfnamefont {A.~R.}\ \bibnamefont {Akhmerov}}, \ and\
	\bibinfo {author} {\bibfnamefont {X.}~\bibnamefont {Waintal}},\ }\href
{http://stacks.iop.org/1367-2630/16/i=6/a=063065} {\bibfield  {journal}
	{\bibinfo  {journal} {New Journal of Physics}\ }\textbf {\bibinfo {volume}
		{16}},\ \bibinfo {pages} {063065} (\bibinfo {year} {2014})}\BibitemShut
{NoStop}%
\bibitem [{\citenamefont {van Wees}\ \emph {et~al.}(1988)\citenamefont {van
		Wees}, \citenamefont {van Houten}, \citenamefont {Beenakker}, \citenamefont
	{Williamson}, \citenamefont {Kouwenhoven}, \citenamefont {van~der Marel},\
	and\ \citenamefont {Foxon}}]{wees:88}%
\BibitemOpen
\bibfield  {author} {\bibinfo {author} {\bibfnamefont {B.~J.}\ \bibnamefont
		{van Wees}}, \bibinfo {author} {\bibfnamefont {H.}~\bibnamefont {van
			Houten}}, \bibinfo {author} {\bibfnamefont {C.~W.~J.}\ \bibnamefont
		{Beenakker}}, \bibinfo {author} {\bibfnamefont {J.~G.}\ \bibnamefont
		{Williamson}}, \bibinfo {author} {\bibfnamefont {L.~P.}\ \bibnamefont
		{Kouwenhoven}}, \bibinfo {author} {\bibfnamefont {D.}~\bibnamefont {van~der
			Marel}}, \ and\ \bibinfo {author} {\bibfnamefont {C.~T.}\ \bibnamefont
		{Foxon}},\ }\href@noop {} {\bibfield  {journal} {\bibinfo  {journal} {Phys.
			Rev. Lett.}\ }\textbf {\bibinfo {volume} {60}},\ \bibinfo {pages} {848}
	(\bibinfo {year} {1988})}\BibitemShut {NoStop}%
\bibitem [{\citenamefont {Wharam}\ \emph {et~al.}(1988)\citenamefont {Wharam},
	\citenamefont {Thornton}, \citenamefont {Newbury}, \citenamefont {Pepper},
	\citenamefont {Ahme}, \citenamefont {Frost}, \citenamefont {Hasko},
	\citenamefont {Peacock},\ and\ \citenamefont {Ritchie}}]{wharam:88}%
\BibitemOpen
\bibfield  {author} {\bibinfo {author} {\bibfnamefont {D.~A.}\ \bibnamefont
		{Wharam}}, \bibinfo {author} {\bibfnamefont {T.~J.}\ \bibnamefont
		{Thornton}}, \bibinfo {author} {\bibfnamefont {R.}~\bibnamefont {Newbury}},
	\bibinfo {author} {\bibfnamefont {M.}~\bibnamefont {Pepper}}, \bibinfo
	{author} {\bibfnamefont {H.}~\bibnamefont {Ahme}}, \bibinfo {author}
	{\bibfnamefont {J.~E.~F.}\ \bibnamefont {Frost}}, \bibinfo {author}
	{\bibfnamefont {D.~G.}\ \bibnamefont {Hasko}}, \bibinfo {author}
	{\bibfnamefont {D.~C.}\ \bibnamefont {Peacock}}, \ and\ \bibinfo {author}
	{\bibfnamefont {D.~A.}\ \bibnamefont {Ritchie}},\ }\href@noop {} {\bibfield
	{journal} {\bibinfo  {journal} {J. Phys. C}\ }\textbf {\bibinfo {volume}
		{21}},\ \bibinfo {pages} {L209} (\bibinfo {year} {1988})}\BibitemShut
{NoStop}%
\bibitem [{\citenamefont {Anderson}(1961)}]{anderson:61}%
\BibitemOpen
\bibfield  {author} {\bibinfo {author} {\bibfnamefont {P.~W.}\ \bibnamefont
		{Anderson}},\ }\href {\doibase 10.1103/PhysRev.124.41} {\bibfield  {journal}
	{\bibinfo  {journal} {Phys. Rev.}\ }\textbf {\bibinfo {volume} {124}},\
	\bibinfo {pages} {41} (\bibinfo {year} {1961})}\BibitemShut {NoStop}%
\bibitem [{\citenamefont {Bruus}\ and\ \citenamefont
	{Flensberg}(2004)}]{bruus-flensberg}%
\BibitemOpen
\bibfield  {author} {\bibinfo {author} {\bibfnamefont {H.}~\bibnamefont
		{Bruus}}\ and\ \bibinfo {author} {\bibfnamefont {K.}~\bibnamefont
		{Flensberg}},\ }\href@noop {} {\emph {\bibinfo {title} {Many-body quantum
			theory in condensed matter physics}}}\ (\bibinfo  {publisher} {Oxford
	University Press},\ \bibinfo {address} {Oxford},\ \bibinfo {year}
{2004})\BibitemShut {NoStop}%
\bibitem [{\citenamefont {Fano}(1961)}]{fano:61}%
\BibitemOpen
\bibfield  {author} {\bibinfo {author} {\bibfnamefont {U.}~\bibnamefont
		{Fano}},\ }\href {\doibase 10.1103/PhysRev.124.1866} {\bibfield  {journal}
	{\bibinfo  {journal} {Phys. Rev.}\ }\textbf {\bibinfo {volume} {124}},\
	\bibinfo {pages} {1866} (\bibinfo {year} {1961})}\BibitemShut {NoStop}%
\bibitem [{\citenamefont {Alhassid}(2000)}]{alhassidreview}%
\BibitemOpen
\bibfield  {author} {\bibinfo {author} {\bibfnamefont {Y.}~\bibnamefont
		{Alhassid}},\ }\href@noop {} {\bibfield  {journal} {\bibinfo  {journal} {Rev.
			Mod. Phys}\ }\textbf {\bibinfo {volume} {72}},\ \bibinfo {pages} {895}
	(\bibinfo {year} {2000})}\BibitemShut {NoStop}%
\bibitem [{\citenamefont {Sakurai}\ and\ \citenamefont
	{Tuan}(1985)}]{sakurai1985}%
\BibitemOpen
\bibfield  {author} {\bibinfo {author} {\bibfnamefont {J.~J.}\ \bibnamefont
		{Sakurai}}\ and\ \bibinfo {author} {\bibfnamefont {S.~F.}\ \bibnamefont
		{Tuan}},\ }\href@noop {} {\emph {\bibinfo {title} {Modern quantum
			mechanics}}},\ Vol.~\bibinfo {volume} {1}\ (\bibinfo  {publisher}
{Addison-Wesley Reading, Massachusetts},\ \bibinfo {year} {1985})\BibitemShut
{NoStop}%
\bibitem [{\citenamefont {Beenakker}(1991)}]{beenakker:91}%
\BibitemOpen
\bibfield  {author} {\bibinfo {author} {\bibfnamefont {C.~W.~J.}\
		\bibnamefont {Beenakker}},\ }\href {\doibase 10.1103/PhysRevB.44.1646}
{\bibfield  {journal} {\bibinfo  {journal} {Phys. Rev. B}\ }\textbf {\bibinfo
		{volume} {44}},\ \bibinfo {pages} {1646} (\bibinfo {year}
	{1991})}\BibitemShut {NoStop}%
\bibitem [{\citenamefont {Korotkov}(1994)}]{Korotkov:1994}%
\BibitemOpen
\bibfield  {author} {\bibinfo {author} {\bibfnamefont {A.~N.}\ \bibnamefont
		{Korotkov}},\ }\href@noop {} {\bibfield  {journal} {\bibinfo  {journal}
		{Phys. Rev. B}\ }\textbf {\bibinfo {volume} {49}},\ \bibinfo {pages} {10381}
	(\bibinfo {year} {1994})}\BibitemShut {NoStop}%
\bibitem [{\citenamefont {Koch}\ \emph {et~al.}(2006)\citenamefont {Koch},
	\citenamefont {von Oppen},\ and\ \citenamefont {Andreev}}]{Koch:2006}%
\BibitemOpen
\bibfield  {author} {\bibinfo {author} {\bibfnamefont {J.}~\bibnamefont
		{Koch}}, \bibinfo {author} {\bibfnamefont {F.}~\bibnamefont {von Oppen}}, \
	and\ \bibinfo {author} {\bibfnamefont {A.~V.}\ \bibnamefont {Andreev}},\
}\href@noop {} {\bibfield  {journal} {\bibinfo  {journal} {Phys. Rev. B}\
}\textbf {\bibinfo {volume} {74}},\ \bibinfo {pages} {205438} (\bibinfo
{year} {2006})}\BibitemShut {NoStop}%
\bibitem [{\citenamefont {Koch}(2006)}]{thesis:Koch}%
\BibitemOpen
\bibfield  {author} {\bibinfo {author} {\bibfnamefont {J.}~\bibnamefont
		{Koch}},\ }\emph {\bibinfo {title} {Quantum transport through single-molecule
		devices}},\ \href@noop {} {Ph.D. thesis},\ \bibinfo  {school} {Freie
	Universit\"at Berlin} (\bibinfo {year} {2006})\BibitemShut {NoStop}%
\bibitem [{\citenamefont {{Ferguson}}\ \emph {et~al.}(2016)\citenamefont
	{{Ferguson}}, \citenamefont {{R{\"o}ssler}}, \citenamefont {{Ihn}},
	\citenamefont {{Ensslin}}, \citenamefont {{Blatter}},\ and\ \citenamefont
	{{Zilberberg}}}]{proceedings}%
\BibitemOpen
\bibfield  {author} {\bibinfo {author} {\bibfnamefont {M.~S.}\ \bibnamefont
		{{Ferguson}}}, \bibinfo {author} {\bibfnamefont {C.}~\bibnamefont
		{{R{\"o}ssler}}}, \bibinfo {author} {\bibfnamefont {T.}~\bibnamefont
		{{Ihn}}}, \bibinfo {author} {\bibfnamefont {K.}~\bibnamefont {{Ensslin}}},
	\bibinfo {author} {\bibfnamefont {G.}~\bibnamefont {{Blatter}}}, \ and\
	\bibinfo {author} {\bibfnamefont {O.}~\bibnamefont {{Zilberberg}}},\
}\href@noop {} {\bibfield  {journal} {\bibinfo  {journal} {ArXiv e-prints}\ }
(\bibinfo {year} {2016})},\ \Eprint {http://arxiv.org/abs/1612.03850}
{arXiv:1612.03850 [cond-mat.mes-hall]} \BibitemShut {NoStop}%
\bibitem [{\citenamefont {Furusaki}\ and\ \citenamefont
	{Matveev}(1995)}]{furusaki1995theory}%
\BibitemOpen
\bibfield  {author} {\bibinfo {author} {\bibfnamefont {A.}~\bibnamefont
		{Furusaki}}\ and\ \bibinfo {author} {\bibfnamefont {K.~A.}\ \bibnamefont
		{Matveev}},\ }\href@noop {} {\bibfield  {journal} {\bibinfo  {journal}
		{Physical Review B}\ }\textbf {\bibinfo {volume} {52}},\ \bibinfo {pages}
	{16676} (\bibinfo {year} {1995})}\BibitemShut {NoStop}%
\bibitem [{\citenamefont {Zumb{\"u}hl}\ \emph {et~al.}(2004)\citenamefont
	{Zumb{\"u}hl}, \citenamefont {Marcus}, \citenamefont {Hanson},\ and\
	\citenamefont {Gossard}}]{zumbuhl_cotunneling_2004}%
\BibitemOpen
\bibfield  {author} {\bibinfo {author} {\bibfnamefont {D.~M.}\ \bibnamefont
		{Zumb{\"u}hl}}, \bibinfo {author} {\bibfnamefont {C.~M.}\ \bibnamefont
		{Marcus}}, \bibinfo {author} {\bibfnamefont {M.~P.}\ \bibnamefont {Hanson}},
	\ and\ \bibinfo {author} {\bibfnamefont {A.~C.}\ \bibnamefont {Gossard}},\
}\href {\doibase 10.1103/PhysRevLett.93.256801} {\bibfield  {journal}
{\bibinfo  {journal} {Physical Review Letters}\ }\textbf {\bibinfo {volume}
	{93}},\ \bibinfo {pages} {256801} (\bibinfo {year} {2004})}\BibitemShut
{NoStop}%
\bibitem [{\citenamefont {Lacroix}(1981)}]{lacroix:81}%
\BibitemOpen
\bibfield  {author} {\bibinfo {author} {\bibfnamefont {C.}~\bibnamefont
		{Lacroix}},\ }\href {http://stacks.iop.org/0305-4608/11/i=11/a=020}
{\bibfield  {journal} {\bibinfo  {journal} {Journal of Physics F: Metal
			Physics}\ }\textbf {\bibinfo {volume} {11}},\ \bibinfo {pages} {2389}
	(\bibinfo {year} {1981})}\BibitemShut {NoStop}%
\bibitem [{\citenamefont {Lacroix}(1982)}]{lacroix:82}%
\BibitemOpen
\bibfield  {author} {\bibinfo {author} {\bibfnamefont {C.}~\bibnamefont
		{Lacroix}},\ }\href {\doibase http://dx.doi.org/10.1063/1.330756} {\bibfield
	{journal} {\bibinfo  {journal} {Journal of Applied Physics}\ }\textbf
	{\bibinfo {volume} {53}},\ \bibinfo {pages} {2131} (\bibinfo {year}
	{1982})}\BibitemShut {NoStop}%
\bibitem [{\citenamefont {Entin-Wohlman}\ \emph {et~al.}(2005)\citenamefont
	{Entin-Wohlman}, \citenamefont {Aharony},\ and\ \citenamefont
	{Meir}}]{entin-wohlman:05}%
\BibitemOpen
\bibfield  {author} {\bibinfo {author} {\bibfnamefont {O.}~\bibnamefont
		{Entin-Wohlman}}, \bibinfo {author} {\bibfnamefont {A.}~\bibnamefont
		{Aharony}}, \ and\ \bibinfo {author} {\bibfnamefont {Y.}~\bibnamefont
		{Meir}},\ }\href {\doibase 10.1103/PhysRevB.71.035333} {\bibfield  {journal}
	{\bibinfo  {journal} {Phys. Rev. B}\ }\textbf {\bibinfo {volume} {71}},\
	\bibinfo {pages} {035333} (\bibinfo {year} {2005})}\BibitemShut {NoStop}%
\bibitem [{\citenamefont {Kashcheyevs}\ \emph {et~al.}(2006)\citenamefont
	{Kashcheyevs}, \citenamefont {Aharony},\ and\ \citenamefont
	{Entin-Wohlman}}]{kashcheyevs:06}%
\BibitemOpen
\bibfield  {author} {\bibinfo {author} {\bibfnamefont {V.}~\bibnamefont
		{Kashcheyevs}}, \bibinfo {author} {\bibfnamefont {A.}~\bibnamefont
		{Aharony}}, \ and\ \bibinfo {author} {\bibfnamefont {O.}~\bibnamefont
		{Entin-Wohlman}},\ }\href {\doibase 10.1103/PhysRevB.73.125338} {\bibfield
	{journal} {\bibinfo  {journal} {Phys. Rev. B}\ }\textbf {\bibinfo {volume}
		{73}},\ \bibinfo {pages} {125338} (\bibinfo {year} {2006})}\BibitemShut
{NoStop}%
\bibitem [{\citenamefont {Van~Roermund}\ \emph {et~al.}(2010)\citenamefont
	{Van~Roermund}, \citenamefont {Shiau},\ and\ \citenamefont
	{Lavagna}}]{Lavagna2010}%
\BibitemOpen
\bibfield  {author} {\bibinfo {author} {\bibfnamefont {R.}~\bibnamefont
		{Van~Roermund}}, \bibinfo {author} {\bibfnamefont {S.-y.}\ \bibnamefont
		{Shiau}}, \ and\ \bibinfo {author} {\bibfnamefont {M.}~\bibnamefont
		{Lavagna}},\ }\href {\doibase 10.1103/PhysRevB.81.165115} {\bibfield
	{journal} {\bibinfo  {journal} {Phys. Rev. B}\ }\textbf {\bibinfo {volume}
		{81}},\ \bibinfo {pages} {165115} (\bibinfo {year} {2010})}\BibitemShut
{NoStop}%
\bibitem [{\citenamefont {Van~Roermund}(2012)}]{thesis:Roermund}%
\BibitemOpen
\bibfield  {author} {\bibinfo {author} {\bibfnamefont {R.}~\bibnamefont
		{Van~Roermund}},\ }\emph {\bibinfo {title} {Theoretical study of
		non-equilibrium transport in Kondo quantum dots}},\ \href@noop {} {Ph.D.
	thesis},\ \bibinfo  {school} {University of Grenoble} (\bibinfo {year}
{2012})\BibitemShut {NoStop}%
\bibitem [{\citenamefont {Meir}\ and\ \citenamefont
	{Wingreen}(1992)}]{Meir1992}%
\BibitemOpen
\bibfield  {author} {\bibinfo {author} {\bibfnamefont {Y.}~\bibnamefont
		{Meir}}\ and\ \bibinfo {author} {\bibfnamefont {N.~S.}\ \bibnamefont
		{Wingreen}},\ }\href {\doibase 10.1103/PhysRevLett.68.2512} {\bibfield
	{journal} {\bibinfo  {journal} {Physical Review Letters}\ }\textbf {\bibinfo
		{volume} {68}},\ \bibinfo {pages} {2512} (\bibinfo {year}
	{1992})}\BibitemShut {NoStop}%
\bibitem [{\citenamefont {Zubarev}(1960)}]{Zubarev1960}%
\BibitemOpen
\bibfield  {author} {\bibinfo {author} {\bibfnamefont {D.~N.}\ \bibnamefont
		{Zubarev}},\ }\href {\doibase 10.1070/PU1960v003n03ABEH003275} {\bibfield
	{journal} {\bibinfo  {journal} {Phys. Usp.}\ }\textbf {\bibinfo {volume}
		{3}},\ \bibinfo {pages} {320} (\bibinfo {year} {1960})}\BibitemShut {NoStop}%
\bibitem [{\citenamefont {Kubo}(1962)}]{Kubo1962}%
\BibitemOpen
\bibfield  {author} {\bibinfo {author} {\bibfnamefont {R.}~\bibnamefont
		{Kubo}},\ }\href {\doibase 10.1143/JPSJ.17.1100} {\bibfield  {journal}
	{\bibinfo  {journal} {Journal of the Physical Society of Japan}\ }\textbf
	{\bibinfo {volume} {17}},\ \bibinfo {pages} {1100} (\bibinfo {year}
	{1962})},\ \Eprint
{http://arxiv.org/abs/http://dx.doi.org/10.1143/JPSJ.17.1100}
{http://dx.doi.org/10.1143/JPSJ.17.1100} \BibitemShut {NoStop}%
\bibitem [{\citenamefont {Goldberg}\ \emph {et~al.}(2005)\citenamefont
	{Goldberg}, \citenamefont {Flores},\ and\ \citenamefont
	{Monreal}}]{Goldberg2005}%
\BibitemOpen
\bibfield  {author} {\bibinfo {author} {\bibfnamefont {E.~C.}\ \bibnamefont
		{Goldberg}}, \bibinfo {author} {\bibfnamefont {F.}~\bibnamefont {Flores}}, \
	and\ \bibinfo {author} {\bibfnamefont {R.~C.}\ \bibnamefont {Monreal}},\
}\href {\doibase 10.1103/PhysRevB.71.035112} {\bibfield  {journal} {\bibinfo
	{journal} {Phys. Rev. B}\ }\textbf {\bibinfo {volume} {71}},\ \bibinfo
{pages} {035112} (\bibinfo {year} {2005})}\BibitemShut {NoStop}%
\bibitem [{\citenamefont {Monreal}\ and\ \citenamefont
	{Flores}(2005)}]{Monreal2005}%
\BibitemOpen
\bibfield  {author} {\bibinfo {author} {\bibfnamefont {R.~C.}\ \bibnamefont
		{Monreal}}\ and\ \bibinfo {author} {\bibfnamefont {F.}~\bibnamefont
		{Flores}},\ }\href {\doibase 10.1103/PhysRevB.72.195105} {\bibfield
	{journal} {\bibinfo  {journal} {Phys. Rev. B}\ }\textbf {\bibinfo {volume}
		{72}},\ \bibinfo {pages} {195105} (\bibinfo {year} {2005})}\BibitemShut
{NoStop}%
\bibitem [{\citenamefont {Wilson}(1975)}]{NRGReview1970}%
\BibitemOpen
\bibfield  {author} {\bibinfo {author} {\bibfnamefont {K.~G.}\ \bibnamefont
		{Wilson}},\ }\href {\doibase 10.1103/RevModPhys.47.773} {\bibfield  {journal}
	{\bibinfo  {journal} {Rev. Mod. Phys.}\ }\textbf {\bibinfo {volume} {47}},\
	\bibinfo {pages} {773} (\bibinfo {year} {1975})}\BibitemShut {NoStop}%
\bibitem [{\citenamefont {Bulla}\ \emph {et~al.}(2008)\citenamefont {Bulla},
	\citenamefont {Costi},\ and\ \citenamefont {Pruschke}}]{NRGReview2008}%
\BibitemOpen
\bibfield  {author} {\bibinfo {author} {\bibfnamefont {R.}~\bibnamefont
		{Bulla}}, \bibinfo {author} {\bibfnamefont {T.~A.}\ \bibnamefont {Costi}}, \
	and\ \bibinfo {author} {\bibfnamefont {T.}~\bibnamefont {Pruschke}},\ }\href
{\doibase 10.1103/RevModPhys.80.395} {\bibfield  {journal} {\bibinfo
		{journal} {Rev. Mod. Phys.}\ }\textbf {\bibinfo {volume} {80}},\ \bibinfo
	{pages} {395} (\bibinfo {year} {2008})}\BibitemShut {NoStop}%
\bibitem [{\citenamefont {White}(1992)}]{White1992}%
\BibitemOpen
\bibfield  {author} {\bibinfo {author} {\bibfnamefont {S.~R.}\ \bibnamefont
		{White}},\ }\href {\doibase 10.1103/PhysRevLett.69.2863} {\bibfield
	{journal} {\bibinfo  {journal} {Phys. Rev. Lett.}\ }\textbf {\bibinfo
		{volume} {69}},\ \bibinfo {pages} {2863} (\bibinfo {year}
	{1992})}\BibitemShut {NoStop}%
\bibitem [{\citenamefont {Schollw\"ock}(2005)}]{DMRGReview2005}%
\BibitemOpen
\bibfield  {author} {\bibinfo {author} {\bibfnamefont {U.}~\bibnamefont
		{Schollw\"ock}},\ }\href {\doibase 10.1103/RevModPhys.77.259} {\bibfield
	{journal} {\bibinfo  {journal} {Rev. Mod. Phys.}\ }\textbf {\bibinfo {volume}
		{77}},\ \bibinfo {pages} {259} (\bibinfo {year} {2005})}\BibitemShut
{NoStop}%
\bibitem [{\citenamefont {Dias~da Silva}\ \emph {et~al.}(2017)\citenamefont
	{Dias~da Silva}, \citenamefont {Lewenkopf}, \citenamefont {Vernek},
	\citenamefont {Ferreira},\ and\ \citenamefont {Ulloa}}]{Ulloa2017}%
\BibitemOpen
\bibfield  {author} {\bibinfo {author} {\bibfnamefont {L.~G. G.~V.}\
		\bibnamefont {Dias~da Silva}}, \bibinfo {author} {\bibfnamefont {C.~H.}\
		\bibnamefont {Lewenkopf}}, \bibinfo {author} {\bibfnamefont {E.}~\bibnamefont
		{Vernek}}, \bibinfo {author} {\bibfnamefont {G.~J.}\ \bibnamefont
		{Ferreira}}, \ and\ \bibinfo {author} {\bibfnamefont {S.~E.}\ \bibnamefont
		{Ulloa}},\ }\href {\doibase 10.1103/PhysRevLett.119.116801} {\bibfield
	{journal} {\bibinfo  {journal} {Phys. Rev. Lett.}\ }\textbf {\bibinfo
		{volume} {119}},\ \bibinfo {pages} {116801} (\bibinfo {year}
	{2017})}\BibitemShut {NoStop}%
\bibitem [{\citenamefont {Langreth}(1966)}]{Langreth1966}%
\BibitemOpen
\bibfield  {author} {\bibinfo {author} {\bibfnamefont {D.~C.}\ \bibnamefont
		{Langreth}},\ }\href {\doibase 10.1103/PhysRev.150.516} {\bibfield  {journal}
	{\bibinfo  {journal} {Phys. Rev.}\ }\textbf {\bibinfo {volume} {150}},\
	\bibinfo {pages} {516} (\bibinfo {year} {1966})}\BibitemShut {NoStop}%
\bibitem [{\citenamefont {Ferguson}\ \emph {et~al.}()\citenamefont {Ferguson},
	\citenamefont {Blatter},\ and\ \citenamefont {Zilberberg}}]{Ferguson2018}%
\BibitemOpen
\bibfield  {author} {\bibinfo {author} {\bibfnamefont {M.~S.}\ \bibnamefont
		{Ferguson}}, \bibinfo {author} {\bibfnamefont {G.}~\bibnamefont {Blatter}}, \
	and\ \bibinfo {author} {\bibfnamefont {O.}~\bibnamefont {Zilberberg}},\
}\href@noop {} {}\bibinfo {note} {In preparation}\BibitemShut {NoStop}%
\bibitem [{\citenamefont {Haldane}(1978)}]{Haldane1978}%
\BibitemOpen
\bibfield  {author} {\bibinfo {author} {\bibfnamefont {F.~D.~M.}\
		\bibnamefont {Haldane}},\ }\href {\doibase 10.1103/PhysRevLett.40.416}
{\bibfield  {journal} {\bibinfo  {journal} {Phys. Rev. Lett.}\ }\textbf
	{\bibinfo {volume} {40}},\ \bibinfo {pages} {416} (\bibinfo {year}
	{1978})}\BibitemShut {NoStop}%
\bibitem [{Note1()}]{Note1}%
\BibitemOpen
\bibinfo {note} {The expression for the exact Kondo temperature \(\protect
	\hspace {20pt} T_{\protect \rm \scriptscriptstyle K}^{\protect \rm ex}
	=(U\Gamma /4)^{1/2}\protect \qopname \relax o{exp}\left [\pi \epsilon
	_\protect \mathrm {d}(\epsilon _\protect \mathrm {d}+U)/(\Gamma U)\right ] \)
	can be found in Ref.~[\protect \rev@citealpnum {Haldane1978}] along with its
	derivation.}\BibitemShut {Stop}%
\bibitem [{\citenamefont {Andrei}(1980)}]{Andrei1980}%
\BibitemOpen
\bibfield  {author} {\bibinfo {author} {\bibfnamefont {N.}~\bibnamefont
		{Andrei}},\ }\href {\doibase 10.1103/PhysRevLett.45.379} {\bibfield
	{journal} {\bibinfo  {journal} {Phys. Rev. Lett.}\ }\textbf {\bibinfo
		{volume} {45}},\ \bibinfo {pages} {379} (\bibinfo {year} {1980})}\BibitemShut
{NoStop}%
\bibitem [{\citenamefont {Wiegmann}\ and\ \citenamefont
	{Tsvelick}(1983)}]{Wiegmann1983}%
\BibitemOpen
\bibfield  {author} {\bibinfo {author} {\bibfnamefont {P.~B.}\ \bibnamefont
		{Wiegmann}}\ and\ \bibinfo {author} {\bibfnamefont {A.~M.}\ \bibnamefont
		{Tsvelick}},\ }\href {http://stacks.iop.org/0022-3719/16/i=12/a=017}
{\bibfield  {journal} {\bibinfo  {journal} {Journal of Physics C: Solid State
			Physics}\ }\textbf {\bibinfo {volume} {16}},\ \bibinfo {pages} {2281}
	(\bibinfo {year} {1983})}\BibitemShut {NoStop}%
\bibitem [{\citenamefont {Tsvelick}\ and\ \citenamefont
	{Wiegmann}(1983)}]{Tsvelick1983}%
\BibitemOpen
\bibfield  {author} {\bibinfo {author} {\bibfnamefont {A.~M.}\ \bibnamefont
		{Tsvelick}}\ and\ \bibinfo {author} {\bibfnamefont {P.~B.}\ \bibnamefont
		{Wiegmann}},\ }\href {http://stacks.iop.org/0022-3719/16/i=12/a=018}
{\bibfield  {journal} {\bibinfo  {journal} {Journal of Physics C: Solid State
			Physics}\ }\textbf {\bibinfo {volume} {16}},\ \bibinfo {pages} {2321}
	(\bibinfo {year} {1983})}\BibitemShut {NoStop}%
\bibitem [{\citenamefont {Keldysh}(1964)}]{Keldysh1964}%
\BibitemOpen
\bibfield  {author} {\bibinfo {author} {\bibfnamefont {L.~V.}\ \bibnamefont
		{Keldysh}},\ }\href@noop {} {\bibfield  {journal} {\bibinfo  {journal} {Zh.
			Eksp. Teor. Fiz.}\ }\textbf {\bibinfo {volume} {47}},\ \bibinfo {pages}
	{1515} (\bibinfo {year} {1964})}\BibitemShut {NoStop}%
\end{thebibliography}
%

\end{document}